\documentclass[a4paper,11pt]{article}
\usepackage{graphicx} 
\usepackage[utf8]{inputenc}
\usepackage[english]{babel}
\usepackage{authblk}
\usepackage{bbm}
\usepackage{caption}
\usepackage{tikz}
\RequirePackage[T1]{fontenc}
\RequirePackage{flushend}
\RequirePackage{color}
\usepackage{changes}
\usepackage{multirow}
\usepackage{soul}
\RequirePackage{hyperref}
\hypersetup{
	linktocpage,
    colorlinks,
    citecolor=blue,
    filecolor=black,
    linkcolor=blue,
    urlcolor=blue,
}
\usepackage{ragged2e} 
\usepackage{subfigure} 
\usepackage{slashed}\usepackage{float}
\usepackage{amsmath,amssymb,lmodern,setspace}

\title{Isospin breaking corrections in $2\pi$ production in tau decays and $e^+e^-$ annihilation: consequences for the muon $g-2$ and CVC tests}
\author[1]{Gabriel L\'opez Castro~\footnote{gabriel.lopez@cinvestav.mx}}
\author[2]{Alejandro Miranda~\footnote{jmiranda@ifae.es}}
\author[1,3]{Pablo Roig~\footnote{pablo.roig@cinvestav.mx}}

\affil[1]{\normalsize Departamento de F\'isica, Centro de Investigaci\'on y de Estudios Avanzados del IPN,
Apdo. Postal 14-740,07000 Ciudad de M\'exico, M\'exico}
\affil[2]{\normalsize Institut de F\'isica d'Altes Energies (IFAE),
The Barcelona Institute of Science and Technology (BIST),
Campus UAB, E-08193 Bellaterra (Barcelona), Spain}
\affil[3]{\normalsize IFIC, Universitat de València – CSIC, Catedrático José Beltrán 2, E-46980 Paterna, Spain}
\date{}

\begin{document}
\maketitle\abstract{We revisit the isospin-breaking corrections relating the $e^+e^-$ hadronic cross-section and the tau decay spectral function, focusing on the di-pion channel, that gives the dominant contribution to the hadronic vacuum polarization piece of the muon $g-2$. We test different types of electromagnetic and weak form factors and show that both, the Gounaris-Sakurai and a dispersive-based approach, describe accurately $\tau$ lepton and $e^+e^-$ data (less when KLOE measurements are included in the fits) and comply reasonably well with analyticity constraints. From these results we obtain the isospin-breaking contribution to the conserved vector current (CVC) prediction of the ${\rm BR}(\tau \to \pi\pi\nu_{\tau})$ and to the $2\pi$ hadronic vacuum polarization (HVP) contribution to the muon $g-2$, in agreement with previous determinations and with similar precision. Our results abound in the 
\textcolor{black}{utility} of using tau data-based results in the updated data-driven prediction of the muon $g-2$ in the Standard Model.
}

\section{Introduction}\label{sec_Intro}
Isospin is an approximate but very powerful flavor symmetry of hadronic particles and their interactions. Isospin breaking (IB) effects are generally small, since in the Standard Model they are induced by the charge and mass difference of $u$ and $d$ quarks. Isospin symmetry and IB play an important role in precision tests of the SM, like the determination of the $|V_{ud}|$ (and $|V_{us}|$) quark mixing CKM matrix element, where the $u$-$d$ mass difference and electromagnetic radiative corrections are crucial to achieve a precision at the $10^{-4}$ ($10^{-3}$) level  \cite{ParticleDataGroup:2024cfk}. Likewise, IB can play a relevant role in the SM prediction of the muon $g-2$ \cite{Alemany:1997tn, Cirigliano:2001er, Cirigliano:2002pv, Davier:2002dy, Davier:2003pw, Davier:2010fmf, Jegerlehner:2011ti} to understand the current anomaly observed with the experimental average \cite{ParticleDataGroup:2024cfk,Muong-2:2006rrc,Muong-2:2021ojo,Muong-2:2023cdq}, which would reach the $5.1$ standard deviations ($\sigma$) if the White Paper SM prediction \cite{Aoyama:2020ynm,Colangelo:2022jxc} (based on Refs.~\cite{Davier:2017zfy,Keshavarzi:2018mgv,Colangelo:2018mtw,Hoferichter:2019mqg,Davier:2019can,Keshavarzi:2019abf,Kurz:2014wya,FermilabLattice:2017wgj,Budapest-Marseille-Wuppertal:2017okr,RBC:2018dos,Giusti:2019xct,Shintani:2019wai,FermilabLattice:2019ugu,Gerardin:2019rua,Aubin:2019usy,Giusti:2019hkz,Melnikov:2003xd,Masjuan:2017tvw,Colangelo:2017fiz,Hoferichter:2018kwz,Gerardin:2019vio,Bijnens:2019ghy,Colangelo:2019uex,Pauk:2014rta,Danilkin:2016hnh,Jegerlehner:2017gek,Knecht:2018sci,Eichmann:2019bqf,Roig:2019reh,Colangelo:2014qya,Blum:2019ugy,Aoyama:2012wk,Aoyama:2019ryr,Czarnecki:2002nt,Gnendiger:2013pva}) is taken at face value. Specifically, the current evaluations of the dominant two-pion contribution to the hadronic vacuum polarization (HVP) in the data-driven approach obtained from the spectral functions measured in electron-positron annihilation \cite{KLOE:2012anl, BaBar:2012bdw, KLOE-2:2017fda,CMD-2:2003gqi,CMD-2:2005mvb,Aulchenko:2006dxz,CMD-2:2006gxt,Achasov:2006vp,SND:2020nwa,KLOE:2008fmq,KLOE:2010qei,BESIII:2015equ} (before CMD-3 data \cite{CMD-3:2023alj, CMD-3:2023rfe}) and tau lepton decay \cite{ALEPH:2005qgp,Belle:2008xpe,CLEO:1999dln,OPAL:1998rrm}, differ by $\sim[2,3]\,\sigma$ after including state-of-the-art IB corrections to tau data \cite{Davier:2010fmf, Flores-Baez:2006yiq, Flores-Baez:2007vnd, Miranda:2020wdg, Masjuan:2023qsp, Davier:2023fpl} (see also \cite{Maltman:2005yk,Maltman:2005qq,Davier:2010nc,Davier:2013sfa,Bruno:2018ono,Narison:2023srj,Esparza-Arellano:2023dps}). This discrepancy and the advent of more precise measurements of the pion form factor in electron-positron collisions, have prevented in recent years the use of tau decay data to obtain this dominant contribution to the muon $g-2$ HVP part.

Nowadays, it is clear that to firmly establish the muon $g-2$ anomaly, it becomes necessary to understand the differences in the cross sections for $\pi\pi$ production in electron-positron collisions as measured by different experiments. Thus, while the HVP prediction based on Babar $2\pi$ data~\cite{BaBar:2012bdw}  implies a SM prediction that differs from the measured muon $g-2$ by $3.5\,\sigma$, that based on KLOE results~\cite{KLOE:2012anl, KLOE-2:2017fda} differs by {\color{black}{$4.5\sigma \ 
$ in the whole 
data set region}}~\cite{Davier:2023fpl}. Recent data on the cross section for this channel measured by CMD-3 \cite{CMD-3:2023alj, CMD-3:2023rfe} gives a prediction which departs from the measured muon $g-2$ value by only $1.0\,\sigma$, which is further supported by the recent mixed (lattice QCD+data-driven) evaluation of Ref.~\cite{Boccaletti:2024guq}. Understanding the origin of these differences is crucial before one can claim a significant anomaly that establishes presence of New Physics in the muon $g-2$.

Long ago \cite{Alemany:1997tn}, it was suggested that one can also use precise $\tau \to\pi\pi\nu_{\tau}$ data to predict the dominant HVP contribution provided IB corrections are properly taken into account to relate the $2\pi$ spectral functions of tau decays and $e^+e^-$ data. Using state-of-the-art IB corrections \cite{Davier:2010fmf, Flores-Baez:2006yiq, Flores-Baez:2007vnd, Miranda:2020wdg, Masjuan:2023qsp, Davier:2023fpl}, the prediction based on $\tau$ decays gives a result that is closer to the measured muon $g-2$ (at $[2,3]\,\sigma$) and in better agreement with Babar and CMD-3 data \cite{BaBar:2012bdw,CMD-3:2023alj,CMD-3:2023rfe}. Furthermore, these results are also in good accord with the most precise prediction reported by Lattice QCD calculations~\cite{Borsanyi:2020mff} (see also the recent \cite{Djukanovic:2024cmq} and \cite{Bazavov:2024eou} results), that departs only $1.9\,\sigma$ from the measured $a_\mu=(g_\mu-2)/2$. 

A different and {\color{black}{complementary}} test of the IB corrections in the $2\pi$ spectral function is provided by the branching ratio of the $\tau \to \pi \pi \nu_{\tau}$ decay \cite{Davier:2010fmf}, whose measurement has achieved so far a precision of 0.35 \% \cite{ParticleDataGroup:2024cfk}. While the HVP contribution to the muon $g-2$ is sensitive mainly to the low energy and $\rho(770)$ resonance region of the  $2\pi$ spectral function, the tau lepton branching ratio probes more precisely the rho-resonance region. IB corrections have to be applied to tau data to predict the HVP muon $g-2$ and, conversely, IB needs to be added to $e^+e^-$ data to predict the two-pion rate of tau decays. This dual procedure provides a consistency check of IB corrections~\footnote{It must be mentioned that a difference between the $e^+e^-$ and $\tau$ data-based predictions could also be due to new physics affecting only the latter, see Refs.~\cite{Miranda:2018cpf,Cirigliano:2018dyk,Gonzalez-Solis:2020jlh,Cirigliano:2021yto} for its interpretation within a low-energy effective field theory (EFT) and its relation to the SMEFT \cite{Buchmuller:1985jz,Grzadkowski:2010es}. It is also interesting to recall the implications of using tau instead of $e^+e^-$ data for the HVP part of $\alpha_{QED}(M_Z)$ in the global electroweak fit, as pointed out in Ref.~\cite{Passera:2008jk}.}. 

In this paper we use $e^+e^- \to \pi^+\pi^-$ data to test the IB corrections in the prediction of the $\tau \to \pi\pi\nu_{\tau}$ branching fraction. Specifically, we focus on the different parametrizations of the electromagnetic and weak form factors of the pion and their impact on the muon $g-2$ and conserved vector current (CVC) tests. We include in our analysis the most precise measurements: the recent data reported by the CMD-3 collaboration \cite{CMD-3:2023alj, CMD-3:2023rfe}, as well as previous measurements from BaBar \cite{BaBar:2012bdw} and KLOE \cite{KLOE:2012anl, KLOE-2:2017fda}, and Belle data \cite{Belle:2008xpe} for the tau decays~\footnote{These are the most precise for the spectrum, while ALEPH \cite{ALEPH:2005qgp} achieved the most accurate branching ratio determination.}. In addition to the usual form factor shapes employed by experiments, we also include in our study a dispersive pion form factor, including the effects of excited rho--meson resonances and the $\rho-\omega-\phi$ mixing, as well as accounting for inelastic effects through a conformal polynomial. Good description of the data is achieved, with analyticity tests favouring the Gounaris-Sakurai \cite{Gounaris:1968mw} and dispersive form factors. From these, we obtain the IB corrections relating the $e^+e^-$ cross-section and the tau decay spectral function and also the IB correction needed to use the tau measurements for the $\pi\pi$ contribution to the HVP part of the muon $g-2$. Our results agree with earlier determinations with comparable uncertainties.

This article is structured as follows. In section \ref{sec_IBCorrs} we review briefly the IB corrections modifying the CVC relation between the tau decay spectral function and the $e^+e^-$ cross-section, and their impact on the $a_\mu^\mathrm{HVP}$ evaluation. In section \ref{sec_FFs} we explain the different form factor parametrizations that we consider and the associated IB corrections. Our results are described in section \ref{sec_Res} and our conclusions drawn in section \ref{sec_Concl}. We confirm the agreement between different evaluations of the IB corrections. This, together with the consistency of tau decay data, {\color{black}{makes relevant and helpful the use of tau-data to predict}} the two-pion contribution to $a_\mu^\mathrm{HVP}$ in updating the White Paper \cite{Aoyama:2020ynm} and the SM prediction for $a_\mu$.

\section{IB corrections to the CVC predictions of \texorpdfstring{$a_{\mu}$}{Lg} and \texorpdfstring{$2\pi$}{Lg} tau decay}\label{sec_IBCorrs}
This section provides a brief summary of the formulas relating the two-pion production in tau decay and electron-positron annihilation including IB effects.

In the data driven approach, at the leading order (LO), the HVP contribution  to the muon magnetic anomaly $a_{\mu}=(g_\mu-2)/2$ 
is given by~\cite{Bouchiat:1961lbg,Durand:1962zzb,Brodsky:1967sr,Gourdin:1969dm}  
\begin{equation}\label{eq.amu_dispersive}
    a_\mu^{\text{HVP, LO}}=\frac{1}{4\pi^3}\int_{s_{\text{thr}}}^{\infty}ds \,K(s)\,\sigma_{e^+e^-\to\text{hadrons}(+\gamma)}^{0}(s),
\end{equation}

\noindent
where $\sigma^{0}_{e^+e^-\to \text{hadrons}(+\gamma)}(s)$ is the bare hadronic cross-section (opening with the $\pi^0\gamma$ channel, at $s_\mathrm{thr}=m_{\pi^0}^2$) with vacuum polarization (VP) effects removed~\cite{Eidelman:1995ny}. $K(s)$ is a smooth kernel concentrated at low energies~\cite{Brodsky:1967sr,Lautrup:1968tdb},
\begin{equation}\begin{split}
    &K(x)=\frac{x^2}{2}(2-x^2)+\frac{(1+x^2)(1+x)^2}{x^2}\left(\log(1+x)-x+\frac{x^2}{2}\right)+\frac{(1+x)}{(1-x)}x^2\log(x),
\end{split}\end{equation}
where $x\equiv x(s)=\frac{\textstyle 1-\beta_\mu(s)}{\textstyle 1+\beta_\mu(s)}$ and $\beta_\mu(s)=\sqrt{1-4m_\mu^2/s}$. 
This kernel enhances the contributions of hadronic channels that open at lower center of mass energies, $\sqrt{s}$, as well as resonances in the cross section with masses below 1 GeV.

The isotopic properties of the electromagnetic and charged weak hadronic vector currents allow to relate the production of an even number of pions in tau decays and in electron-positron collisions. Long ago, Alemany, Davier and H\"ocker \cite{Alemany:1997tn} proposed that semileptonic tau decays into two and four pions could be used as an additional input to  evaluate Eq.~(\ref{eq.amu_dispersive}), provided the needed IB effects were accounted for. In the following, we will focus on the predictions of the dominant ($\pi\pi$) contributions to $ a_\mu^{\text{HVP, LO}}$ and to the branching ratio of tau decays that stem from the IB corrected spectral functions.

After including IB corrections, the {\it bare} hadronic $e^+e^-$ cross-section  $\sigma_{\pi\pi(\gamma)}^0$  is related to the {\it observed} invariant-mass distribution of two pions in $\tau$ decays   $d\Gamma_{\pi\pi[\gamma]}/ds$ via~\cite{Cirigliano:2001er,Cirigliano:2002pv}
\begin{equation}\label{eq:pipi_cross_section}
\left.\sigma^0_{\pi\pi(\gamma)}\right\vert_\text{CVC}=\left[\frac{K_\sigma(s)}{K_\Gamma(s)}\frac{d\Gamma_{\pi\pi[\gamma]}}{ds}\right]\times\frac{R_{\text{IB}}(s)}{S^{\pi\pi}_{\text{EW}}},
\end{equation}
where
\begin{equation}\begin{split}
    K_\Gamma(s)&=\frac{G_F^2\vert V_{ud}\vert^2 m_\tau^3}{384\pi^3}\left(1-\frac{s}{m_\tau^2}\right)^2\left(1+\frac{2s}{m_\tau^2}\right),\\[1ex]
    K_{\sigma}(s)&=\frac{\pi \alpha^2}{3s} \ .
\end{split}\end{equation}
The IB corrections are encoded in the product of  several energy-dependent factors
\begin{equation}\label{RIB}
    R_{\text{IB}}(s)=\frac{\text{FSR}(s)}{G_{\text{EM}}(s)}\frac{\beta^3_{\pi^{+}\pi^{-}}(s)}{\beta^3_{\pi^+\pi^0}(s)}\left\vert\frac{F_V(s)}{f_{+}(s)}\right\vert^2,
\end{equation}
where $s\beta_{PP'}(s)=\sqrt{[s-(m_P-m_{P'})^2][s-(m_P+m_{P'})^2]}$. 
Eq. (\ref{eq:pipi_cross_section}) includes the universal short-distance electroweak radiative corrections in $S_{\text{EW}}$~\cite{Sirlin:1974ni,Sirlin:1977sv,Sirlin:1981ie,Marciano:1985pd,Marciano:1988vm,Marciano:1993sh,Braaten:1990ef,Erler:2002mv,Cirigliano:2023fnz}. Final-State-Radiation corrections to the $\pi^+\pi^-$ channel~\cite{Schwinger:1989ix,Drees:1990te} originate the FSR(s) factor~\footnote{FSR was absent in $R_{\text{IB}}$ before Ref.~\cite{Davier:2010fmf}.}, whereas the $G_{\text{EM}}(s)$ factor incorporates the long-distance QED corrections to the $\tau^-\to\pi^-\pi^0\nu_\tau$ decay with virtual plus real photon radiation, currently taken at $O(\alpha)$~\footnote{This factor was the focus of Refs.~\cite{Cirigliano:2001er, Cirigliano:2002pv,Flores-Baez:2006yiq,Flores-Tlalpa:2006snz,Miranda:2020wdg}, obtained either using vector meson dominance or Resonance Chiral Theory~\cite{Ecker:1988te,Ecker:1989yg,Cirigliano:2006hb,Kampf:2011ty,Roig:2013baa}. After thorough discussions and comparisons with the Orsay group we find agreement in this correction, as shown in Table \ref{tab:IB_BR}.}. The phase space correction due to the $\pi^\pm - \pi^0$ mass difference gives rise to the $\beta^{3}_{\pi^-\pi^+}(s)/\beta^{3}_{\pi^-\pi^0}(s)$ factor. The last factor in Eq. (\ref{RIB}) amounts to the ratio between the electromagnetic ($F_V(s)$) and the weak ($f_{+}(s)$) pion form factors, including one of the leading IB effects, the $\rho^0-\omega$ mixing contribution.  This factor, in which we focus in this paper, is model-dependent and incorporates IB in the shapes of the form factors. Below $\sqrt{s}=1$ GeV, this source of IB stems from the mass difference of the charged and neutral $\pi$ and $\rho(770)$ mesons, the difference of decay widths of the latter, as well as  the rho-omega mixing. 

The IB corrections to $a_\mu^\text{HVP, LO}$ using $\tau$ data in the dominant $\pi\pi$ channel~\cite{GomezDumm:2013sib,Gonzalez-Solis:2019iod} can be assessed by the following quantity~\cite{Davier:2010fmf}
\begin{equation}\begin{split}\label{eq:Delta_a_mu}
    \Delta a_\mu^{\text{HVP, LO}}[\pi\pi,\tau]&=\frac{1}{4\pi^3}\int_{4m_\pi^2}^{m_\tau^2}ds\, K(s)\left[\frac{K_\sigma(s)}{K_\Gamma(s)}\frac{d\Gamma_{\pi\pi[\gamma]}}{ds}\right]\left(\frac{R_{\text{IB}}(s)}{S^{\pi\pi}_{\text{EW}}}-1\right)\\[1ex]
    &=\frac{\alpha^2m_\tau^2}{6\pi^2\vert V_{ud}\vert^2}\frac{\mathcal{B}_{\pi\pi}}{\mathcal{B}_e}\int_{4m_\pi^2}^{m_\tau^2}ds\frac{K(s)}{s}\left(1-\frac{s}{m_\tau^2}\right)^{-2}\left(1+\frac{2s}{m_\tau^2}\right)^{-1}\frac{1}{N_{\pi\pi}}\frac{dN_{\pi\pi}}{ds}\\[1ex]
    &\qquad\times\left(\frac{R_{\text{IB}}(s)}{S_{\text{EW}}}-1\right),
\end{split}\end{equation}
which measures the difference between $\sigma_{\pi\pi(\gamma)}^{0}$ and the naive Conserved Vector Current (CVC) approximation, with $S_{\text{EW}}\equiv S_\mathrm{EW}^{\pi\pi}/S_\mathrm{EW}^e=1$ and $R_{\text{IB}}(s)=1$. We note that this definition of $S_\mathrm{EW}$ takes into account that the di-pion tau decay spectrum is normalized to the electron tau decay mode~\cite{Belle:2008xpe}.

Another important {\color{black}{complementary}} cross-check of the IB corrections is provided by the  prediction of the branching ratio $\mathcal{B}_{\pi\pi^0}=\Gamma(\tau\to\pi\pi\nu_\tau(\gamma))/\Gamma_\tau$, based on CVC. Taking into account the IB corrections, it can be evaluated from 
the $e^+e^-\to\pi^+\pi^-(\gamma)$ cross section data~\footnote{\color{black}IB by the mass difference of the $u$ and $d$ quarks ($m_u\neq m_d$) generates an isoscalar $I=0$ part, which leads to the $\rho^0-\omega$ mixing.}. The corresponding branching fraction is then given by~\cite{Davier:2010fmf,Jegerlehner:2017gek} 
\begin{equation}\label{bcvc}
    \mathcal{B}_{\pi\pi^0}^\text{CVC}=\mathcal{B}_e\int_{\textcolor{black}{4m_\pi^2}}^{m_\tau^2}ds\,\sigma^{0}_{\pi^+\pi^-(\gamma)}(s)\mathcal{N}(s)\frac{S_\text{EW}}{R_{\text{IB}}(s)},
\end{equation}
with
\begin{equation}
    \mathcal{N}(s)=\frac{3\vert V_{\text{ud}}\vert^2}{2\pi\,\alpha^2\,m_\tau^2}s\left(1-\frac{s}{m_\tau^2}\right)^2\left(1+\frac{2s}{m_\tau^2}\right).
\end{equation}
In analogy with Eq. (\ref{eq:Delta_a_mu}), the IB corrections to $\mathcal{B}_{\pi\pi^0}^\text{CVC}$ can be evaluated from
\begin{equation}\label{ibtpbr}
     \Delta\mathcal{B}_{\pi\pi^0}^\text{CVC}=\mathcal{B}_e\int_{4m_\pi^2}^{m_\tau^2}ds\,\sigma^{0}_{\pi^+\pi^-(\gamma)}(s)\mathcal{N}(s)\left(\frac{S_\text{EW}}{R_{\text{IB}}(s)}-1\right),
\end{equation}
where the numerical values of $\mathcal{B}_e\equiv\mathcal{B}(\tau^-\to e^-\bar{\nu}_e\nu_\tau)$, $m_\tau$ and $|V_{\text{ud}}|$ are taken from the PDG~\cite{ParticleDataGroup:2024cfk}.  

We would like to emphasize that while the $K(s)$ kernel in Eq. (\ref{eq:Delta_a_mu}) tests the IB corrections and the weak pion form factor at low energy and close to the $\rho(770)$ resonance region, Eq. (\ref{bcvc}) probes the IB corrections and the pion electromagnetic form factor close to the resonance region. Therefore, the branching fraction of two-pion tau decays and the HVP contribution to $a_{\mu}$ provide {\color{black}{a complementary}} test of IB corrections in this channel.

\section{Form factor parametrizations and IB corrections}\label{sec_FFs}
The electromagnetic, $F_V(s)$, and weak, $f_+(s)$, form factors of the pion introduced in the previous section are defined, respectively, in the following equations
\begin{eqnarray}
\langle \pi^+(p_+)\pi^-(p_-)|j_{\mu}^{\rm em}| 0 \rangle &=& (p_++p_-)_{\mu} F_{V}(s)\ , \\
\langle \pi^+(p_+)\pi^0(p_0)|j_{\mu}^{\rm weak}| 0 \rangle  &=&\sqrt{2}\Big[ (p_++p_0)_{\mu} f_+(s)+ (p_+-p_0)_{\mu} f_-(s) \Big]\ ,
\end{eqnarray}  
where $s$ denotes the squared invariant-mass of the two-pion system. In the limit of exact isospin symmetry, $f_+(s)=F_V(s)$, $f_-(s)=0$ and, in particular, $f_+(0)=F_V(0)=1$ \footnote{IB induces a small correction to the weak charge of the pion which is of second order in the $u-d$ quark mass difference owing to the Ademollo-Gatto theorem \cite{Ademollo:1964sr}, namely $f_+(0)=1-O((m_d-m_u)^2)$, which we will neglect in the following.}.

 Including the first order IB effects arising from the $\rho-\omega$ and $\rho-\phi$ mixing,  the electromagnetic form factor can be parametrized in the Vector-Meson Dominance (VMD) model as follows~\cite{BaBar:2012bdw}:
\begin{equation}\label{eq:KSGS}\begin{split}
    F_V(s)&=
    \bigg[BW_{\rho}(s,m_\rho,\Gamma_\rho)\bigg(1+\delta_{\rho\omega}\frac{s}{m_{\omega}^2}\,BW_{\omega}^{NR}(s,m_\omega,\Gamma_\omega)+\delta_{\rho\phi}\frac{s}{m_{\phi}^2}\,BW_{\phi}^{NR}(s,m_\phi,\Gamma_\phi)\bigg)\\[1ex]
   &\quad+c_{\rho^\prime}\,BW_{\rho^\prime}(s,m_{\rho^\prime},\Gamma_{\rho^\prime})+c_{\rho^{\prime\prime}}\,BW_{\rho^{\prime\prime}}(s,m_{\rho^{\prime\prime}},\Gamma_{\rho^{\prime\prime}})\bigg]/(1+c_{\rho^\prime}+c_{\rho^{\prime\prime}}),
\end{split}\end{equation}
where $m_{\rho,\rho'\cdots}$ ($\Gamma_{\rho,\rho',\cdots}$) are the masses (widths) of the neutral isotriplet vector mesons. A similar expression holds for the weak form factor $f_+(s)$, with charged rho mesons parameters replacing the neutral ones and vanishing $\delta_{\rho\omega/\rho\phi}$ mixing parameters.

The superscript NR in Eq.~(\ref{eq:KSGS}) refers to the isoscalar $\omega/\phi$ narrow resonances that mix with the isovector $\rho(770)$ resonance through the mixing parameter $\delta_{\rho\omega,\rho\phi}$, which quantifies IB in these mixings. These contributions are usually parametrized by the simplest Breit-Wigner (BW) forms
\begin{equation}
   BW^{NR}_V(s,m_V,\Gamma_V)=\frac{m_V^2}{m_V^2-s-im_V\Gamma_V} \, ,
\end{equation}
where $m_V$ and $\Gamma_V$ are the mass and on-shell decay width of the resonance.

The experimental groups typically perform their analyses of the resonant shapes in terms of two model-dependent parametrizations. In the Gounaris-Sakurai (GS) model~\cite{Gounaris:1968mw}, the rho resonance is described by the following resonant shape

\begin{equation}
    BW^{GS}(s,m,\Gamma)=\frac{m^2\left[1+d(m)\Gamma/m\right]}{m^2-s+f(s,m,\Gamma)-im\Gamma(s,m,\Gamma)},
\end{equation}
with the energy-dependent width given as 
\begin{equation}
    m\Gamma(s,m,\Gamma)=\Gamma\,\frac{s}{m}\left(\frac{\beta_{\pi\pi}(s)}{\beta_{\pi\pi}(m^2)}\right)^3\theta(s-4m_\pi^2),
\end{equation}
where $\beta_{\pi\pi}(s)=\sqrt{1-4m_\pi^2/s}$. The expressions for $f(s,m,\Gamma)$ and the $d$-parameter can be found in Ref. \cite{Gounaris:1968mw}. The on-shell width corresponds to $\Gamma=\Gamma(m^2,m,\Gamma)$.

Another parametrization commonly employed by experiments to describe the isovector resonances is the K\"uhn-Santamar\'ia (KS) model~\cite{Kuhn:1990ad}. In this case, the resonant shape reads:
\begin{equation}
   BW^{KS}(s,m,\Gamma)=\frac{m^2}{m^2-s-im\Gamma(s,m,\Gamma)}.
\end{equation}

 Yet another description of the electromagnetic form factor is provided by the Guerrero-Pich (GP) resummation of the chiral loop functions~\cite{Guerrero:1997ku}, which includes only one isovector resonance for energies below 1 GeV. According to it, the expressions that include all the relevant sources of IB at first order and the dominant $\rho$ exchange are given by~\cite{Cirigliano:2001er}:

\begin{equation}\label{eq:FFGP}
    F_V(s)=\frac{m_{\rho^0}^2}{m_{\rho^0}^2-s-i m_{\rho^0}\Gamma_{\rho^0}(s)}\,\left\lbrace\exp{\left[2\tilde{H}_{\pi^+\pi^-}(s)+\tilde{H}_{K^+K^-}(s)\right]}+\delta_{\rho\omega}\frac{s}{m_\omega^2-s-i\,m_\omega\,\Gamma_\omega}\right\rbrace,
\end{equation}

\begin{equation}\label{eq:FFGP2}
    f_{+}(s)=\frac{m_{\rho^+}^2}{m_{\rho^+}^2-s-i m_{\rho^+}\Gamma_{\rho^+}(s)}\,\exp{\left\lbrace2\tilde{H}_{\pi^+\pi^0}(s)+\tilde{H}_{K^+K^0}(s)\right\rbrace}+f^{\text{elm}}_{\text{local}}+\cdots,
\end{equation}
where the $\tilde{H}_{PP'}(s)$ functions (from the real part of the chiral loops) can be found e.g. in Ref. \cite{Cirigliano:2001er};  $f^{\text{elm}}_{\text{local}}\sim \mathcal{O}(10^{-4})$ represents the structure-dependent electromagnetic corrections to the low-$s$ part of the spectrum~\cite{Cirigliano:2001er,Cirigliano:2002pv}

\begin{equation}
    f^{\text{elm}}_{\text{local}}=\frac{\alpha}{4\pi}\left[-\frac{3}{2}-\frac{1}{2}\log\frac{m_\tau^2}{\mu^2}-\log\frac{m_\pi^2}{\mu^2}+2\log\frac{m_\tau^2}{m_\rho^2}-(4\pi)^2\left(-2K_{12}^r(\mu)+\frac{2}{3}X_1+\frac{1}{2}\tilde{X}_6^r(\mu)\right)\right] \ .
\end{equation}
The energy-dependent width in this case is given by \cite{Guerrero:1997ku}
\begin{equation}\label{eq:width_GP}
    \Gamma_\rho(s)=\frac{m_\rho\,s}{96\pi f_\pi^2}\left[\beta^3_{\pi\pi}(s)\theta(s-4m_\pi^2)+\frac{1}{2}\beta_{KK}^3(s)\theta(s-4m_K^2)\right].
\end{equation}
From the Guerrero-Pich (GP) expression, Eq. (\ref{eq:width_GP}), we get

\begin{equation}\begin{split}
    m_\rho\,\Gamma_\rho(s)&=\frac{m_\rho^3\,\beta_{\pi\pi}^3(m_\rho^2)}{96\pi f_\pi^2}\frac{s}{m_\rho}\left[\frac{\beta^3_{\pi\pi}(s)}{\beta_{\pi\pi}^3(m_\rho^2)}\theta(s-4m_\pi^2)+\frac{1}{2}\frac{\beta_{KK}^3(s)}{\beta_{\pi\pi}^3(m_\rho^2)}\theta(s-4m_K^2)\right]\\[1ex]
    &\simeq \Gamma^{\text{GP}}_\rho(m_\rho)\frac{s}{m_\rho}\frac{\beta^3_{\pi\pi}(s)}{\beta_{\pi\pi}^3(m_\rho^2)}\theta(s-4m_\pi^2). \label{GPdw}
\end{split}\end{equation}
We note that -neglecting the Kaon contribution for GP- the energy-dependence of the GS, KS and GP widths is the same. 
The second line in Eq. (\ref{GPdw}) represents a very good approximation below the $1\,\mathrm{GeV}$ region. We can define the following on-shell width of the $\rho(770)$ 
\begin{equation}\label{eq:rho_width_GP}
    \Gamma^{\text{GP}}_\rho(m_\rho)=\frac{m_\rho^3\,\beta_{\pi\pi}^3(m_\rho^2)}{96\pi f_\pi^2}.
\end{equation}
This rho meson width is a prediction of the GP model (once $m_{\rho}$ is known) and not a free parameter to be fitted to data, unlike in the GS and KS models. Besides, $\Gamma_\rho$ is slightly different whether we use the neutral or charged $\rho(770)$ vector meson, depending on IB on the pion and rho masses.

We will also use a dispersive form factor\footnote{$\rho-\omega-\phi$ mixing in this dispersive construction is discussed below, from Eq.~(\ref{eq:disp02}) on.}, whose phase-shift seed will be constructed from~(see \cite{GomezDumm:2013sib,Gonzalez-Solis:2019iod}~\footnote{The $\mathrm{Re}\,A_{P(P)}(s)$ functions are trivially related to the to the $\tilde{H}_{PP}(s)$ functions in Eqs.~(\ref{eq:FFGP}) and (\ref{eq:FFGP2}).})
\begin{eqnarray}\label{eq:FVseed}
F_{V}(s)^{\mathrm{seed}} & = & \frac{m_\rho^2+s(\gamma\mathrm{e}^{i\phi_1}+\delta\mathrm{e}^{i\phi_2})}{m_\rho^2-s-im_\rho\Gamma_\rho(s)}\mathrm{exp}\left\lbrace\mathrm{Re}\left[-\frac{s}{96\pi^2F_\pi^2}\left(A_\pi(s)+\frac{1}{2}A_K(s)\right)\right]\right\rbrace\nonumber\\[1ex]
&-&\gamma\frac{s\mathrm{e}^{i\phi_1}}{m_{\rho^\prime}^2-s-im_{\rho^\prime}\Gamma_{\rho^\prime}(s)}\mathrm{exp}\left\lbrace\frac{s\Gamma_{\rho^\prime}(m_{\rho^\prime}^2)}{\pi m_{\rho^\prime}^3\beta_{\pi\pi}^3(m_{\rho^\prime}^2)}\mathrm{Re}A_\pi(s)\right\rbrace
\nonumber\\[1ex]
&-&\delta\frac{s\mathrm{e}^{i\phi_2}}{m_{\rho^{\prime\prime}}^2-s-im_{\rho^{\prime\prime}}\Gamma_{\rho^{\prime\prime}}(s)}\mathrm{exp}\left\lbrace\frac{s\Gamma_{\rho^{\prime\prime}}(m_{\rho^{\prime\prime}}^2)}{\pi m_{\rho^{\prime\prime}}^3\beta_{\pi\pi}^3(m_{\rho^{\prime\prime}}^2)}\mathrm{Re}A_\pi(s)\right\rbrace\,,
\end{eqnarray}
where $\Gamma_\rho(s)=\Gamma_\rho^{\mathrm{GP}}(s)$ and we use
\begin{equation}
\Gamma_{\rho^\prime,\rho^{\prime\prime}}(s)=\Gamma_{\rho^\prime,\rho^{\prime\prime}}\frac{s}{m_{\rho^\prime,\rho^{\prime\prime}}^2}\left[\frac{\beta_{\pi\pi}^3(s)}{\beta_{\pi\pi}^3(m_{\rho^\prime,\rho^{\prime\prime}}^2)}\theta(s-4m_\pi^2)+\frac{1}{2}\frac{\beta_{KK}^3(s)}{\beta_{KK}^3(m_{\rho^\prime,\rho^{\prime\prime}}^2)}\theta(s-4m_K^2)\right]\,
\end{equation}
to model the energy-dependence of the $\rho'=\rho(1450),\rho''=\rho(1700)$ excitations. 

From $F_{V}(s)^{\mathrm{seed}}$ we define $\tan\delta^1_1(s)=\frac{\mathrm{Im}F_V(s)^{\mathrm{seed}}}{\mathrm{Re}F_V(s)^{\mathrm{seed}}}$, which enters the dispersion relation (see below).
For the dispersive parametrization of the pion form factor we follow Ref.~\cite{Colangelo:2018mtw}  (see also the recent Refs.~\cite{RuizArriola:2024gwb,Kirk:2024oyl}), where the pion VFF is given by a product of three functions,
\begin{equation}\label{eq:disp02}
    F_\pi^V(s)=\Omega_1^1(s)G_\omega(s)G_\text{in}^N(s),
\end{equation}
where
\begin{equation}\label{eq_Omnes}
    \Omega_1^1(s)=\exp{\left\lbrace\frac{s}{\pi}\int_{4m_\pi^2}^\infty ds^\prime\frac{\delta_1^1(s^\prime)}{s^\prime(s^\prime-s)} \right\rbrace}
\end{equation}
is the usual Omn\`es function~\cite{Omnes:1958hv} with $\delta_1^1(s)$ the isospin $I=1$ elastic $\pi\pi$ phase shift in the isospin-symmetric limit. We will use $\tan\delta^1_1(s)=\frac{\mathrm{Im}F_V(s)^{\mathrm{seed}}}{\mathrm{Re}F_V(s)^{\mathrm{seed}}}$ in the Omn\`es function (\ref{eq_Omnes}).

The most important IB effect, the $\rho-\omega$ mixing, is accounted for by the $G_\omega(s)$ function, which is enhanced by the small mass difference between the $\rho$ and $\omega$ resonances. The full parametrization that implements the correct threshold behavior of the discontinuity~\cite{Leutwyler:2002hm}, is given by
\begin{equation}
    G_\omega(s)=1+\frac{s}{\pi}\int_{9m_\pi^2}^\infty ds^\prime \frac{\mathrm{Im}\,g_\omega(s^\prime)}{s^\prime(s^\prime-s)}\left(\frac{1-\frac{9m_\pi^2}{s^\prime}}{1-\frac{9m_\pi^2}{m_\omega^2}}\right)^4,
\end{equation}
with (here we stick to the notation for the isospin mixing used in Ref.~\cite{Colangelo:2018mtw}; we note that $\epsilon_{\omega}\approx \delta_{\rho\omega}$)
\begin{equation}
    g_\omega(s)=1+\epsilon_\omega \frac{s}{(m_\omega-\frac{i}{2}\Gamma_\omega)^2-s}.
\end{equation}
As was already pointed out in Ref.~\cite{Colangelo:2018mtw}~\footnote{
See also Refs.~\cite{Colangelo:2022prz,Hoferichter:2023sli,Hoferichter:2023bjm,Stoffer:2023gba}, showing that the IB corrections due to radiative channels produce a small $\rho-\omega$ mixing phase that is largely cancelled by the $3\pi$ channel, and including the related effect of CMD-3 data on this phase.}
, it is possible to replace $G_\omega(s)$ by $g_\omega(s)$ with almost no significant impact in the energy range of interest, thanks to its strong localization around the $\omega$ resonance. To take into account the $\rho-\phi$ mixing when the CMD-3 data is used, we replace $g_\omega(s)$ by 
\begin{equation}
    g_{\omega\phi}(s)=1+\epsilon_\omega \frac{s}{(m_\omega-\frac{i}{2}\Gamma_\omega)^2-s}+\epsilon_\phi \frac{s}{(m_\phi-\frac{i}{2}\Gamma_\phi)^2-s}.
\end{equation}
The last factor in Eq.~(\ref{eq:disp02}),  $G^N_\mathrm{in}(s)$, is analytic in the complex $s$-plane with a cut on the real axis starting at $s=16m_\pi^2$~\footnote{Although the inelasticities below $s=(m_\omega+m_{\pi^0})^2$ are extremely weak~\cite{Colangelo:2018mtw}, we consider the $4\pi$ ones in this analysis.}. This function is described by a conformal polynomial~\cite{Colangelo:2018mtw} 
\begin{equation}
    G_\mathrm{in}^N(s)=1+\sum^N_{k=1}c_k\left(z^k(s)-z^k(0)\right),
\end{equation}
where the conformal variable is 
\begin{equation}
    z(s)=\frac{\sqrt{s_\mathrm{in}-s_c}-\sqrt{s_\mathrm{in}-s}}{\sqrt{s_\mathrm{in}-s_c}+\sqrt{s_\mathrm{in}-s}}.
\end{equation}
The conformal polynomial generates a branch-cut singularity at $s=s_\mathrm{in}$ and conserves the charge since $G_\mathrm{in}^N(0)=1$. In order to have the $P$-wave behavior at the inelastic threshold, i.e. $G_\mathrm{in}^N(s)\propto (s_\mathrm{in}-s)^{3/2}$ close to $s_\mathrm{in}$, there is an additional restriction~\cite{Colangelo:2018mtw}
\begin{equation}\label{eq:in_rest}
    c_1=-\sum_{k=2}^Nk\,c_k.
\end{equation}
The point $s_c$ is a free parameter in the conformal polynomial, we take $s_c=-1\,\mathrm{GeV}^2$, characteristic of the onset of inelastic effects in this channel. The charge radius of the pion, $\langle r^2\rangle_V^\pi$, is defined by the derivative of the form factor
\begin{equation}\label{eq:radius2}
    \langle r^2\rangle_V^\pi=6\left.\frac{dF_\pi^V(s)}{ds}\right\vert_{s=0}=\frac{6}{\pi}\int_{4m_\pi^2}^\infty ds\frac{\mathrm{Im} F_\pi^V(s)}{s^2},
\end{equation}
that is evaluated via a dispersion relation.

\section{Results}\label{sec_Res}
The branching fraction of the $\tau^-\to\pi^-\pi^0\nu_\tau$ decays can be predicted from the $e^+e^- \to \pi^+\pi^-$ cross section data using the CVC hypothesis and including IB corrections (see for example, \cite{Gilman:1984ry,Sobie:1995kp,Bernicha:1995rh}). Using state-of-the-art IB corrections, as done in \cite{Davier:2010fmf},  a deficit of $2.1\,\sigma$ has been observed with respect to the measured rate (based on 2010 world averages for both sets of data). In view of the recent CMD-3 measurement \cite{CMD-3:2023alj,CMD-3:2023rfe}, we re-evaluate the IB-corrected rate of di-pion tau decays using the different pion form factor models to estimate the effects of the last factor in Eq.~(\ref{RIB}). Conversely, the two-pion tau decay data can be used to predict the dominant HVP contribution to the muon anomalous magnetic moment provided IB corrections are taken into account. The effects of IB in the CVC-based prediction of the tau branching fraction (HVP contribution to $a_{\mu}$) are quantified using Eq.~(\ref{ibtpbr}) (respectively, Eq.~(\ref{eq:Delta_a_mu})). Let us emphasize that our method to evaluate the data driven integrals defined in Section \ref{sec_IBCorrs} is based on the fits to experimental data, in contrast to Ref. \cite{Davier:2010fmf} where the integration is performed directly using the binned data points.

We start by comparing the IB contributions to $a_\mu^{\mathrm{HVP,\,LO}}$ and $\mathcal{B}(\tau^-\to\pi^-\pi^0\nu_\tau)$ using the GP model \cite{Guerrero:1997ku} form factors with the ones obtained in Ref. \cite{Davier:2010fmf}. The results are summarized in Table \ref{tab:IB_BR}. In our evaluations we use $S_\text{EW}=1.0233(3)$ \cite{Alemany:1997tn}, {$|V_{ud}|=0.97367(32)$ \cite{ParticleDataGroup:2024cfk}}, $G_{\text{EM}}(s)$ from Resonance Chiral Theory~\footnote{This approach has also been employed successfully in the computation of other hadronic contributions to the muon $g-2$, both in the HVP \cite{Miranda:2020wdg, Wang:2023njt,Masjuan:2023qsp, Qin:2024ulb} and hadronic light-by-light \cite{Roig:2014uja, Guevara:2018rhj, Roig:2019reh,Estrada:2024cfy} pieces.} \cite{Miranda:2020wdg} (which yields the shift $\Delta a_\mu^{\mathrm{HVP,\,LO}[\pi\pi,\tau]}=\left(-1.67^{+0.60}_{-1.39}\right)\times10^{-10}$ according to our reference determination, in good agreement with the previous results of Refs.~\cite{Cirigliano:2002pv, Davier:2010fmf}), and $\text{FSR}(s)$ from Ref.~\cite{Drees:1990te}. The values in the first four rows of Table \ref{tab:IB_BR}, corresponding to $S_\mathrm{EW}$ and the first two factors in the RHS of Eq.~(\ref{RIB}), exhibit very little model-dependence and are in good agreement with Ref. \cite{Davier:2010fmf}.
We use different input values of the IB parameters entering the pion form factors in the last ratio of Eq. (\ref{RIB}). For the GP model, we use the same input as in previous works~\cite{Cirigliano:2002pv,Miranda:2020wdg,Masjuan:2023qsp}: $\theta_{\rho\omega}=-3m_\rho^2\delta_{\rho\omega}=(-3.5\pm0.7)\cdot 10^{-3}\text{ GeV}^2$ (assumed to be real), $\Delta M_\rho=(0.7\pm0.8)\text{ MeV}$ and $\Delta\Gamma_\rho=(1.5\pm1.3)\text{ MeV}$ 
\cite{ParticleDataGroup:2024cfk} \footnote{This choice of parameters correspond to the FF1 set in Ref.~\cite{Miranda:2020wdg}.} 
. Although some differences with respect to Ref. \cite{Davier:2010fmf} are observed in the central values of IB corrections stemming from rho-meson mass difference and rho-omega mixing (7th and 8th rows in Table \ref{tab:IB_BR}), our numerical results are consistent with Davier \textit{et al.}~\cite{Davier:2010fmf} within the quoted uncertainties.

\begin{table}[ht]
    \centering
    \resizebox{0.85\textwidth}{!}{\begin{tabular}{ccccccc}
    \hline
    Source & \multicolumn{3}{c}{$\Delta a_\mu^\text{HVP, LO}[\pi\pi,\tau]\,(10^{-10})$} & \multicolumn{3}{c}{$\Delta \mathcal{B}_{\pi\pi}^{\text{CVC}}\,(10^{-2})$} \\[0.3ex]
     & GS & KS & GP & GS & KS & GP \\[0.3ex]
      & \multicolumn{2}{c}{Davier \textit{et al.}} & FF1 & \multicolumn{2}{c}{Davier \textit{et al.}} & FF1 \\[0.3ex]
    \hline
    $S_\text{EW}$ & \multicolumn{2}{c}{$-12.21(0.15)$} & $-12.16(0.15)$ & \multicolumn{2}{c}{$+0.57(1)$} & $+0.57(1)$ \\[0.3ex]
    $G_\text{EM}$ & \multicolumn{2}{c}{$-1.92(0.90)$} & $-1.67^{+0.60}_{-1.39}$ & \multicolumn{2}{c}{$-0.07(17)$} &  $-0.09(^{3}_{1})$ \\[0.3ex]
    FSR & \multicolumn{2}{c}{$+4.67(0.47)$} & $+4.62(0.46)$ & \multicolumn{2}{c}{$-0.19(2)$} & $-0.19(2)$ \\[0.3ex]
    $m_{\pi^\pm}-m_{\pi^0}$ effect on $\sigma$ & \multicolumn{2}{c}{$-7.88$} & $-7.52$ & \multicolumn{2}{c}{$+0.19$} & $+0.20$ \\[0.3ex]
    $m_{\pi^\pm}-m_{\pi^0}$ effect on $\Gamma$ & $+4.09$ & $+4.02$ & $+4.11$ & \multicolumn{2}{c}{$-0.22$} & $-0.22$  \\[0.3ex]
    $m_{K^\pm}-m_{K^0}$ effect on $\Gamma$ & $-$ & $-$ & $+0.37$ & $-$ & $-$ & $-0.02$ \\[0.3ex]
    $m_{\rho^\pm}-m_{\rho^0}$ & $+0.20^{+0.27}_{-0.19}$ & $+0.11^{+0.19}_{-0.11}$ & $+1.15^{+1.37}_{-1.31}$  & $+0.08(8)$ & $+0.09(8)$ & $-0.00(0)$ \\[0.3ex]
    $\rho-\omega$ interference & $+2.80(0.19)$ & $+2.80(0.15)$ & $+3.55^{+0.84}_{-0.80}$ & $-0.01(1)$ & $-0.02(1)$ & $-0.10(1)$ \\[0.3ex]
    $\pi\pi\gamma$ & $-5.91(0.59)$ & $-6.39(0.64)$ & $-5.19^{+4.50}_{-4.45}$ & $+0.34(3)$ & $+0.37(4)$ & $+0.29(25)$ \\[0.3ex]
    \hline
    TOTAL & $-16.07(1.22)$ & $-16.70(1.23)$ & $-12.74^{+4.84}_{-4.93}$ & $+0.69(19)$ & $+0.72(19)$ & $+0.44(25)$ \\[0.3ex]
    \hline
    \end{tabular}}
    \captionsetup{width=0.88\linewidth}
    \caption{Comparison of Isospin Breaking (IB) corrections $\Delta a_\mu^{\text{HVP, LO}}[\pi\pi,\tau](10^{-10})$ and $\Delta \mathcal{B}_{\pi\pi}^\text{CVC}(10^{-2})$ obtained using different calculations of IB effects and form factor models: Davier \textit{et al.}  \cite{Davier:2010fmf} and GP \cite{Guerrero:1997ku}. 
    }
    \label{tab:IB_BR}
\end{table}

\begin{figure}[ht!]
    \centering
    \includegraphics[width=0.44\textwidth]{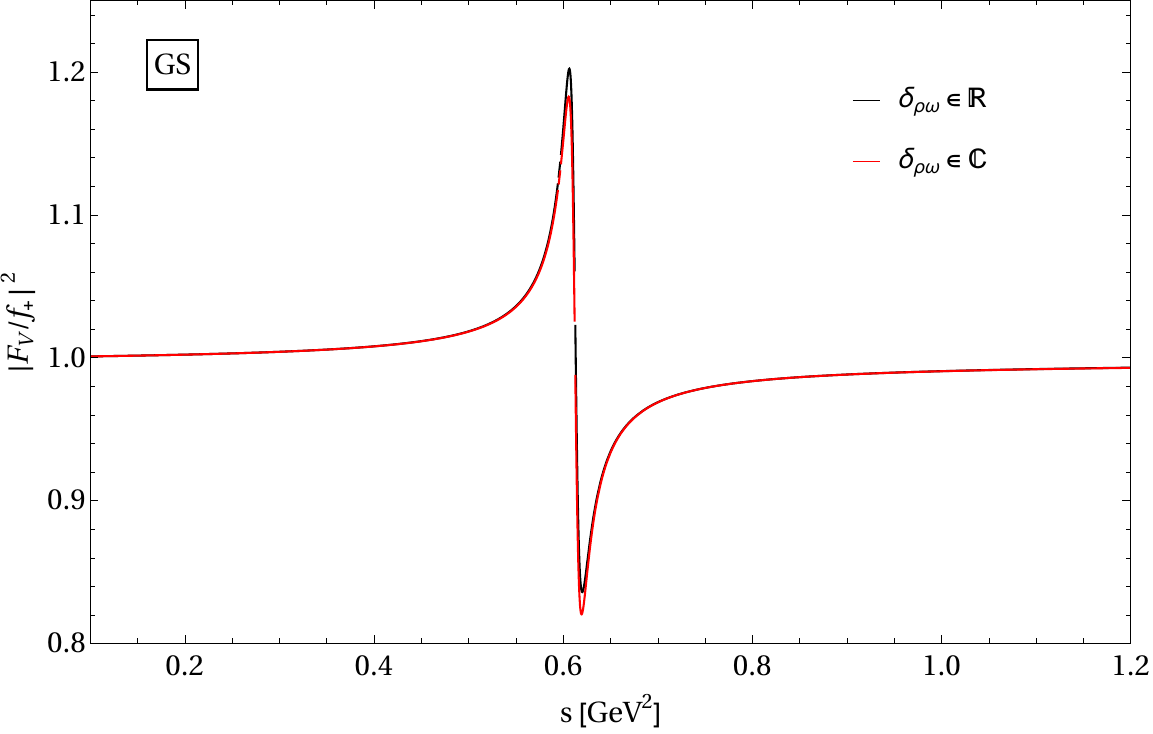}
    \includegraphics[width=0.44\textwidth]{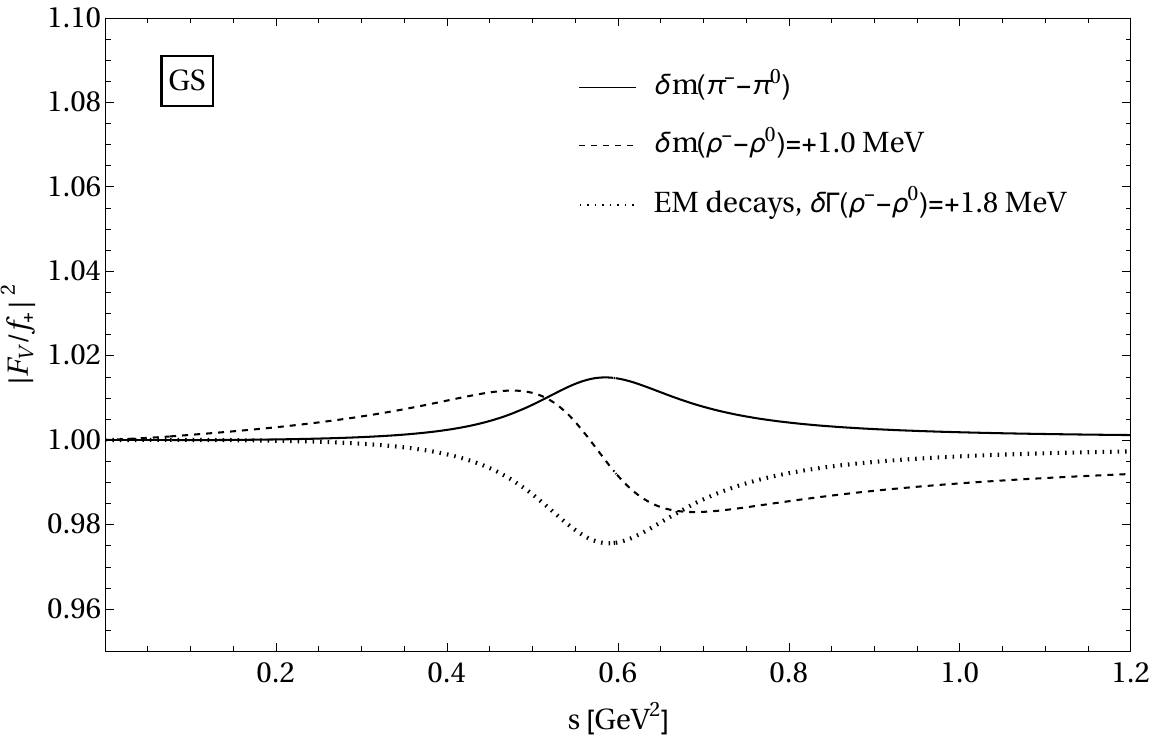}
    \includegraphics[width=0.44\textwidth]{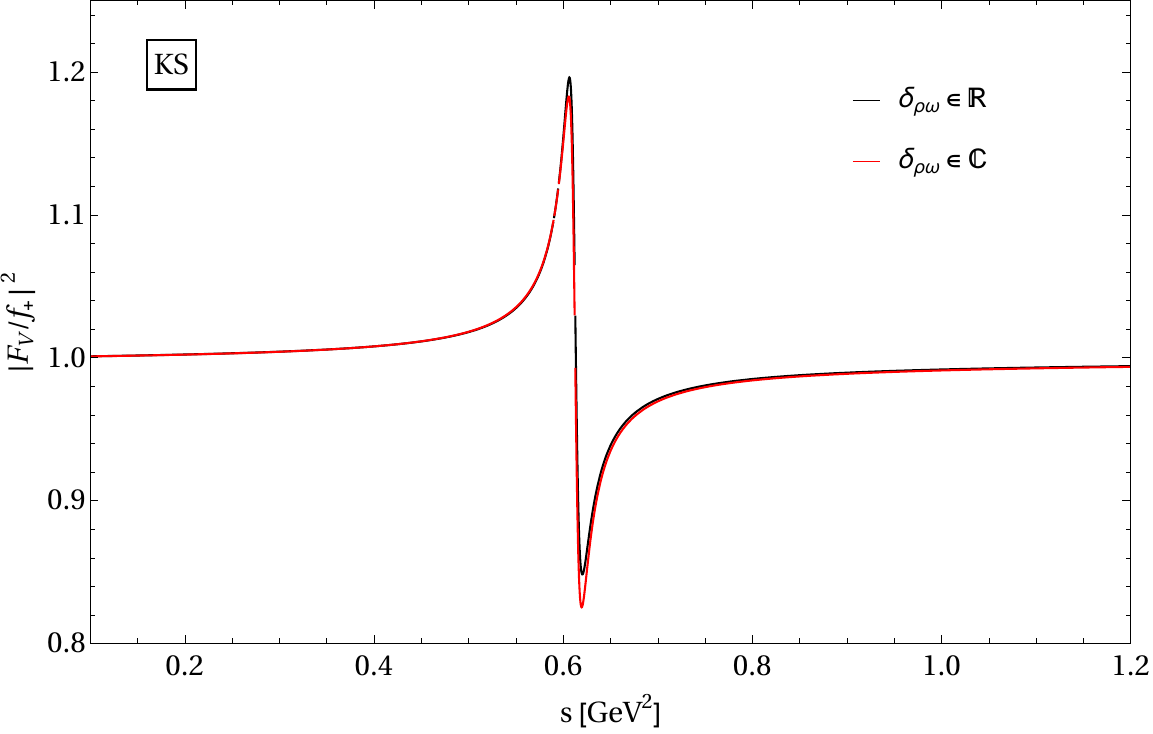}
    \includegraphics[width=0.44\textwidth]{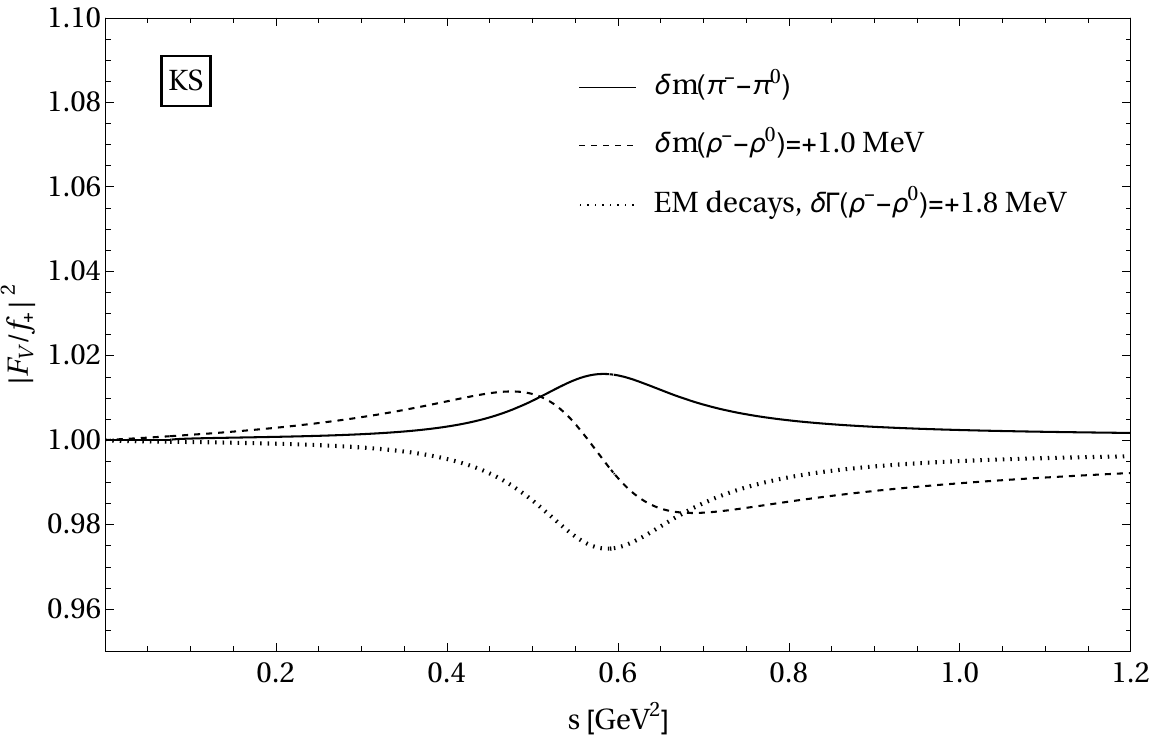}
    \includegraphics[width=0.44\textwidth]{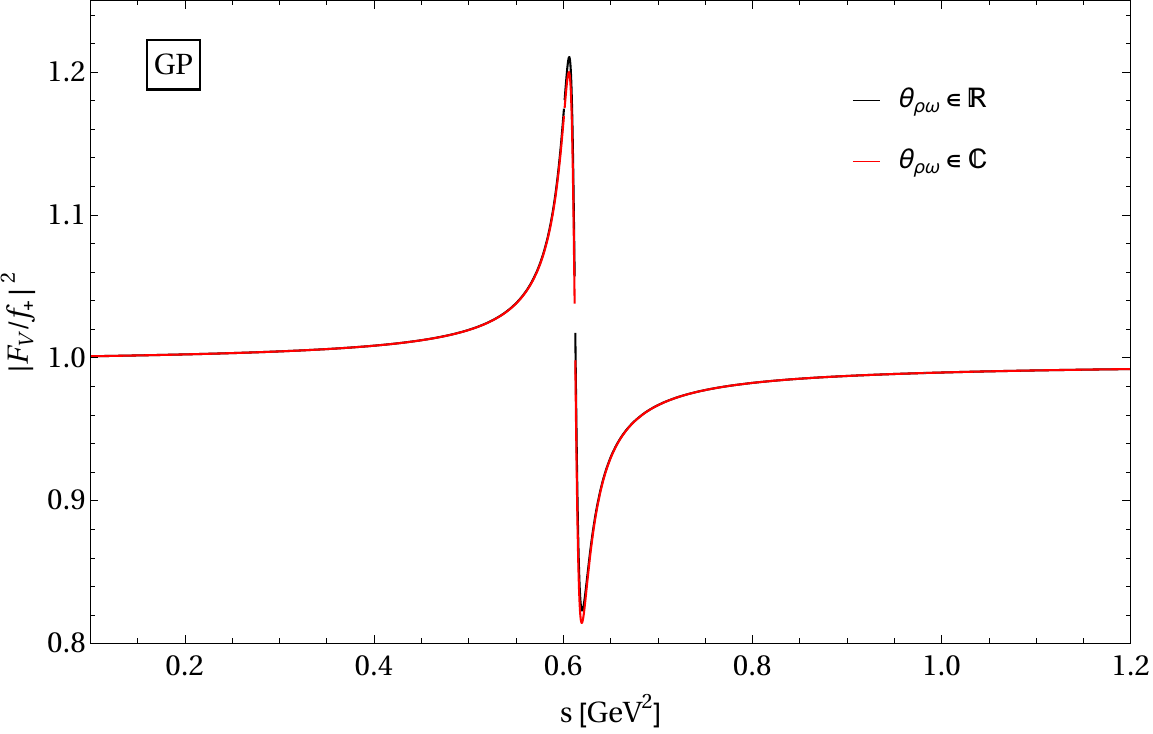}
    \includegraphics[width=0.44\textwidth]{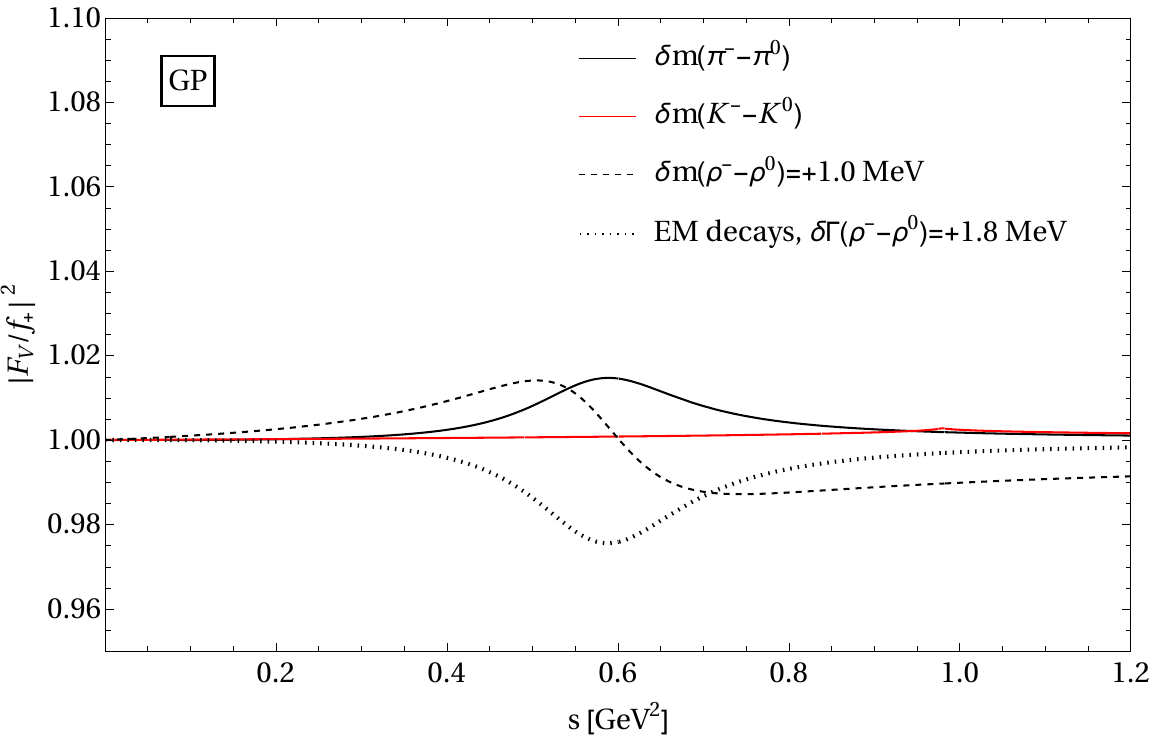}
    \includegraphics[width=0.44\textwidth]{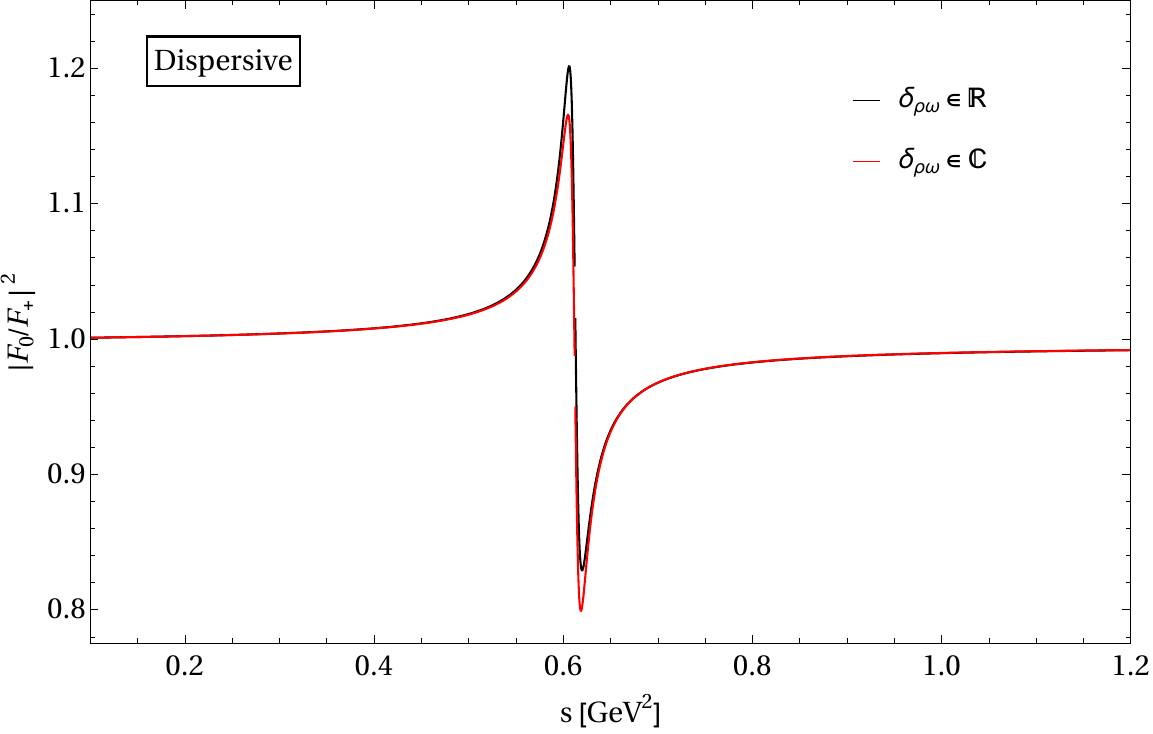}
    \includegraphics[width=0.44\textwidth]{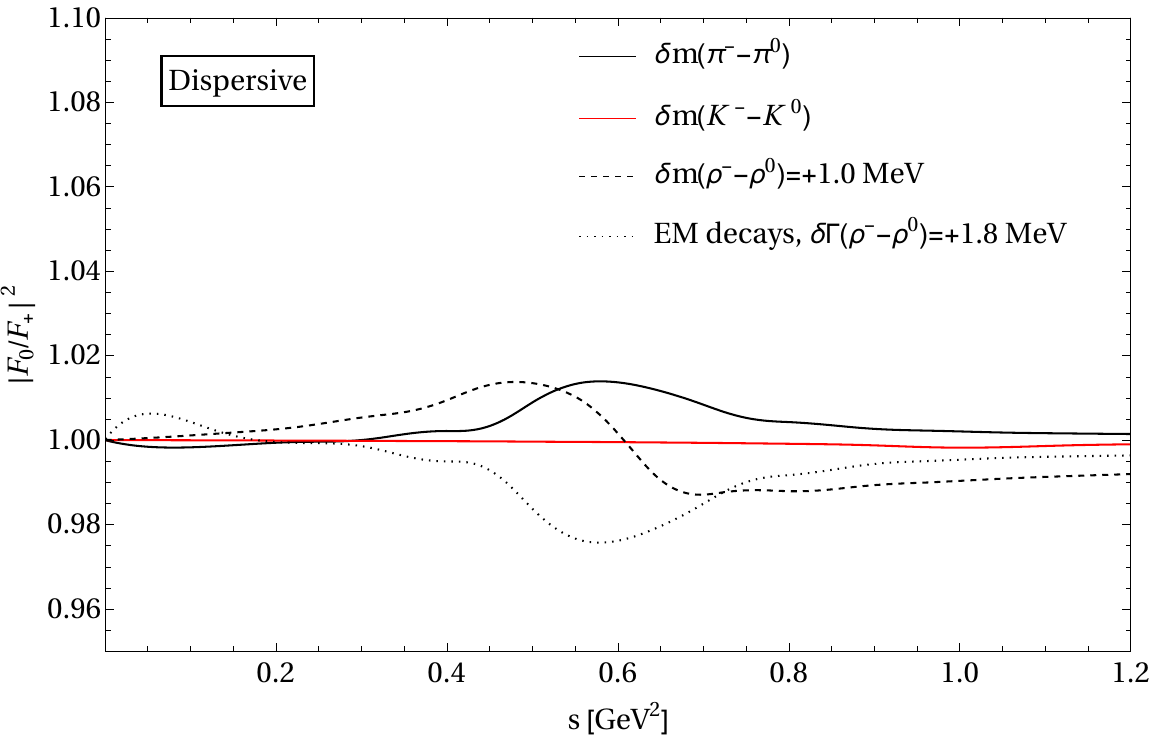}
    \captionsetup{width=0.88\linewidth}
    \caption{IB corrections the ratio of the form factors $\vert F_V(s)/f_+(s)\vert$ in the $\sqrt{s}\leq 1.2$ GeV region. Left: IB corrections due to the $\rho-\omega$ mixing taking into account either a real or a complex $\rho-\omega$ mixing parameter $\delta_{\rho\omega}$ defined in Eq. (4) of Ref.~\cite{Davier:2010fmf}. Right: IB corrections due to the $\pi$ mass splitting (also owing to the $K$ mass splitting, for GP and dispersive), the $\rho$ mass splitting, and the difference in the $\rho$ meson widths. The GS, KS, GP and dispersive results are shown from top to bottom in both cases.}
    \label{fig:IBCtoFF}
\end{figure}

In order to compare the IB effects in the form factor models with the ones of Ref. \cite{Davier:2010fmf}, we use the following set of inputs \cite{Davier:2010fmf}: $\Delta\Gamma_\rho=\Gamma_{\rho^0}-\Gamma_{\rho^\pm}=(1.8\pm 0.2) \text{ MeV}$ due to radiative corrections, $\Delta\Gamma_\rho=-1.07 \text{ MeV}$ due to $\Delta m_\pi$ in phase-space, $\Delta M_\rho=m_{\rho^\pm}-m_{\rho^0}=(1.0\pm0.9)\text{ MeV}$~, $\vert \delta_{\rho\omega}^\text{GS}\vert=(1.98\pm0.04 )\cdot 10^{-3}$, $\text{arg}(\delta_{\rho\omega}^\text{GS})=(9.86\pm1.0)^\circ$, $\vert \delta_{\rho\omega}^\text{KS}\vert=(1.84\pm0.04 )\cdot 10^{-3}$ and $\text{arg}(\delta_{\rho\omega}^\text{KS})=(10.9\pm1.1)^\circ$~\footnote{$\delta_{\rho\omega}$ is extracted from a fit to the KLOE~\cite{KLOE:2012anl} and BaBar~\cite{BaBar:2012bdw} data.}~\footnote{The corresponding parameters in Ref.~\cite{Davier:2023fpl} are $(1.99\pm0.03)\cdot10^{-3}$ and $(3.8\pm1.8)^\circ$, respectively.}. The superscript GS (KS) refers to the fitted values obtained using the Gounaris-Sakurai (K\"uhn-Santamar\'ia) BW shapes for the rho-meson. The IB corrections to the ratio of pion form factors are shown in Fig.~\ref{fig:IBCtoFF}. Good overall consistency among different approaches (and with e.g.~\cite{Davier:2010fmf}) is generally seen. 

\normalsize
To assess the complex character of the rho-omega mixing parameter in the GP model, we perform a global fit to the $e^+e^-$ data from BABAR~\cite{BaBar:2012bdw} and KLOE~\cite{KLOE:2012anl} using Eq.~(\ref{eq:FFGP}) for the electromagnetic and Eq.~(\ref{eq:FFGP2}) for the weak form factors, respectively. The outcomes are summarized in Tables \ref{tab:GP_Fit}, \ref{tab:GP_Fit1} and \ref{tab:GP_Fit2}, respectively, when all BABAR and KLOE datapoints are considered, or cutting them above $\sqrt{s_{\rm max}}=1$ or $0.9$ GeV. In those fits, $\delta_{\rho\omega}$ is a real $\rho-\omega$ mixing parameter in Fits 1, 2; while it is a complex one in Fits 3, 4. The $\dagger$ symbol means that a parameter was fixed in a given fit (the choice of fixed parameters differentiates the four fits shown), a notation that we will use throughout. Clearly, the quality of the fits is improved (smaller $\chi^2/{\rm d.o.f.}$) when the range for $\sqrt{s}$ is restricted closer to the $\rho(770)$ resonance region because the GP model includes only this single resonance.

From the results in Table \ref{tab:GP_Fit} (our conclusion does not change using Tables \ref{tab:GP_Fit1} or \ref{tab:GP_Fit2}), we get $\vert \delta_{\rho\omega}^\text{GP}\vert= (2.28\pm0.04)\cdot 10^{-3}$ and $\text{arg}(\delta_{\rho\omega}^\text{GP})=(6.3\pm0.9)^\circ$, which can be translated to $\vert \theta_{\rho\omega}\vert= 3m_\rho^2 \vert\delta_{\rho\omega}\vert=(4.1\pm0.1)\cdot 10^{-3}\text{ GeV}^2$ and $\text{arg}(\theta_{\rho\omega})=(186.3\pm0.9)^\circ$. Using $\vert\theta_{\rho\omega}\vert=(3.5\pm0.7)\cdot 10^{-3}\text{ GeV}^2$ as in Ref.~\cite{Cirigliano:2002pv} and our former estimation of $\text{arg}(\theta_{\rho\omega})$, we get $\Delta \mathcal{B}^\text{CVC}_{\pi\pi,\text{Mix}}=-0.05(^{2}_{1})\cdot10^{-2}$ instead of $\Delta \mathcal{B}^\text{CVC}_{\pi\pi,\text{Mix}}=-0.10(1)\cdot10^{-2}$ as in Table \ref{tab:IB_BR}. Therefore, the agreement between the GP and the GS and KS models regarding the $\rho-\omega$ interference is improved when $\delta_{\rho\omega}$ is a complex parameter. This effect, which is depicted in the left-hand side of Fig. \ref{fig:IBCtoFF} for the three models, moves down the curve near $s\sim m_\rho^2$.

\begin{table}[ht]
    \centering
    \resizebox{0.80\textwidth}{!}{\begin{tabular}{|c|cccc|}
    \hline
         & Fit 1 & Fit 2 & Fit 3 & Fit 4 \\[0.3ex]
         data points & $337+60$ & $337+60$ & $337+60$ & $337+60$ \\[0.3ex]
         $\chi^2$ & $4132.1$ & $4286.2$ & $4066.7$ & $4239.9$ \\[0.3ex]
         \hline
        $\chi^2/\text{d.o.f}$ & $10.5$ & $10.8$ & $10.3$ & $10.7$ \\[0.3ex]
        $m_\rho$ & $776.1\pm0.1\text{ MeV}$ & $775^\dagger\text{ MeV}$ & $776.2\pm0.1\text{ MeV}$ & $775^\dagger\text{ MeV}$\\[0.3ex] 
        Re[$\delta_{\rho\omega}$] & $(2.49\pm0.04)\cdot 10^{-3}$ & $(2.26\pm0.04)\cdot 10^{-3}$ & $(2.51\pm0.04)\cdot 10^{-3}$ & $(2.26\pm0.04)\cdot 10^{-3}$ \\[0.3ex]
        Im[$\delta_{\rho\omega}$] & $0^\dagger$ & $0^\dagger$ & $(0.30\pm0.04)\cdot 10^{-3}$ & $(0.25\pm0.04)\cdot 10^{-3}$ \\[0.3ex]
        \hline
        $\vert\delta_{\rho\omega}\vert$ & $(2.49\pm0.04)\cdot 10^{-3}$ & $(2.26\pm0.04)\cdot 10^{-3}$ & $(2.53\pm0.04)\cdot 10^{-3}$ & $(2.28\pm0.04)\cdot 10^{-3}$ \\[0.3ex]
        arg[$\delta_{\rho\omega}$] & $0^\circ$ & $0^\circ$ & $(6.7\pm0.8)^\circ$ & $(6.3\pm0.9)^\circ$ \\[0.3ex]
        \hline
    \end{tabular}}
    \captionsetup{width=0.88\linewidth}
    \caption{Fit results for the GP model using both 60 datapoints from KLOE 12~\cite{KLOE:2012anl} and 337 datapoints from BABAR 12~\cite{BaBar:2012bdw}. In Fit 1 $\delta_{\rho\omega}$ is a real parameter, while in Fit 3 it is a complex one. Fits 2 and 4 are similar to Fits 1 and 3, but with $m_\rho$ fixed to 775 MeV.}
    \label{tab:GP_Fit}
\end{table}

\begin{table}[ht]
    \centering
    \resizebox{0.80\textwidth}{!}{\begin{tabular}{|c|cccc|}
    \hline
         & Fit 1 & Fit 2 & Fit 3 & Fit 4 \\[0.3ex]
         data points & $270+60$ & $270+60$ & $270+60$ & $270+60$ \\[0.3ex]
         $\chi^2$ & $1538.6$ & $1658.8$ & $1475.2$ & $1612.4$ \\[0.3ex]
         \hline
        $\chi^2/\text{d.o.f}$ & $4.7$ & $5.0$ & $4.5$ & $4.9$ \\[0.3ex]
        $m_\rho$ & $776.0\pm0.1\text{ MeV}$ & $775^\dagger\text{ MeV}$ & $776.1\pm0.1\text{ MeV}$ & $775^\dagger\text{ MeV}$\\[0.3ex] 
        Re[$\delta_{\rho\omega}$] & $(2.47\pm0.04)\cdot 10^{-3}$ & $(2.27\pm0.04)\cdot 10^{-3}$ & $(2.50\pm0.04)\cdot 10^{-3}$ & $(2.27\pm0.04)\cdot 10^{-3}$ \\[0.3ex]
        Im[$\delta_{\rho\omega}$] & $0^\dagger$ & $0^\dagger$ & $(0.29\pm0.04)\cdot 10^{-3}$ & $(0.25\pm0.04)\cdot 10^{-3}$ \\[0.3ex]
        \hline
        $\vert\delta_{\rho\omega}\vert$ & $(2.47\pm0.04)\cdot 10^{-3}$ & $(2.27\pm0.04)\cdot 10^{-3}$ & $(2.51\pm0.04)\cdot 10^{-3}$ & $(2.29\pm0.04)\cdot 10^{-3}$ \\[0.3ex]
        arg[$\delta_{\rho\omega}$] & $0^\circ$ & $0^\circ$ & $(6.7\pm0.8)^\circ$ & $(6.2\pm0.9)^\circ$ \\[0.3ex]
        \hline
    \end{tabular}}
    \captionsetup{width=0.88\linewidth}
    \caption{Analog to Table \ref{tab:GP_Fit} but using data only below $1\text{ GeV}$.}
    \label{tab:GP_Fit1}
\end{table}

\begin{table}[ht]
    \centering
    \resizebox{0.80\textwidth}{!}{\begin{tabular}{|c|cccc|}
    \hline
         & Fit 1 & Fit 2 & Fit 3 & Fit 4 \\[0.3ex]
         data points & $220+46$ & $220+46$ & $220+46$ & $220+46$ \\[0.3ex]
         $\chi^2$ & $759.0$ & $764.7$ & $714.5$ & $717.1$ \\[0.3ex]
         \hline
        $\chi^2/\text{d.o.f}$ & $2.9$ & $2.9$ & $2.7$ & $2.7$ \\[0.3ex]
        $m_\rho$ & $774.8\pm0.1\text{ MeV}$ & $775^\dagger\text{ MeV}$ & $774.8\pm0.1\text{ MeV}$ & $775^\dagger\text{ MeV}$\\[0.3ex] 
        Re[$\delta_{\rho\omega}$] & $(2.32\pm0.04)\cdot 10^{-3}$ & $(2.37\pm0.04)\cdot 10^{-3}$ & $(2.34\pm0.04)\cdot 10^{-3}$ & $(2.37\pm0.04)\cdot 10^{-3}$ \\[0.3ex]
        Im[$\delta_{\rho\omega}$] & $0^\dagger$ & $0^\dagger$ & $(0.24\pm0.04)\cdot 10^{-3}$ & $(0.25\pm0.04)\cdot 10^{-3}$ \\[0.3ex]
        \hline
        $\vert\delta_{\rho\omega}\vert$ & $(2.32\pm0.04)\cdot 10^{-3}$ & $(2.37\pm0.04)\cdot 10^{-3}$ & $(2.36\pm0.04)\cdot 10^{-3}$ & $(2.39\pm0.04)\cdot 10^{-3}$ \\[0.3ex]
        arg[$\delta_{\rho\omega}$] & $0^\circ$ & $0^\circ$ & $(6.0\pm0.9)^\circ$ & $(6.1\pm0.9)^\circ$ \\[0.3ex]
        \hline
    \end{tabular}}
    \captionsetup{width=0.88\linewidth}
    \caption{Analog to Table \ref{tab:GP_Fit} but using data only below $0.9\text{ GeV}$. \textcolor{black}{All two-sided p-values are smaller than $10^{-4}$.}}
    \label{tab:GP_Fit2}
\end{table}

Now, in order to test the consistency among the different $e^+e^-$ and $\tau$ data sets, we perform several fits using the various $e^+e^-$ data from KLOE~\cite{KLOE:2012anl,KLOE-2:2017fda}, BaBar~\cite{BaBar:2012bdw} and CMD-3~\cite{CMD-3:2023alj,CMD-3:2023rfe}~\footnote{In order to make use of the CMD-3 data, the physical form factor $F_V(s)$, which includes VP effects, has to be corrected accordingly, $\vert F_V^{(0)}(s)\vert^2=\vert F_V(s)\vert^2\vert \alpha(0)/\alpha(s)\vert^2=\vert F_V(s)\vert^2\vert 1-\Pi(s)\vert^2$~\cite{Jegerlehner:2017gek}, where $\Pi(s)$ is the polarization operator computed~\cite{Arbuzov:1997pj} from the measured $e^+e^-\to$ hadrons cross section~\cite{web:Fedor}.}, and the $\tau$ data from Belle~\cite{Belle:2008xpe}, using Eq.~(\ref{eq_chi2JointFit})~\footnote{Concerning tau data, in this work we decided to consider only Belle's while in e.g. Refs.~\cite{Davier:2010fmf,Miranda:2020wdg} also ALEPH, CLEO and OPAL data were used. Our main conclusions are barely affected if these data are also included.}

\begin{equation}\begin{split}\label{eq_chi2JointFit}
    \chi^2&=\sum_{k}^\text{KLOE}\left(\frac{\sigma_k^\mathrm{th}({e^+e^-\to\pi^+\pi^-(\gamma)})-\sigma_k^\mathrm{exp}({e^+e^-\to\pi^+\pi^-(\gamma)})}{\delta\sigma_k^\mathrm{exp}({e^+e^-\to\pi^+\pi^-(\gamma)})}\right)^2+\sum_{k}^\mathrm{CMD-3}\left(\frac{\vert F_V^{\pi,\mathrm{th}}\vert_k^{2}-\vert F_V^{\pi,\mathrm{exp}}\vert_k^{2}}{\delta\vert F_V^{\pi,\mathrm{exp}}\vert_k^{2}}\right)^2\\
    &+\sum_{k}^\text{BABAR}\left(\frac{\sigma_k^\mathrm{th}({e^+e^-\to\pi^+\pi^-(\gamma)})-\sigma_k^\mathrm{exp}({e^+e^-\to\pi^+\pi^-(\gamma)})}{\delta\sigma_k^\mathrm{exp}({e^+e^-\to\pi^+\pi^-(\gamma)})}\right)^2+\sum_{k}^\mathrm{Belle}\left(\frac{\vert f_+^{\pi,\mathrm{th}}\vert_k^{2}-\vert f_+^{\pi,\mathrm{exp}}\vert_k^{2}}{\delta\vert f_+^{\pi,\mathrm{exp}}\vert_k^{2}}\right)^2\,.
\end{split}\end{equation}

\begin{figure}[ht]
    \centering
    \includegraphics[width=0.44\textwidth]{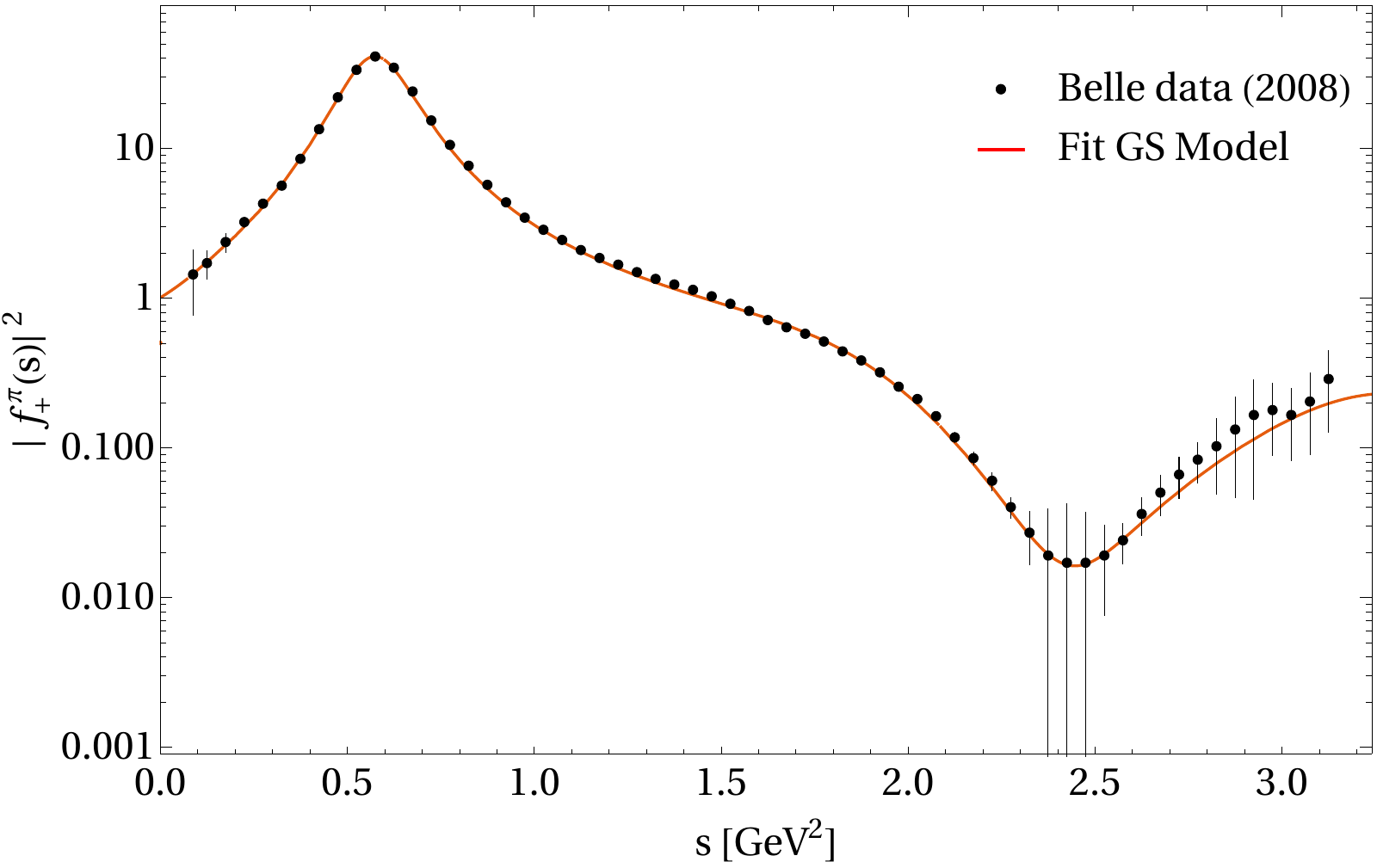}
    \includegraphics[width=0.44\textwidth]{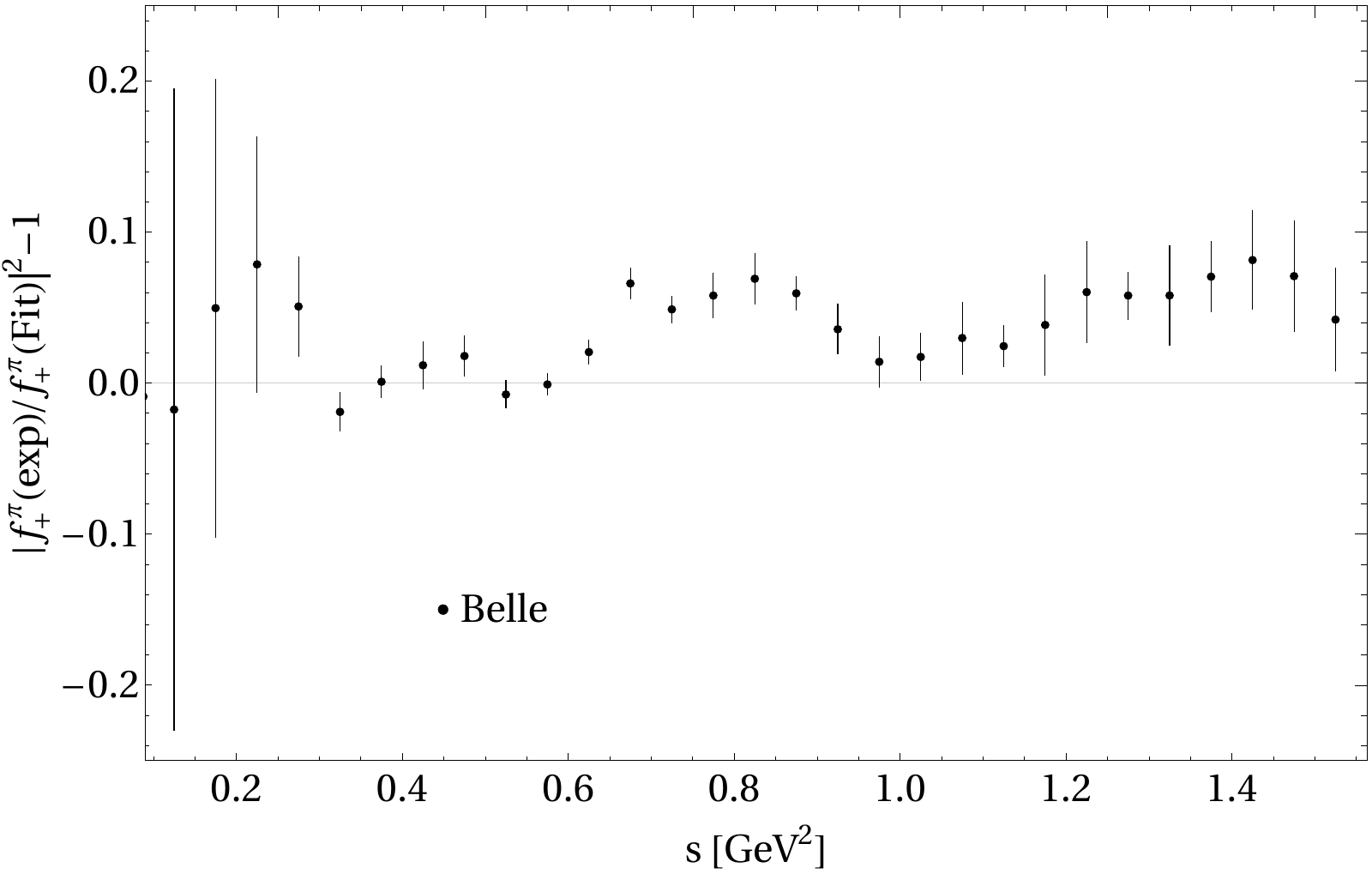}
    \includegraphics[width=0.44\textwidth]{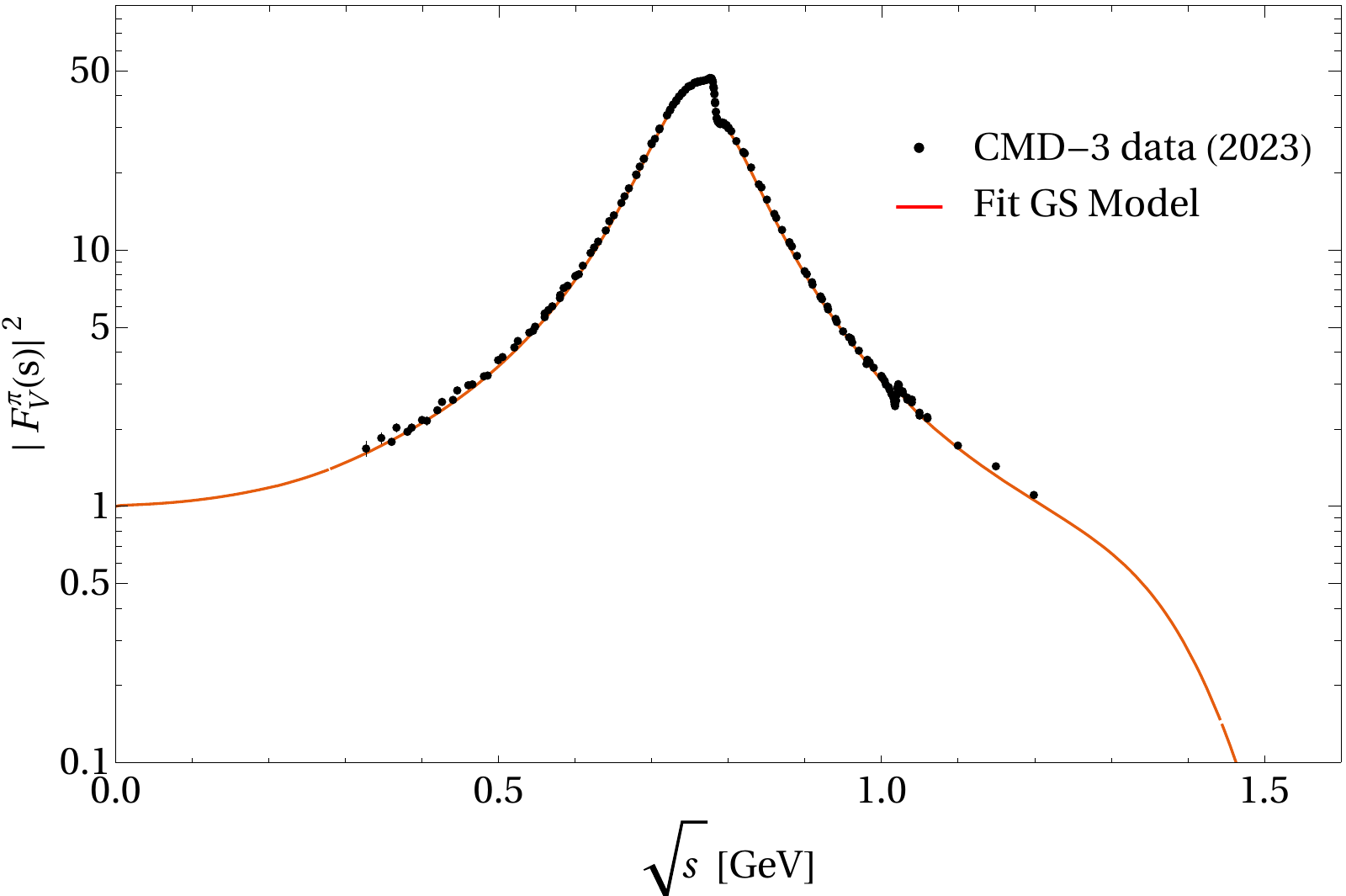}
    \includegraphics[width=0.44\textwidth]{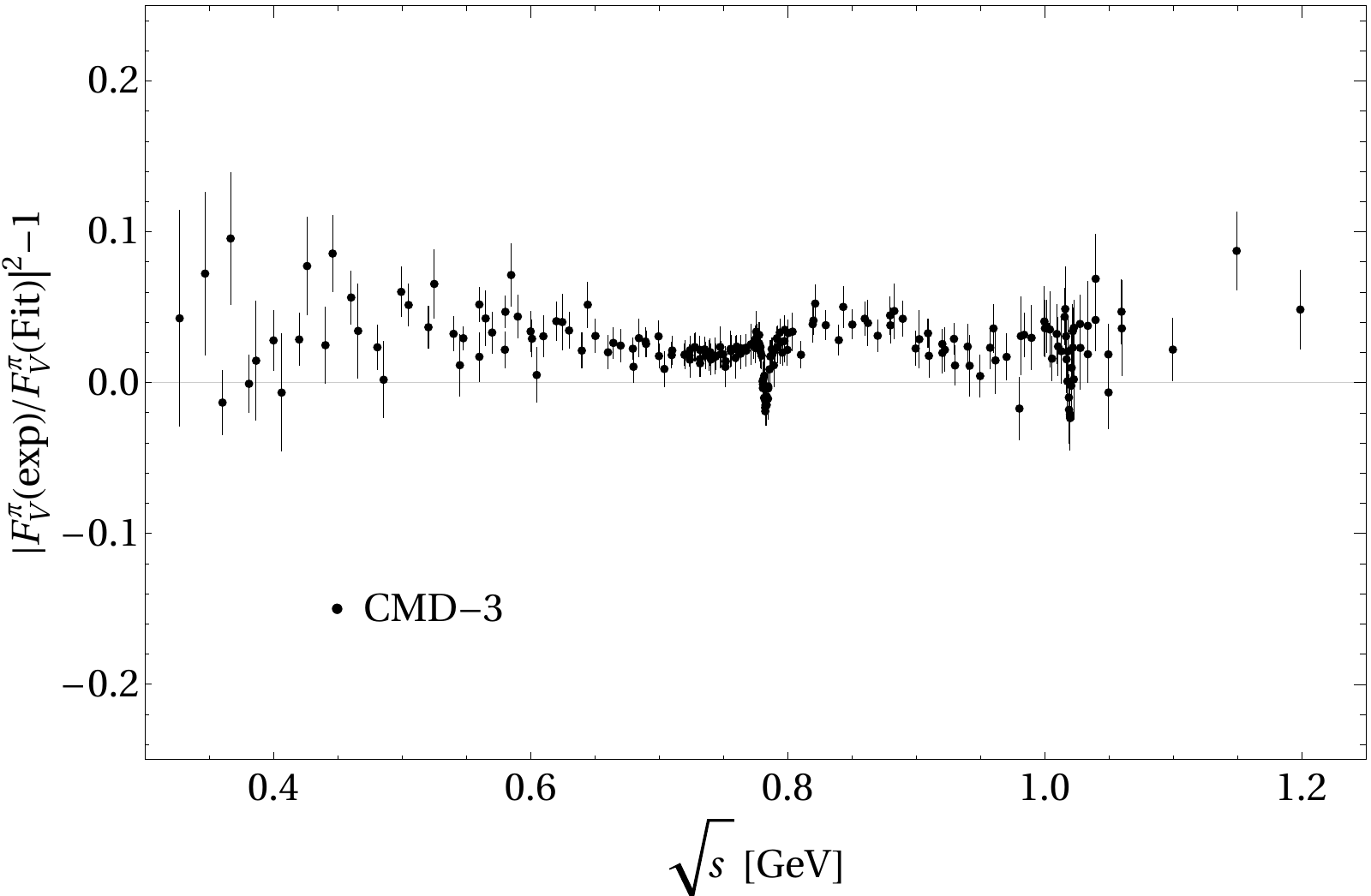}
    \includegraphics[width=0.44\textwidth]{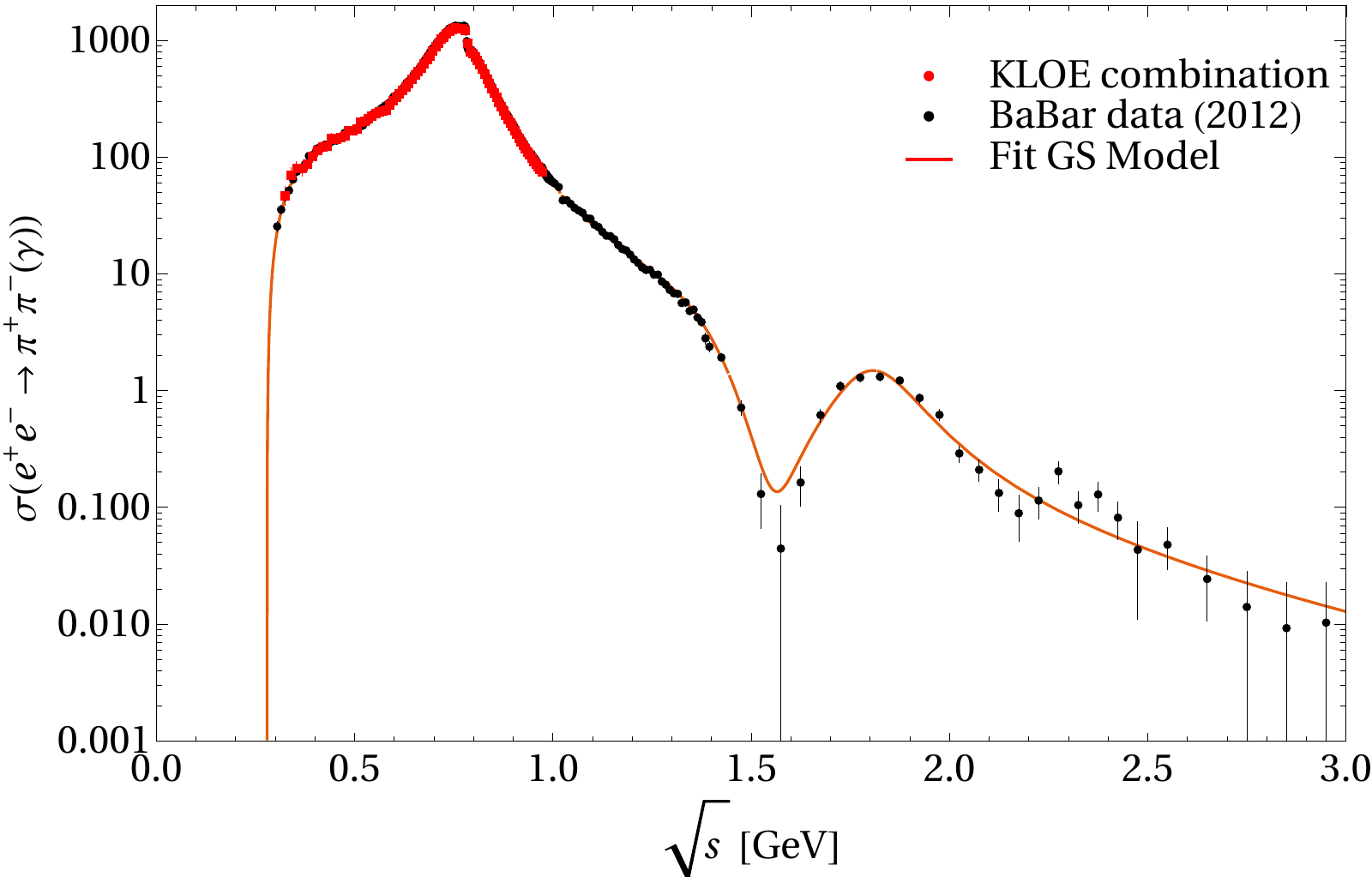}
    \includegraphics[width=0.44\textwidth]{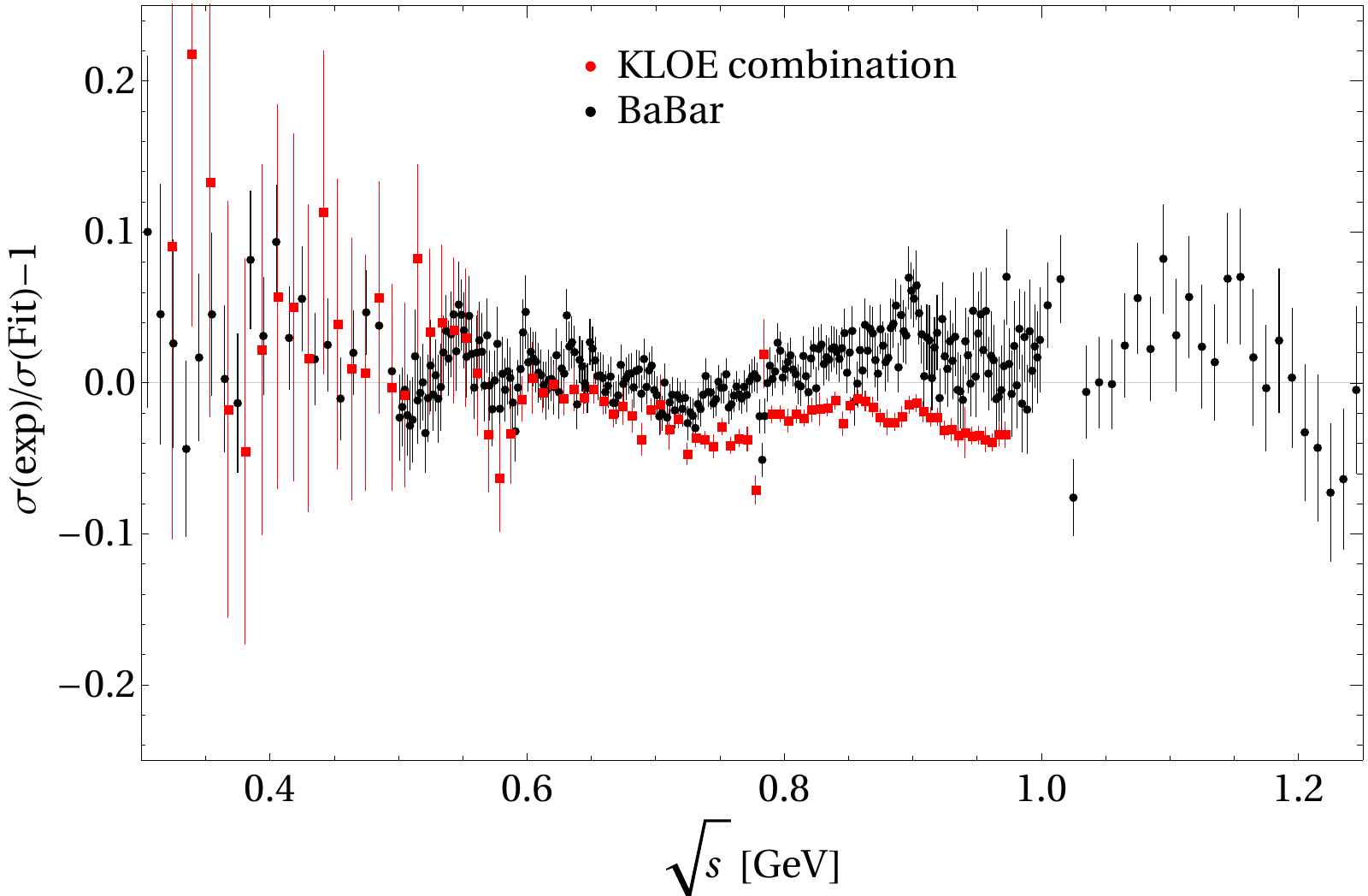}
    \captionsetup{width=0.88\linewidth}
    \caption{Global fit results for the GS model versus data. 
    }
    \label{fig:Fit2_GS}
\end{figure}

\begin{figure}[ht]
    \centering
    \includegraphics[width=0.44\textwidth]{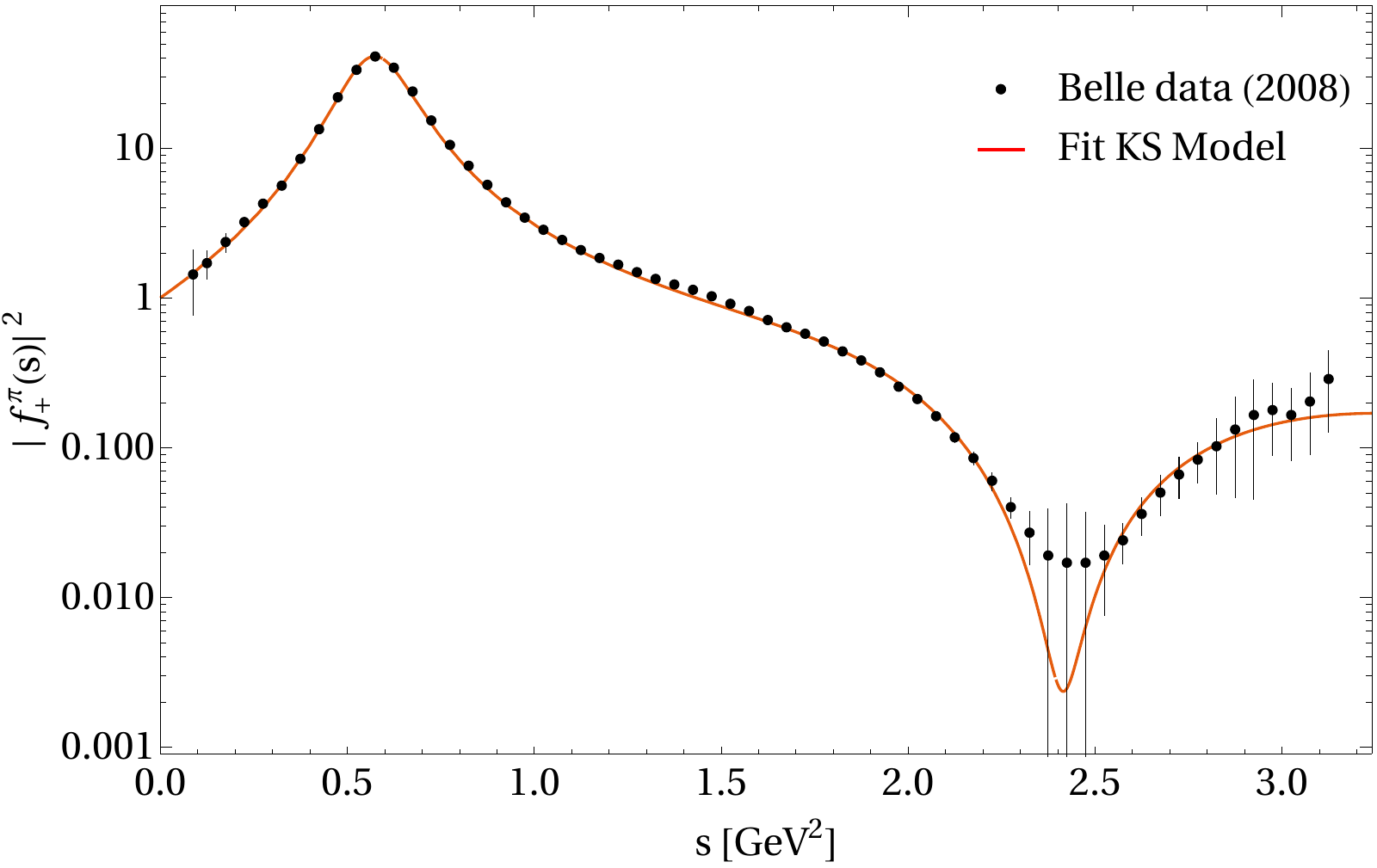}
    \includegraphics[width=0.44\textwidth]{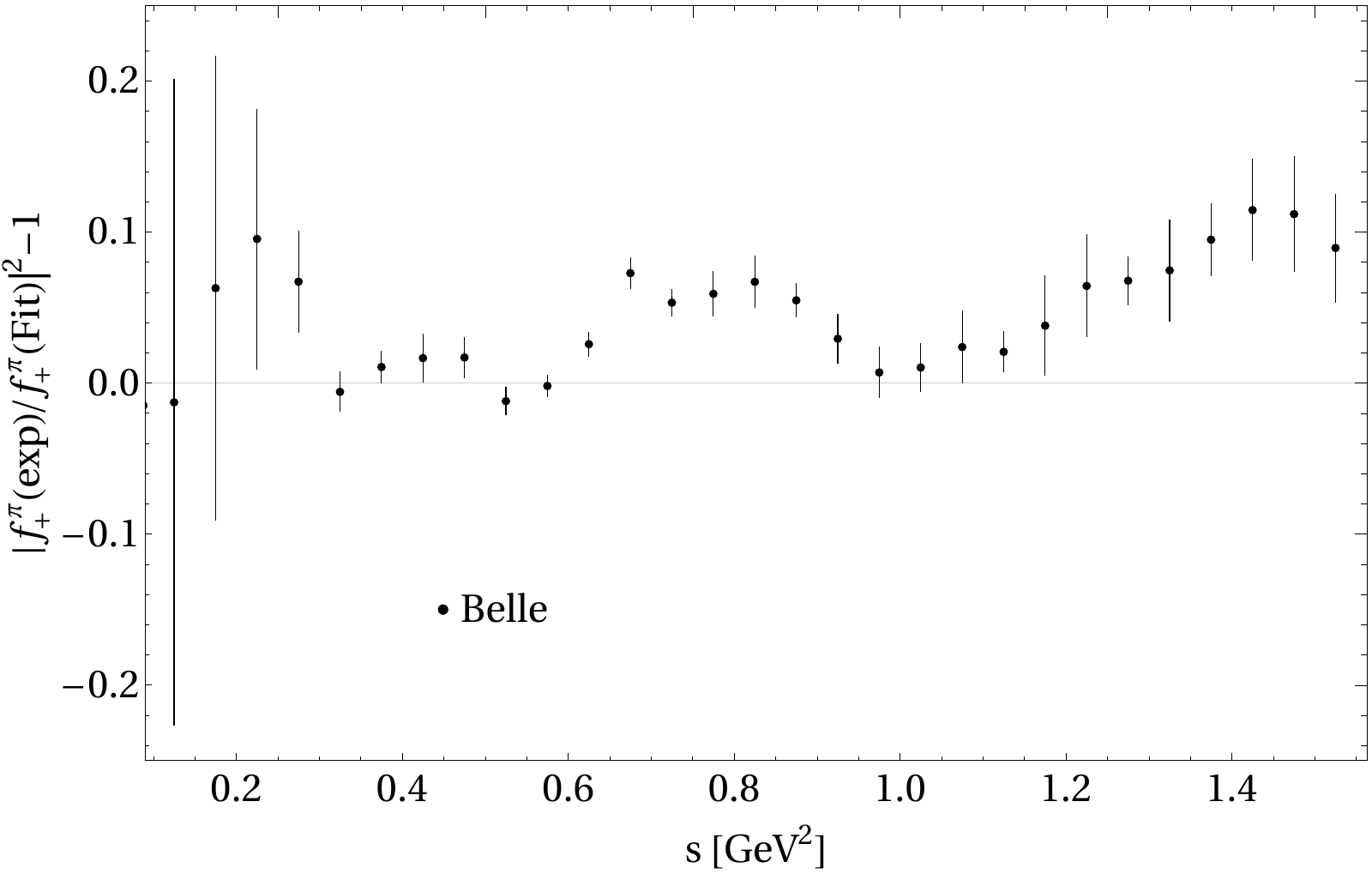}
    \includegraphics[width=0.44\textwidth]{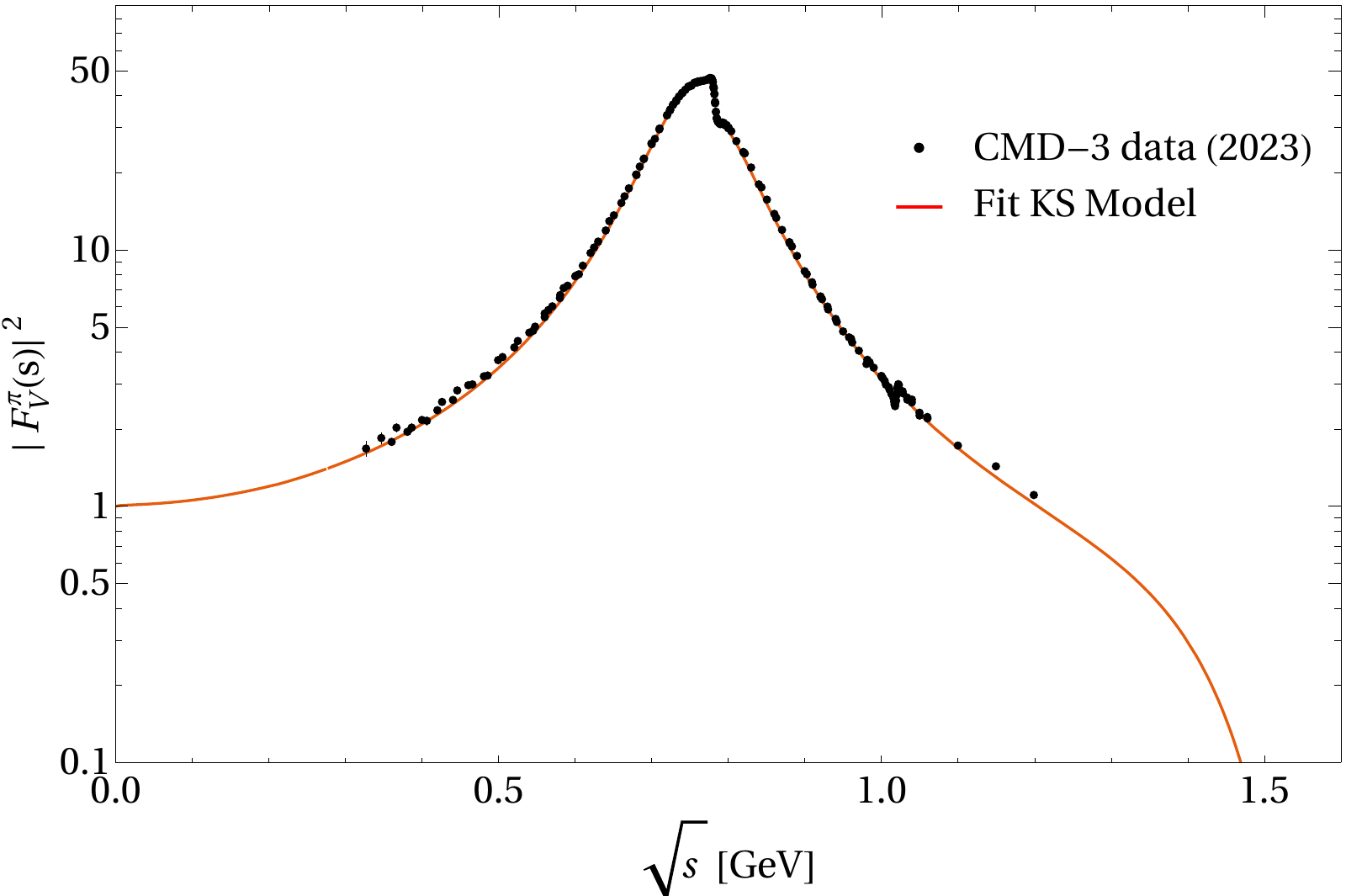}
    \includegraphics[width=0.44\textwidth]{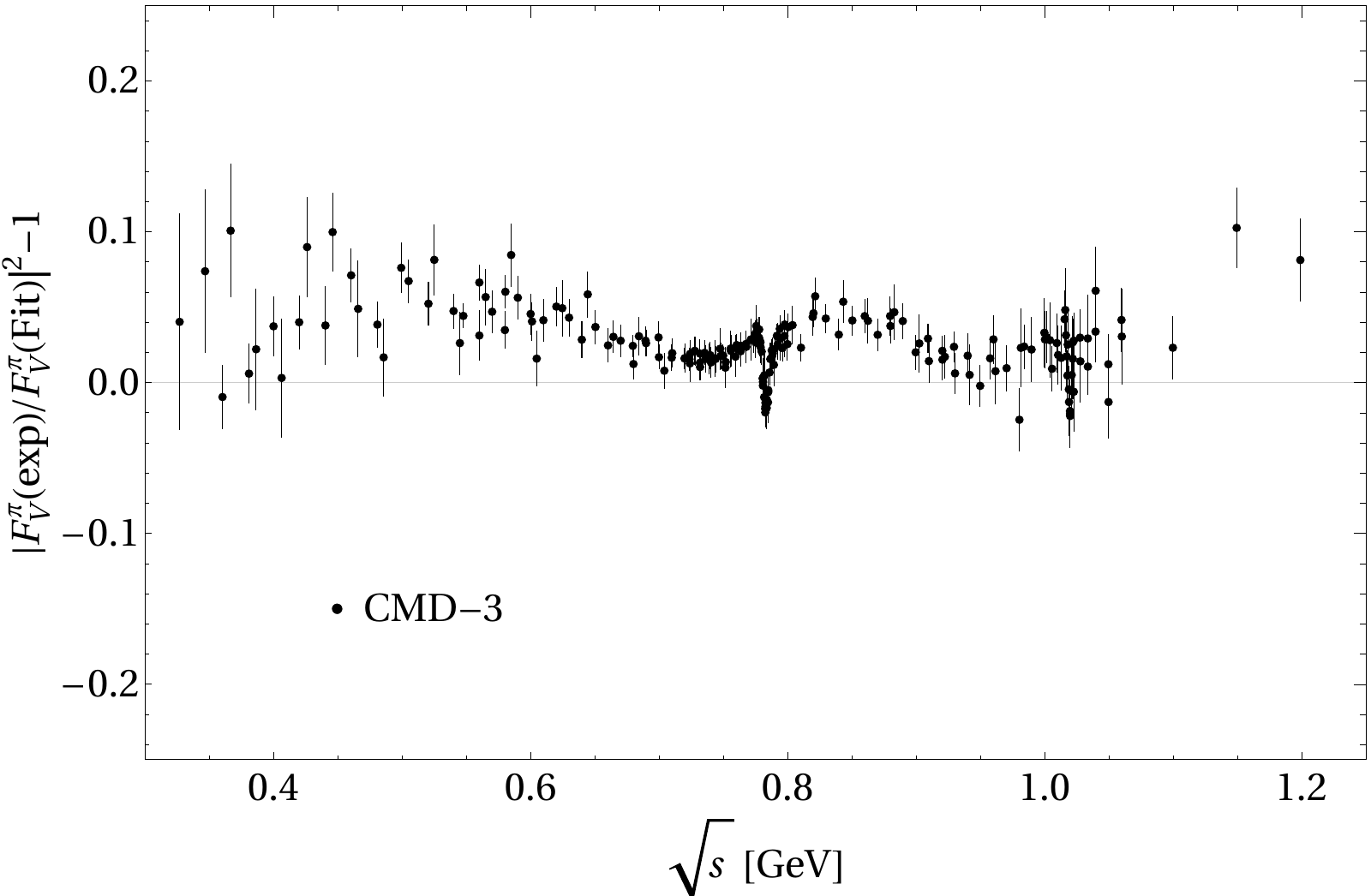}
    \includegraphics[width=0.44\textwidth]{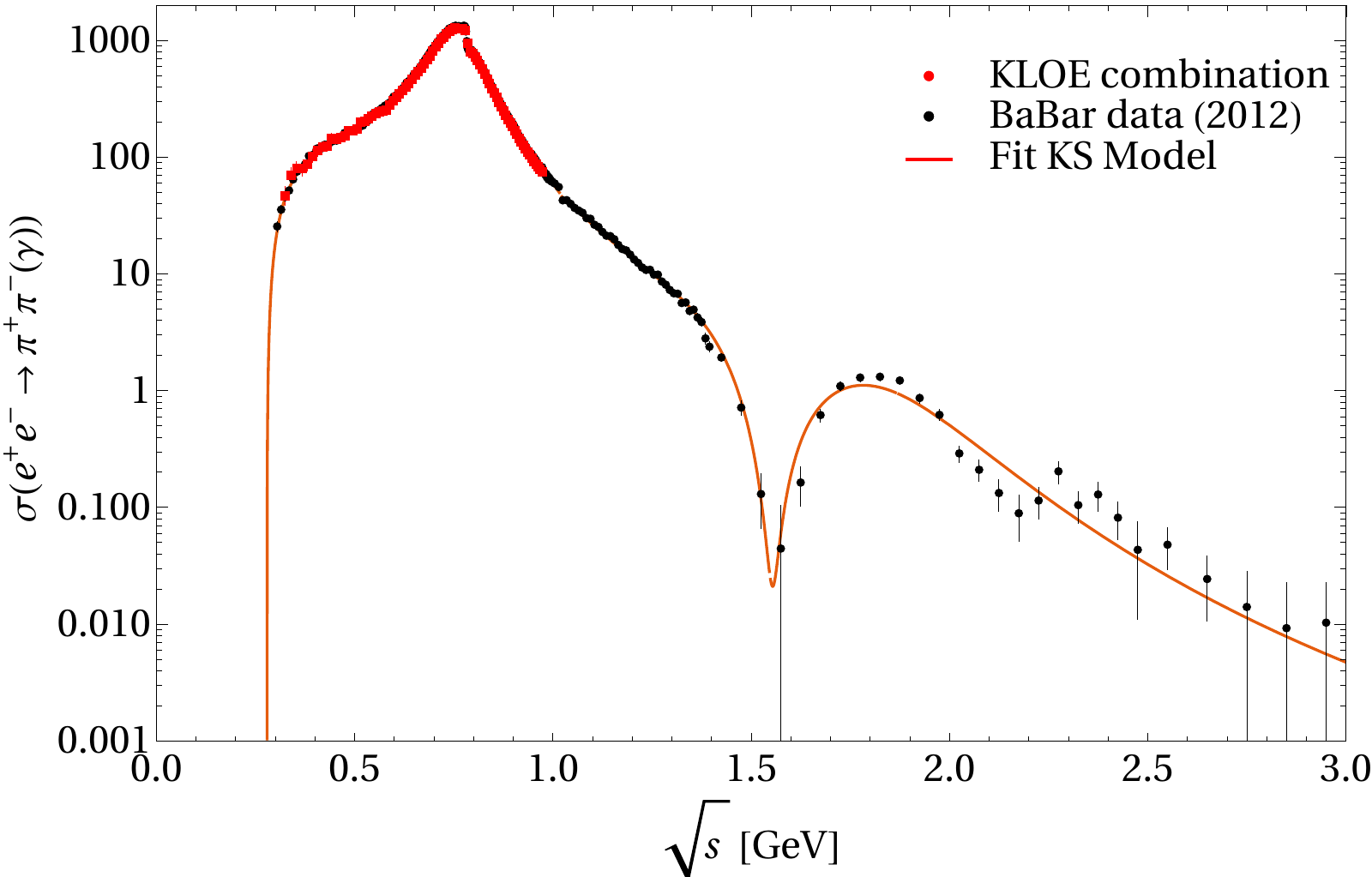}
    \includegraphics[width=0.44\textwidth]{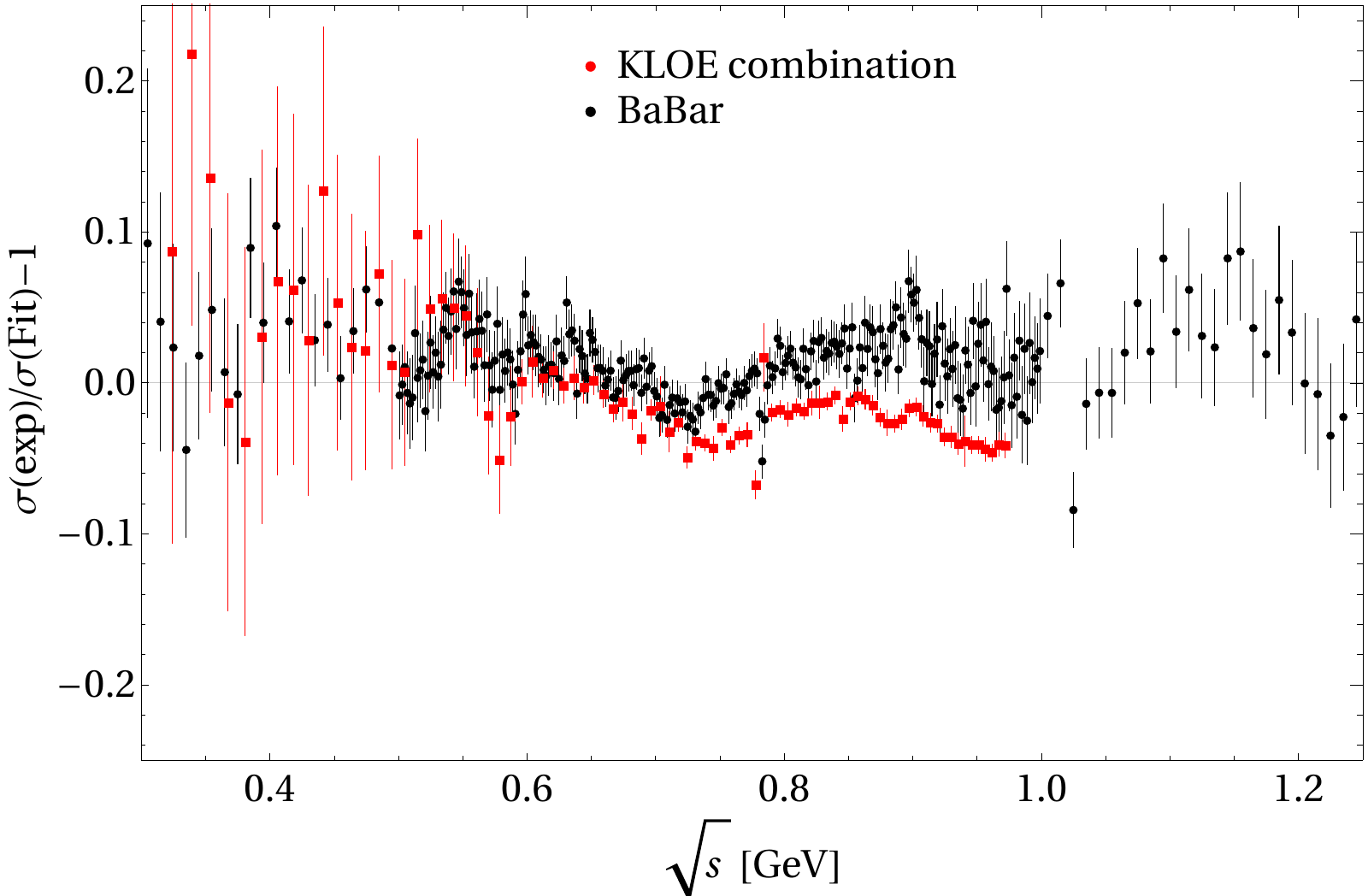}
    \captionsetup{width=0.88\linewidth}
    \caption{Global fit results for the KS model versus data. 
    }
    \label{fig:Fit_KS}
\end{figure}

\begin{figure}[ht]
    \centering
    \includegraphics[width=0.44\textwidth]{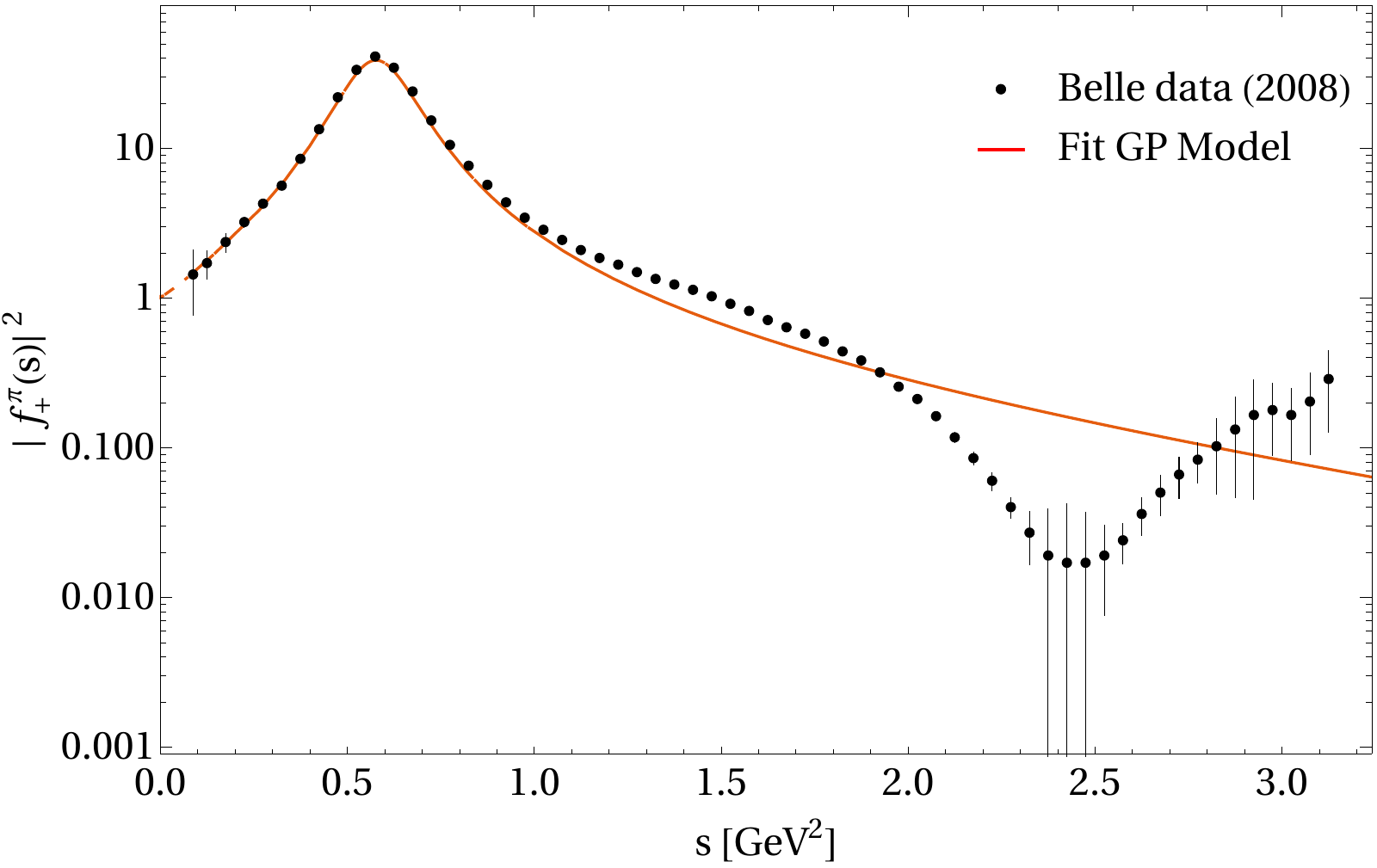}
    \includegraphics[width=0.44\textwidth]{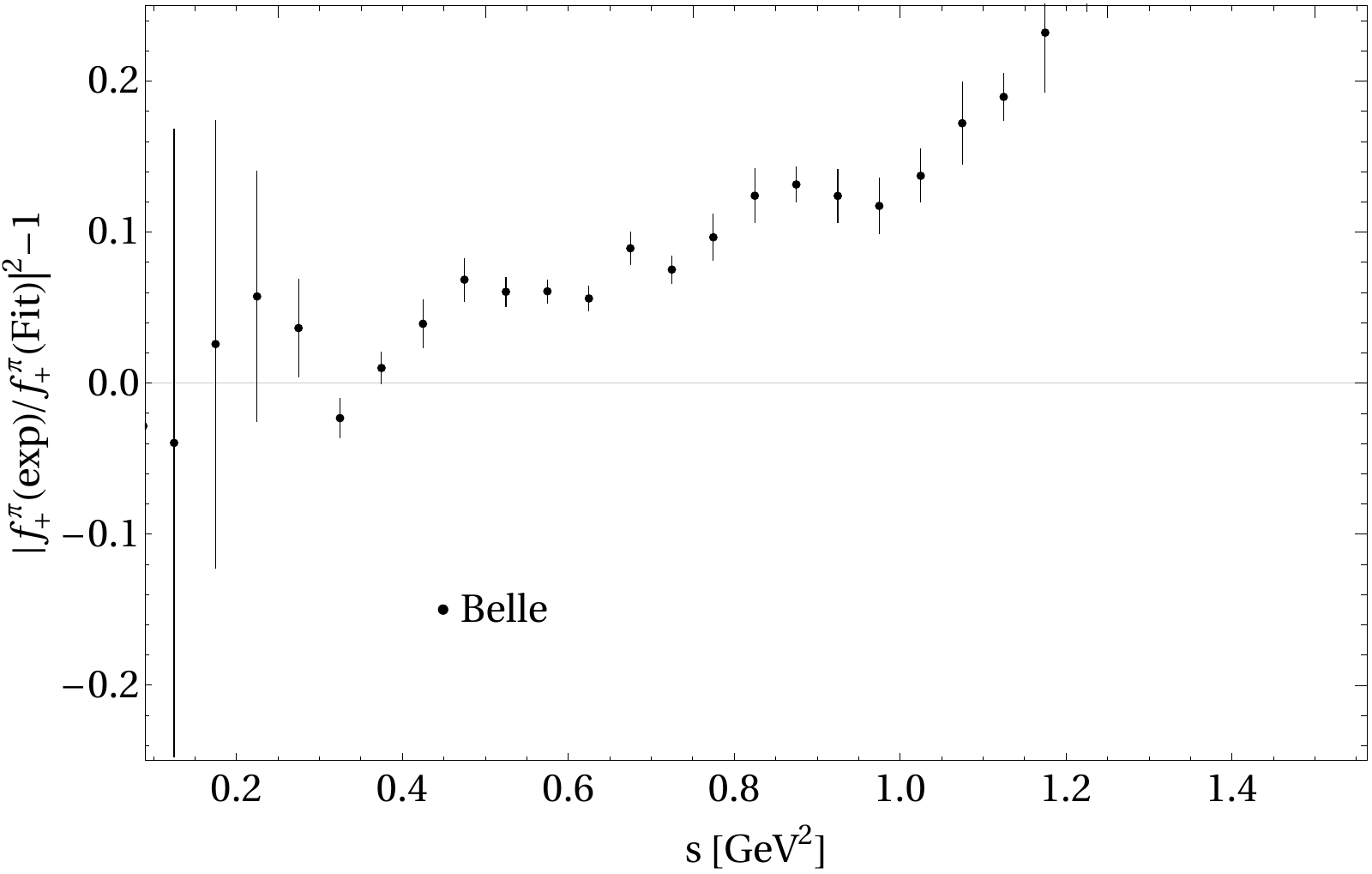}
    \includegraphics[width=0.44\textwidth]{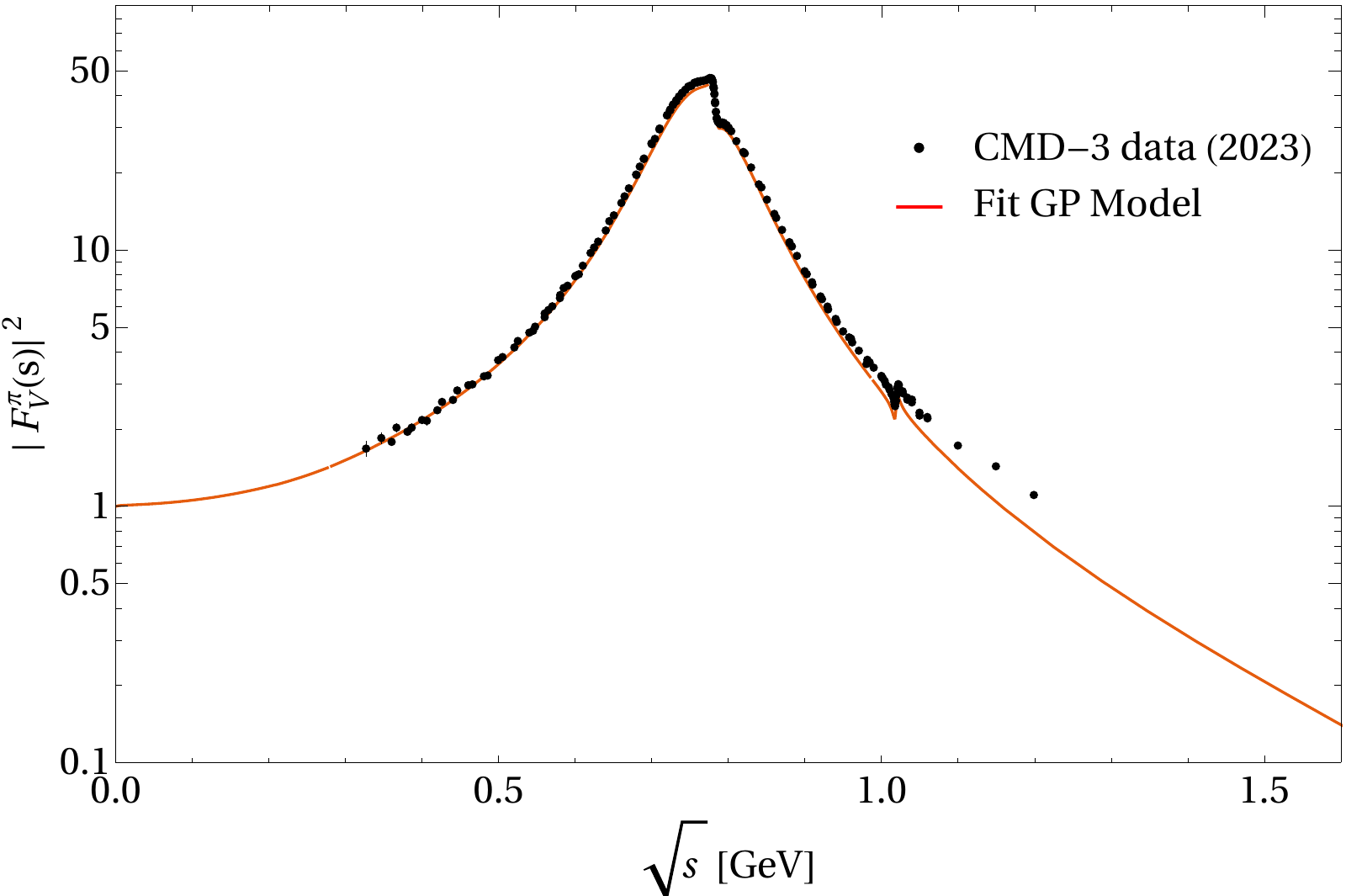}
    \includegraphics[width=0.44\textwidth]{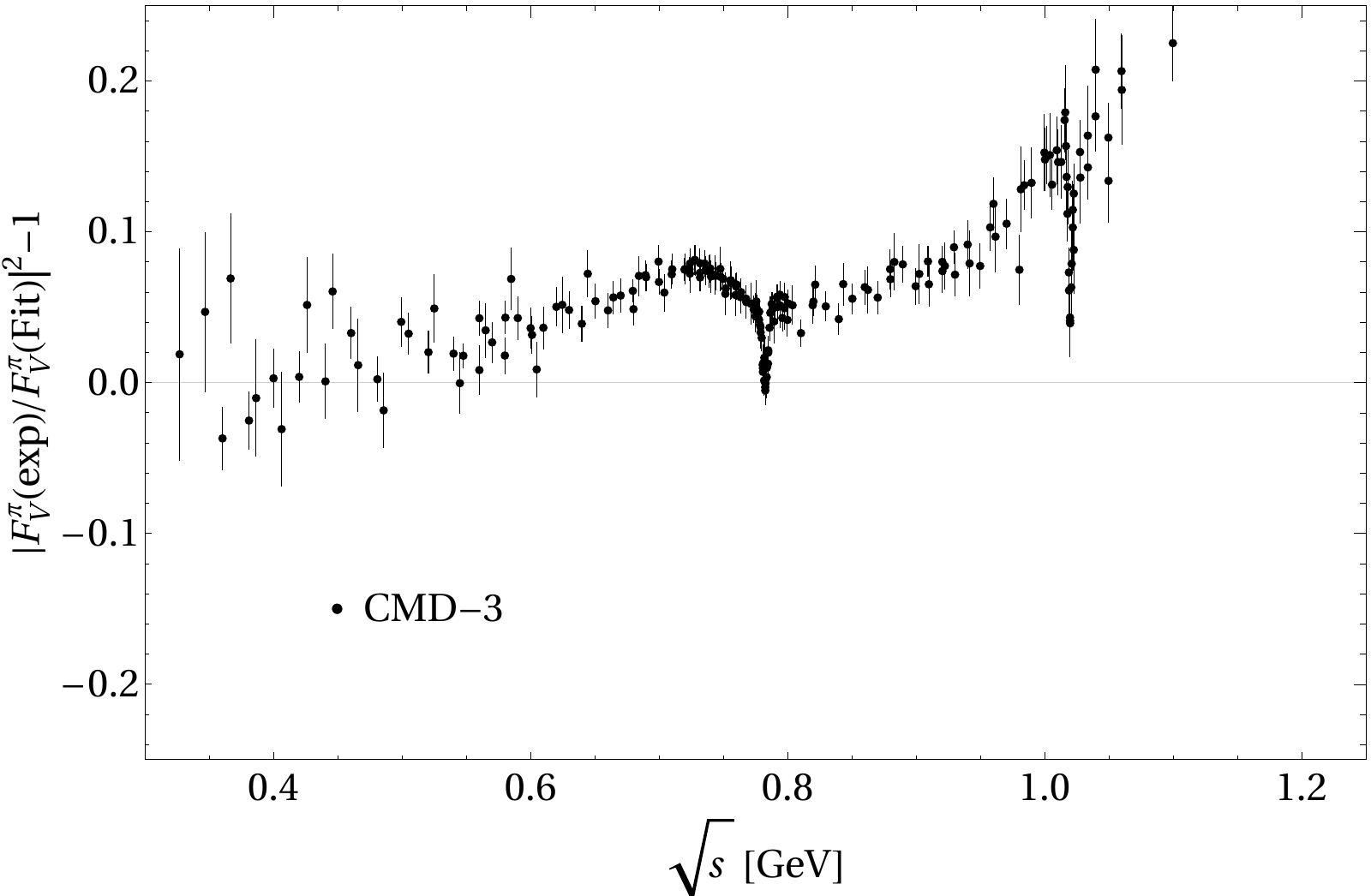}
    \includegraphics[width=0.44\textwidth]{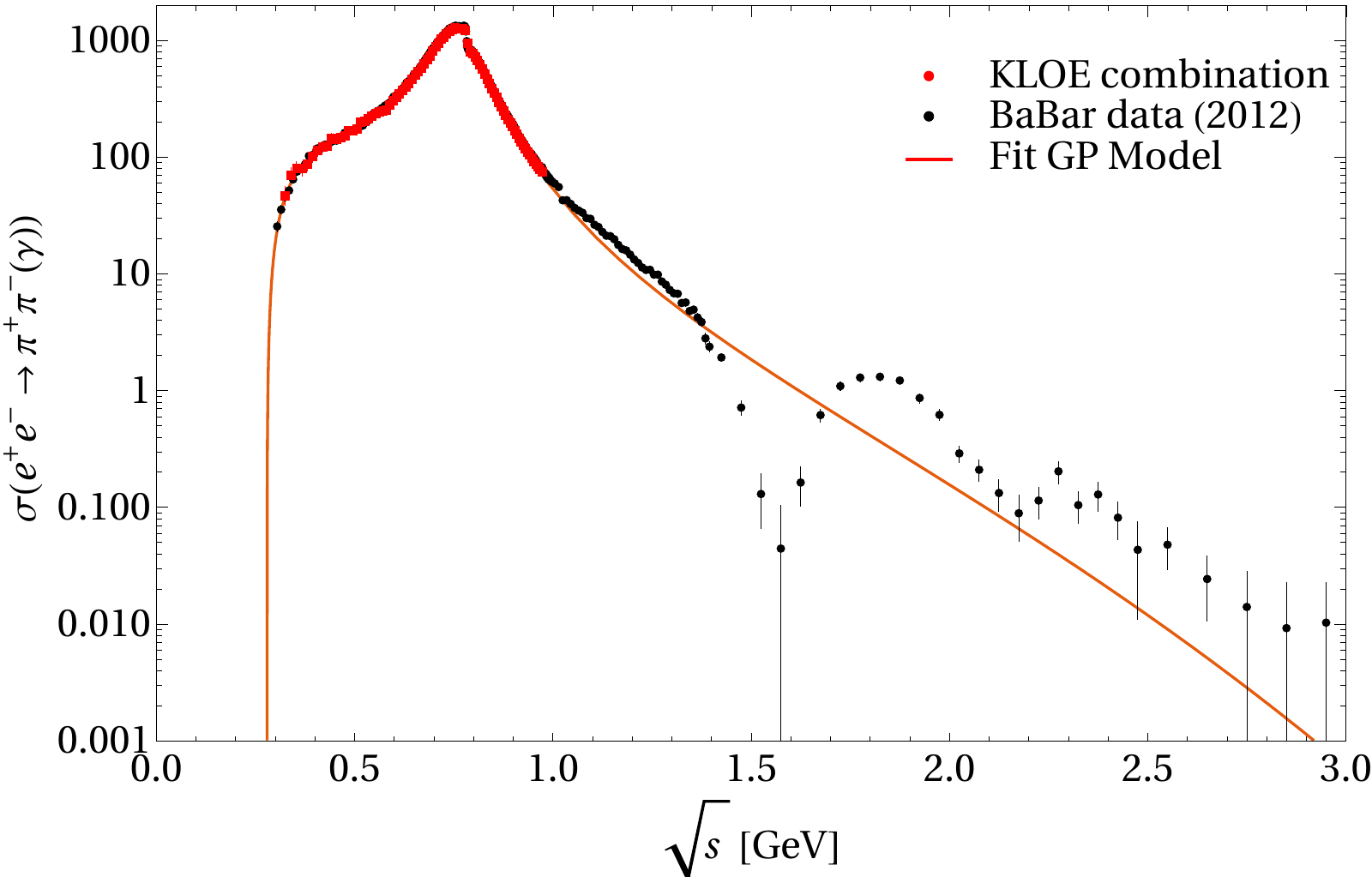}
    \includegraphics[width=0.44\textwidth]{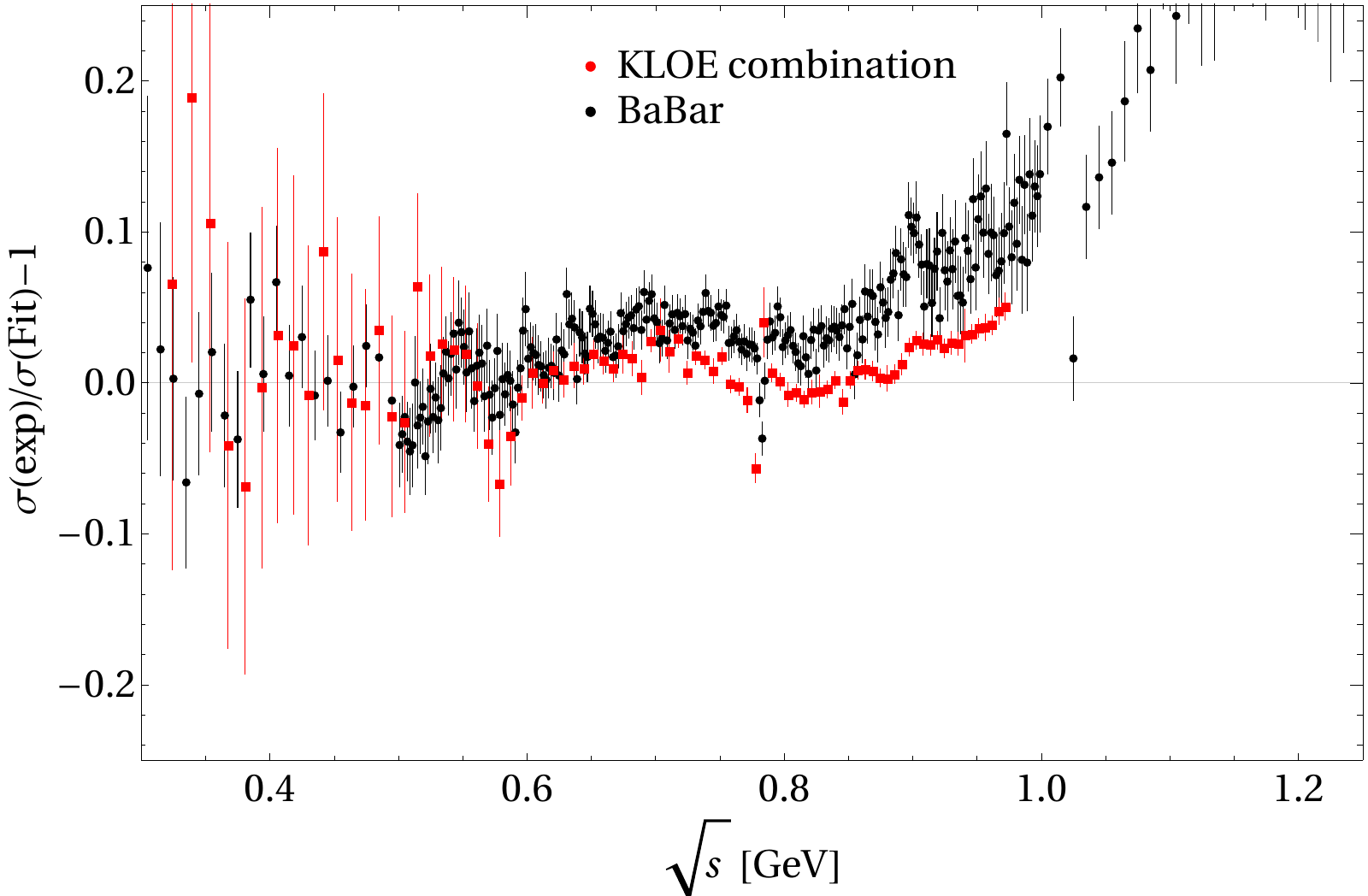}
    \captionsetup{width=0.88\linewidth}
    \caption{ Global fit results for the GP model versus data. 
    }
    \label{fig:Fit_GP}
\end{figure}

\begin{figure}[ht]
    \centering
    \includegraphics[width=0.44\textwidth]{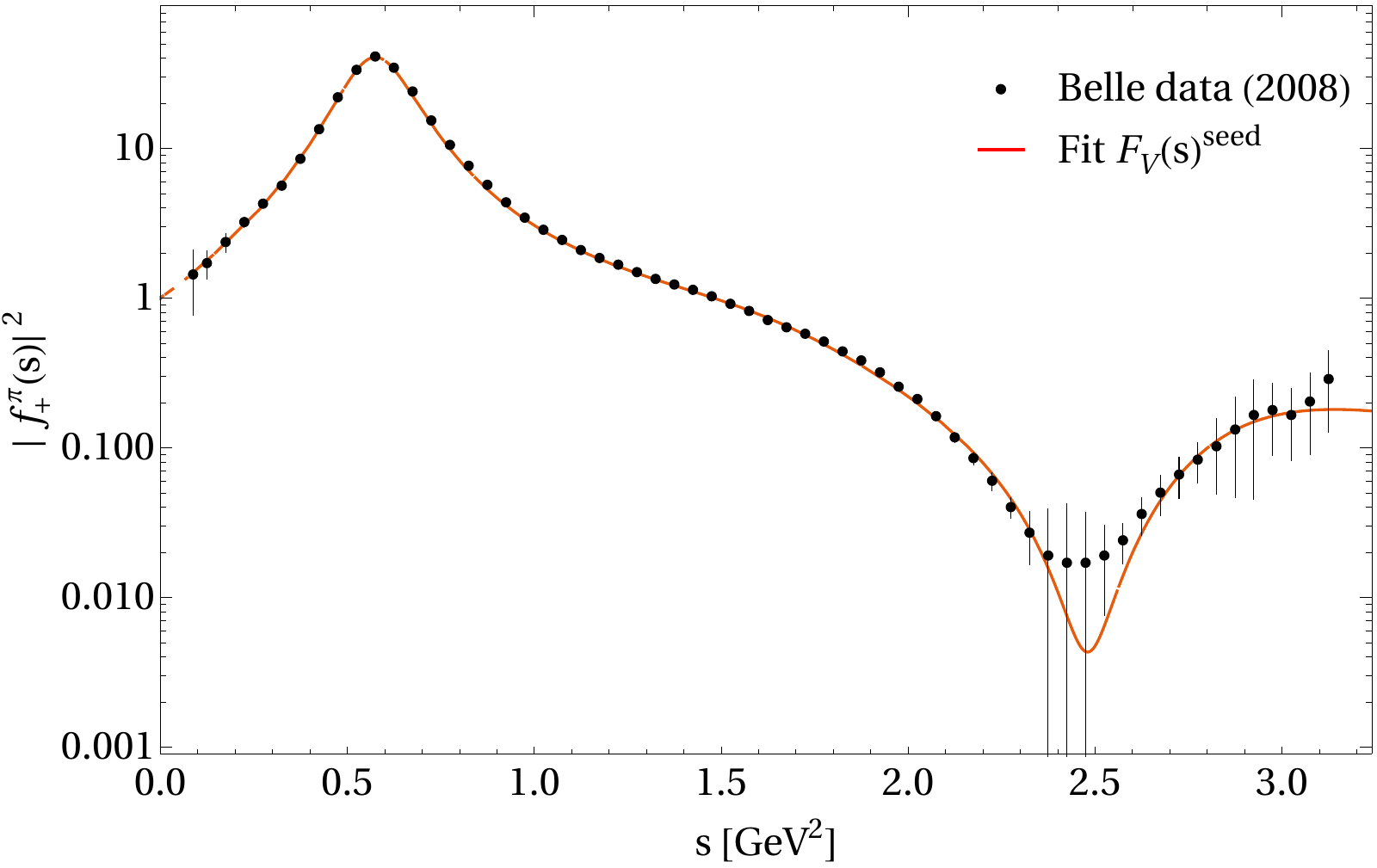}
    \includegraphics[width=0.44\textwidth]{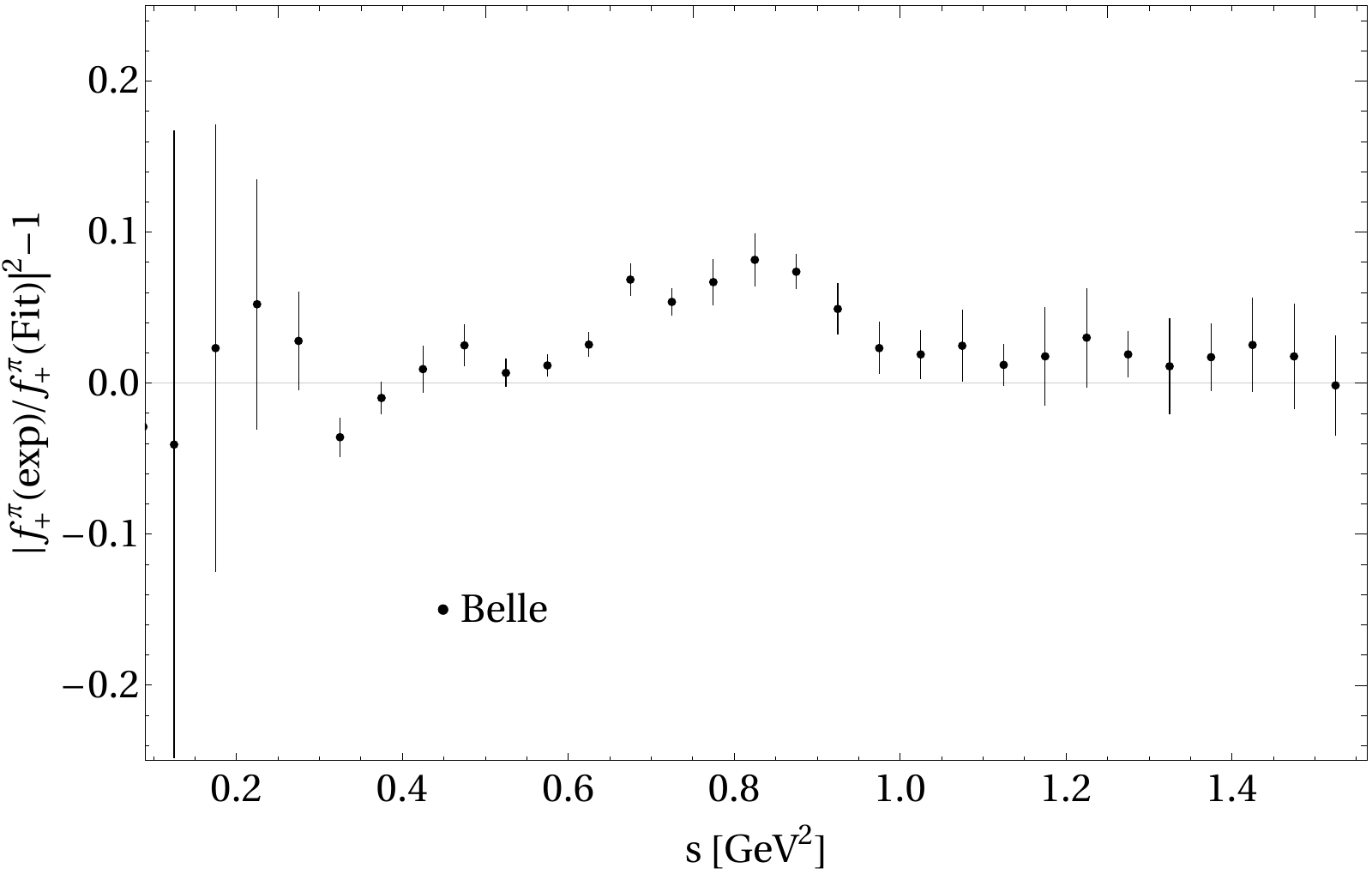}
    \includegraphics[width=0.44\textwidth]{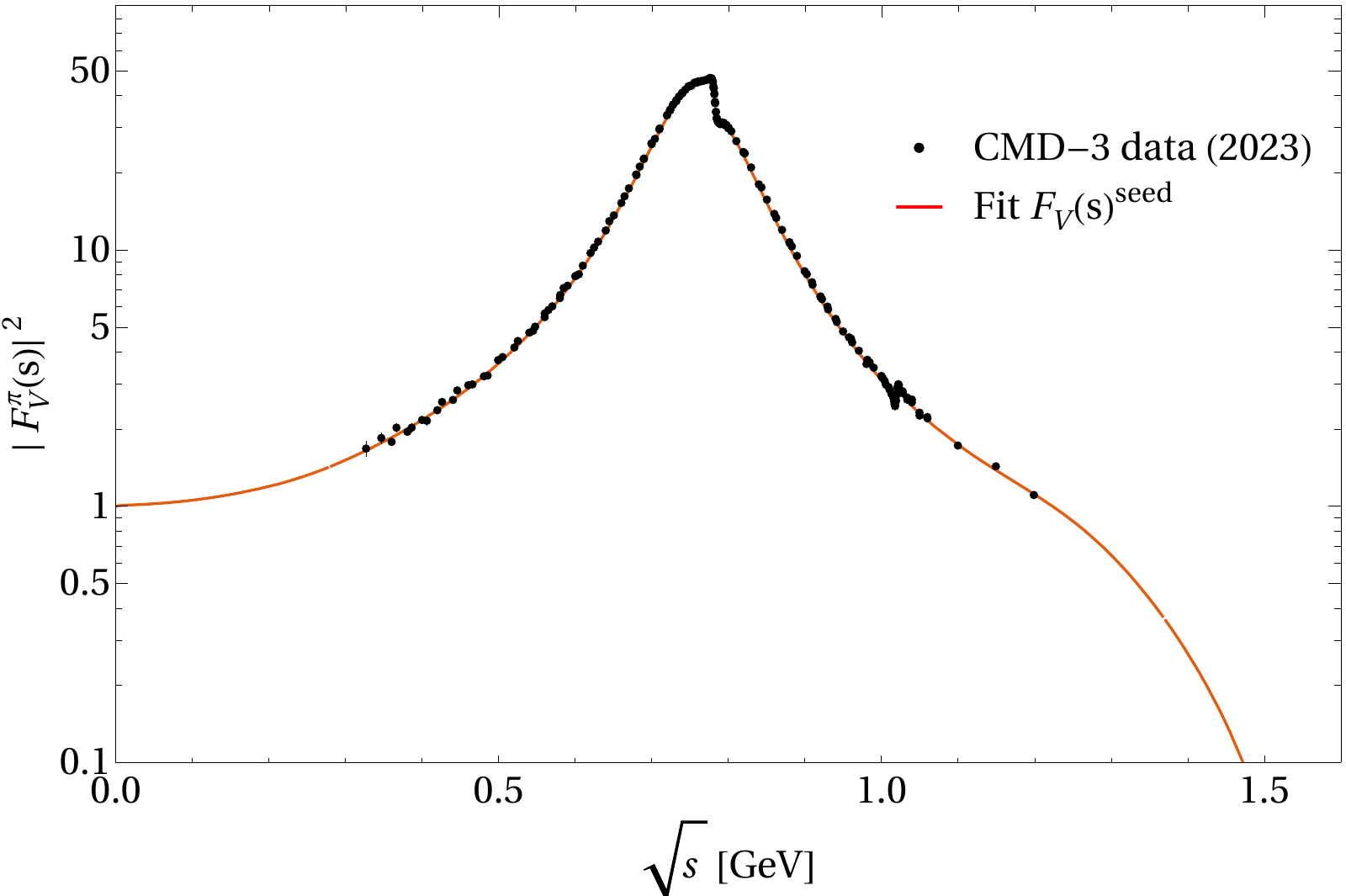}
    \includegraphics[width=0.44\textwidth]{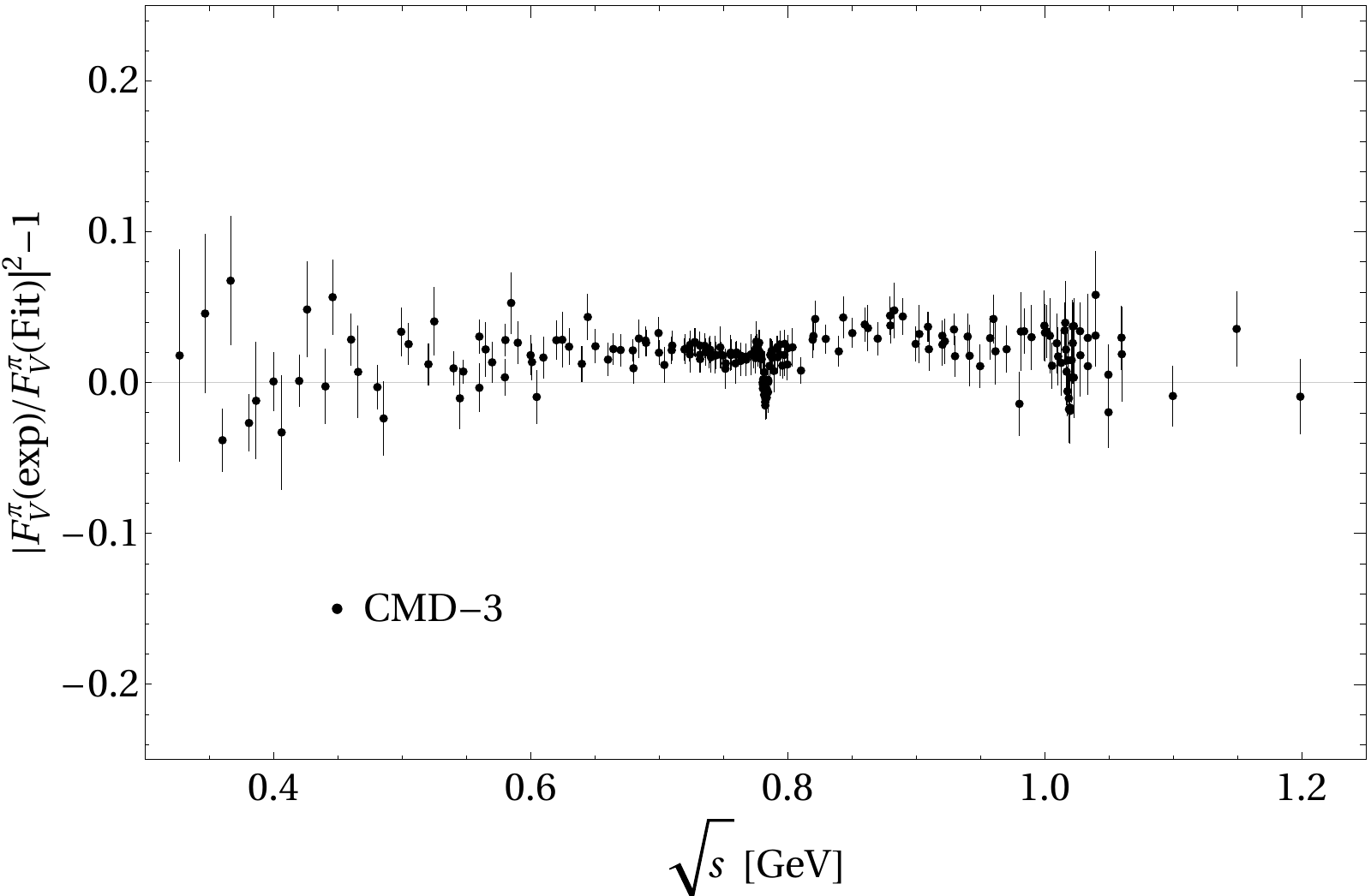}
    \includegraphics[width=0.44\textwidth]{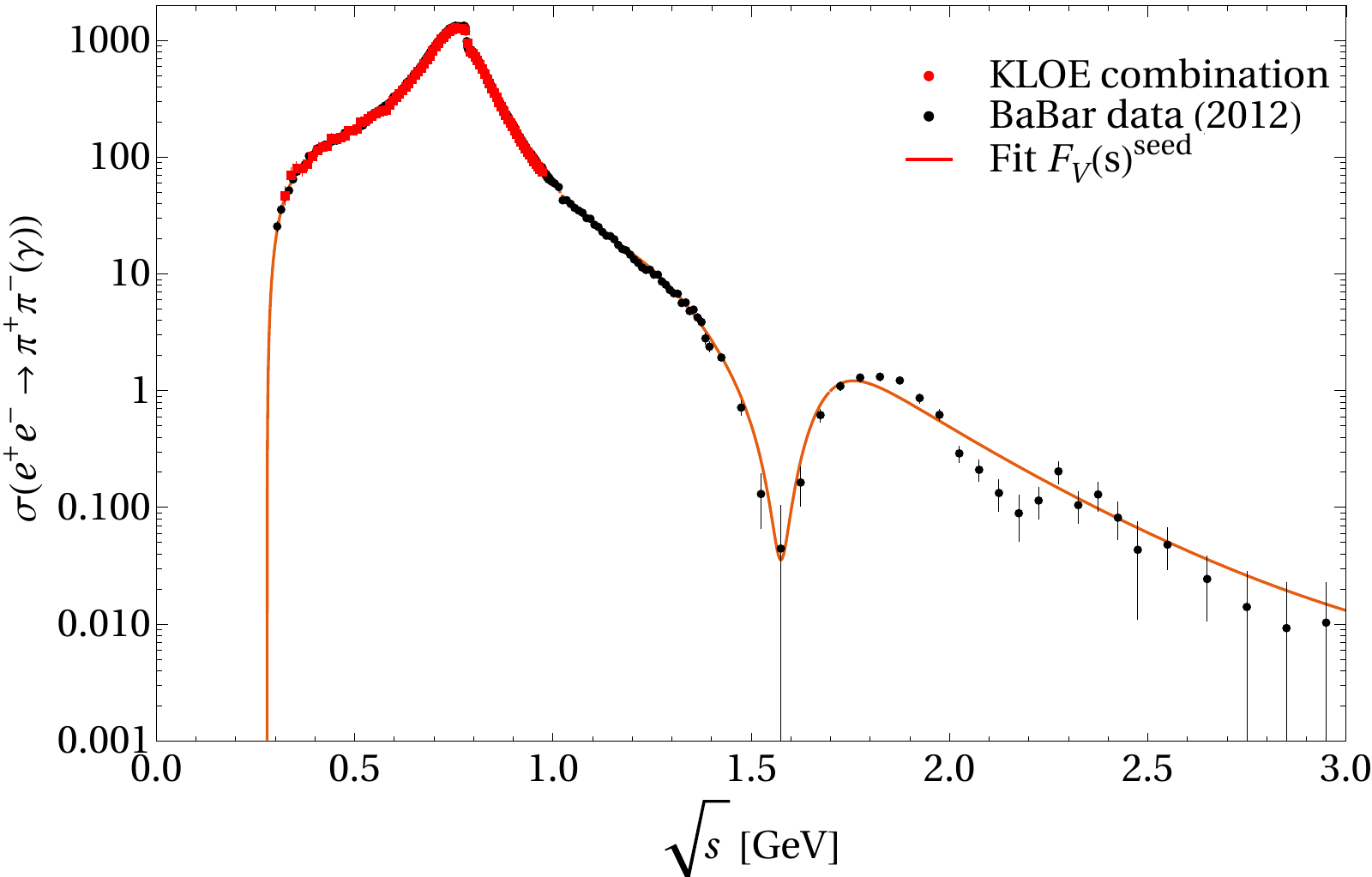}
    \includegraphics[width=0.44\textwidth]{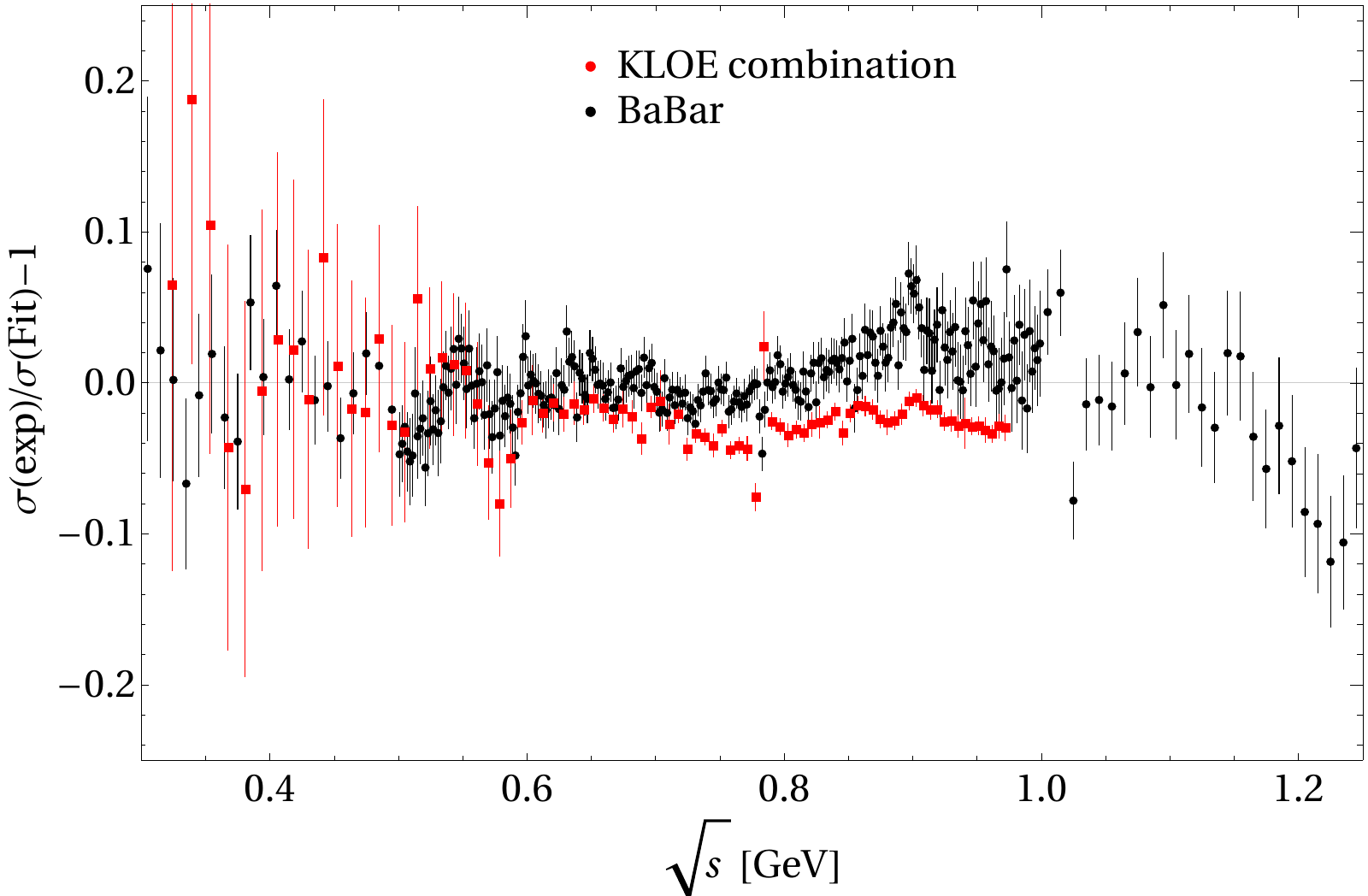}
    \captionsetup{width=0.88\linewidth}
    \caption{ Global fit results for $F_V(s)^{\mathrm{seed}}$ versus data. 
    }
    \label{fig:Fit2_GPp}
\end{figure}

The corresponding results are outlined in Tables \ref{tab:GS_Fit}, \ref{tab:KS_Fit}, \ref{tab:GP_Fit_All} and \ref{tab:GPg_Fit}~\footnote{GP is only shown for comparison as its reliability is limited to $s\lesssim 0.8\,\mathrm{GeV}^2$, given the absence of the $\rho^{\prime(\prime)}$ resonances.}. The first four columns show the outcome of a fit to either BaBar, KLOE or CMD-3 $e^+e^-$ measurements and the tau data from Belle (the $F_V(s)$ and $f_+(s)$ form factors, describing the structure-dependent part of the $e^+e^-$ and $\tau$ data, are equal up to the IB corrections that we have discussed~\footnote{We neglect IB in the masses and widths of the excited isovector resonances, to which data are not yet sensitive.}). The best agreement is found between BaBar and Belle, while the CMD-3/Belle agreement is slightly better than that from KLOE/Belle. The last two columns are the results from a global fit using the KLOE 12 data~\cite{KLOE:2012anl} or a combined KLOE measurement for $\sigma(e^+e^-\to\pi^+\pi^-\gamma(\gamma))$~\cite{KLOE-2:2017fda}, respectively. The main difference between experiments seems to be related to $\text{arg}[\delta_{\rho\omega}]$. The results from the global fit using the combined KLOE data are depicted in Figs. \ref{fig:Fit2_GS}, \ref{fig:Fit_KS}, \ref{fig:Fit_GP} and \ref{fig:Fit2_GPp}. We observe that the comparisons of the best fit results to the different data sets are quite consistent for the GS, KS and Seed parametrizations (GP is only shown for completeness, as it cannot describe data beyond $s\sim 0.8$ GeV$^2$). 

\begin{figure}[ht]
    \centering
    \includegraphics[width=0.44\textwidth]{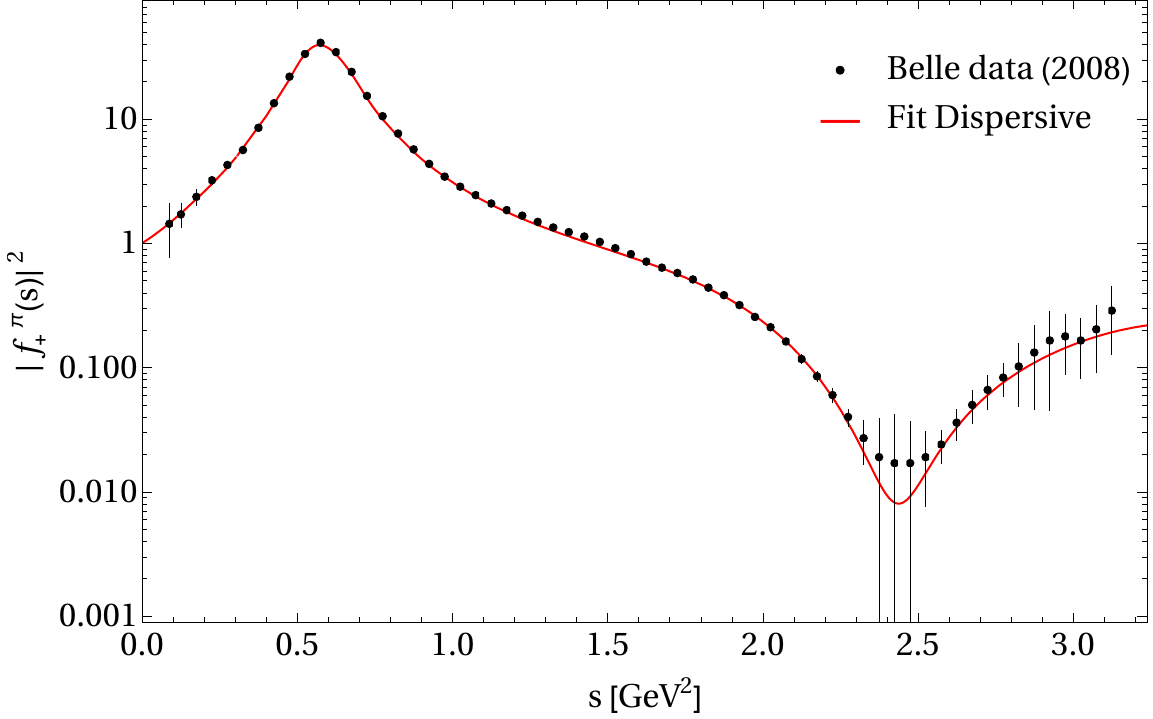}
    \includegraphics[width=0.44\textwidth]{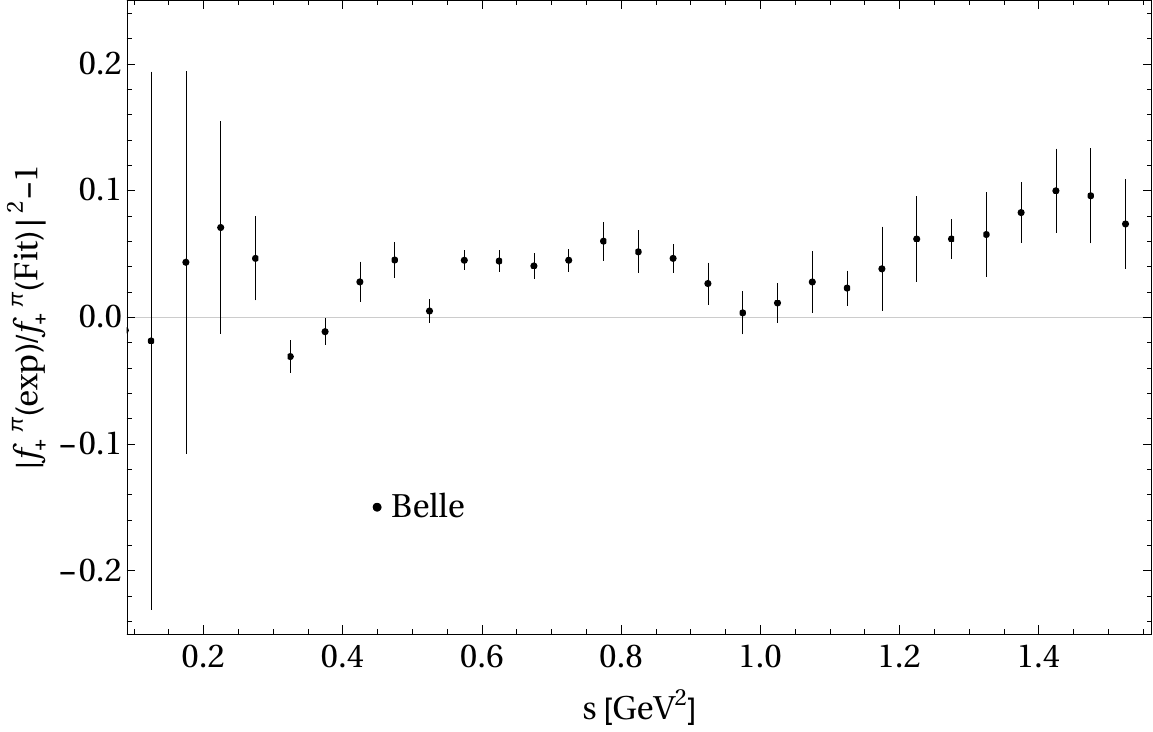}
    \includegraphics[width=0.44\textwidth]{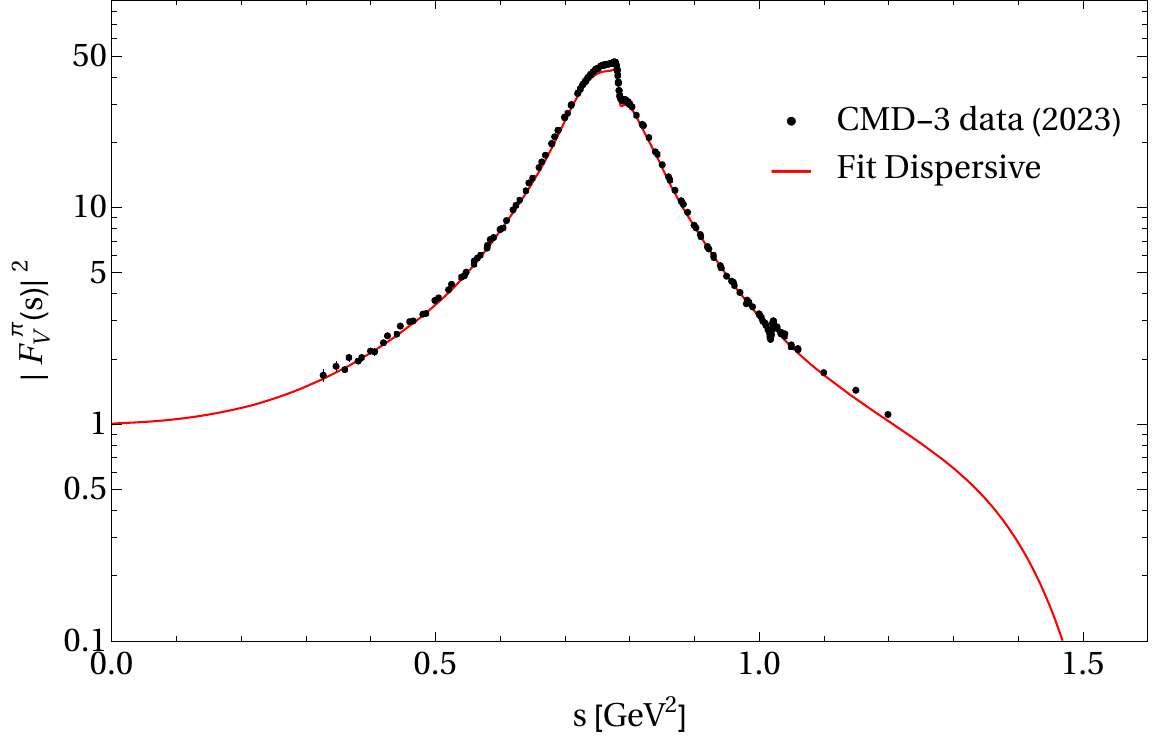}
    \includegraphics[width=0.44\textwidth]{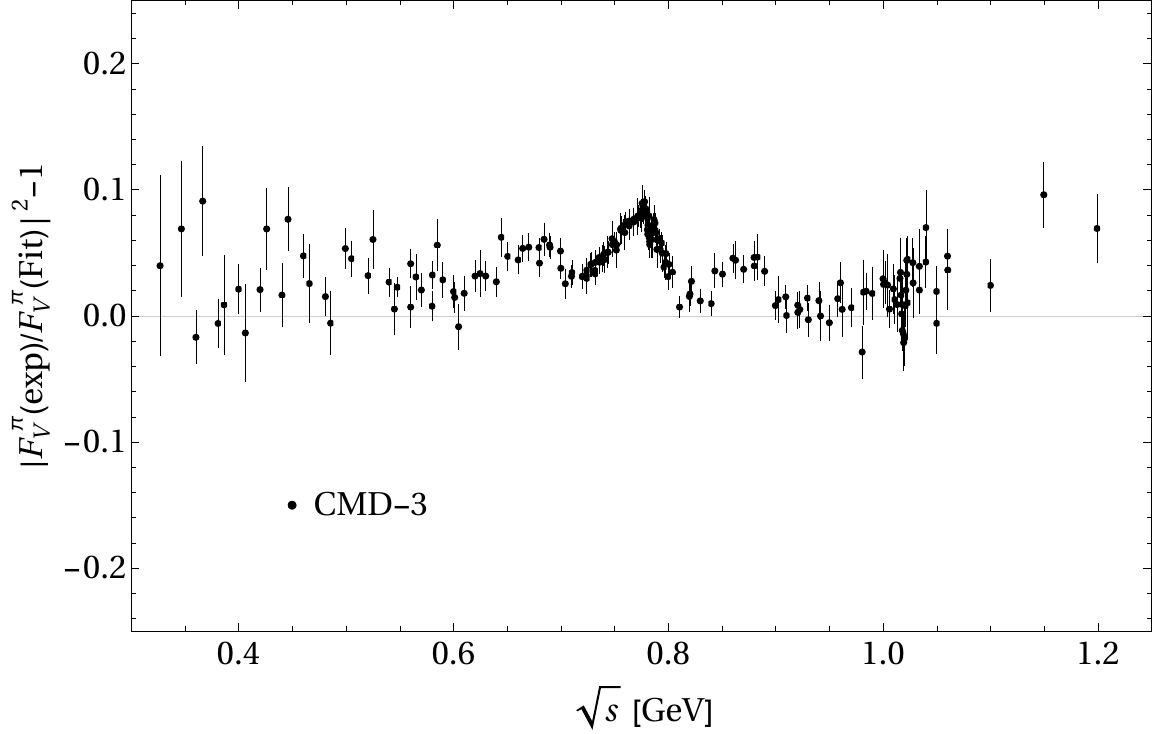}
    \includegraphics[width=0.44\textwidth]{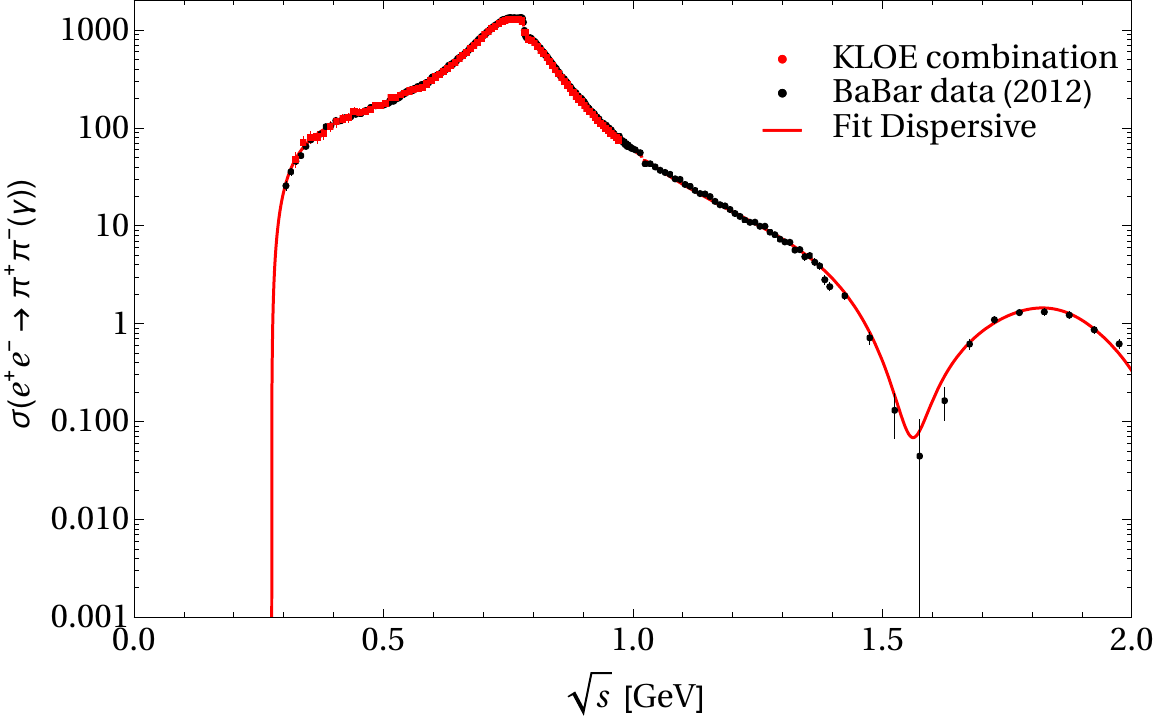}
    \includegraphics[width=0.44\textwidth]{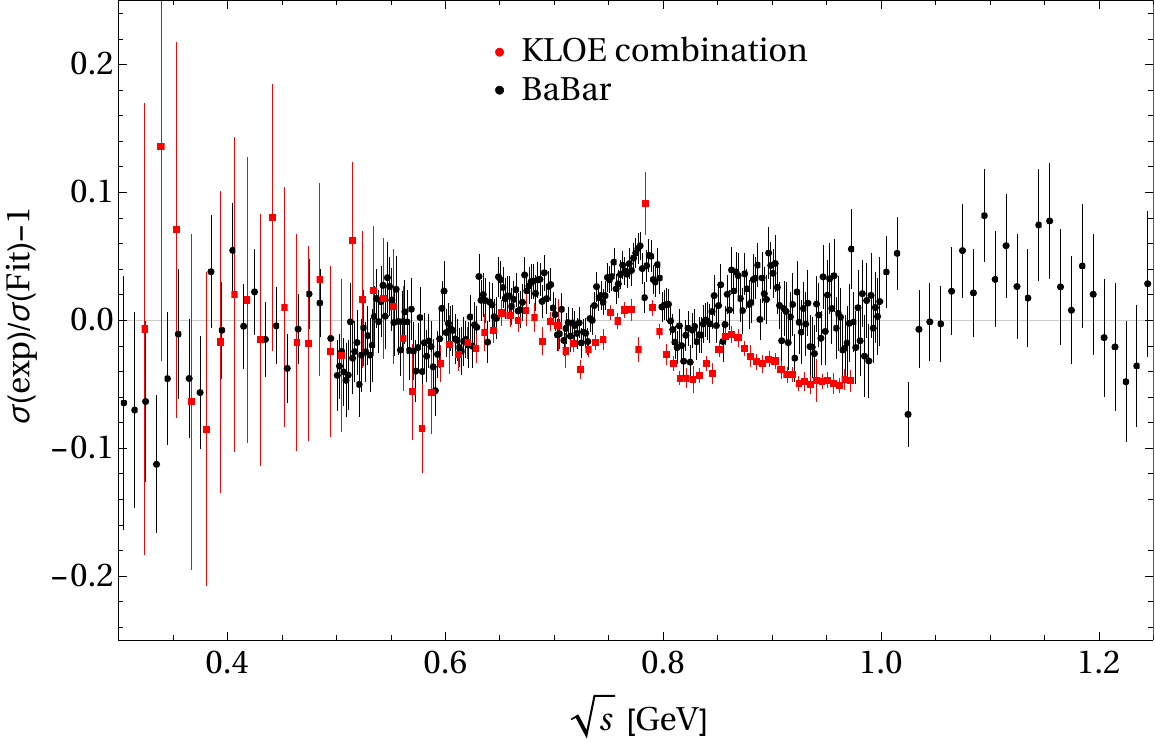}
    \captionsetup{width=0.88\linewidth}
    \caption{ Global fit results for the $p_{4-1}$ dispersive model versus data.
    }
    \label{fig:Fit2_Dispersivep4v4}
\end{figure}

\begin{figure}[ht]
    \centering
    \includegraphics[width=0.44\textwidth]{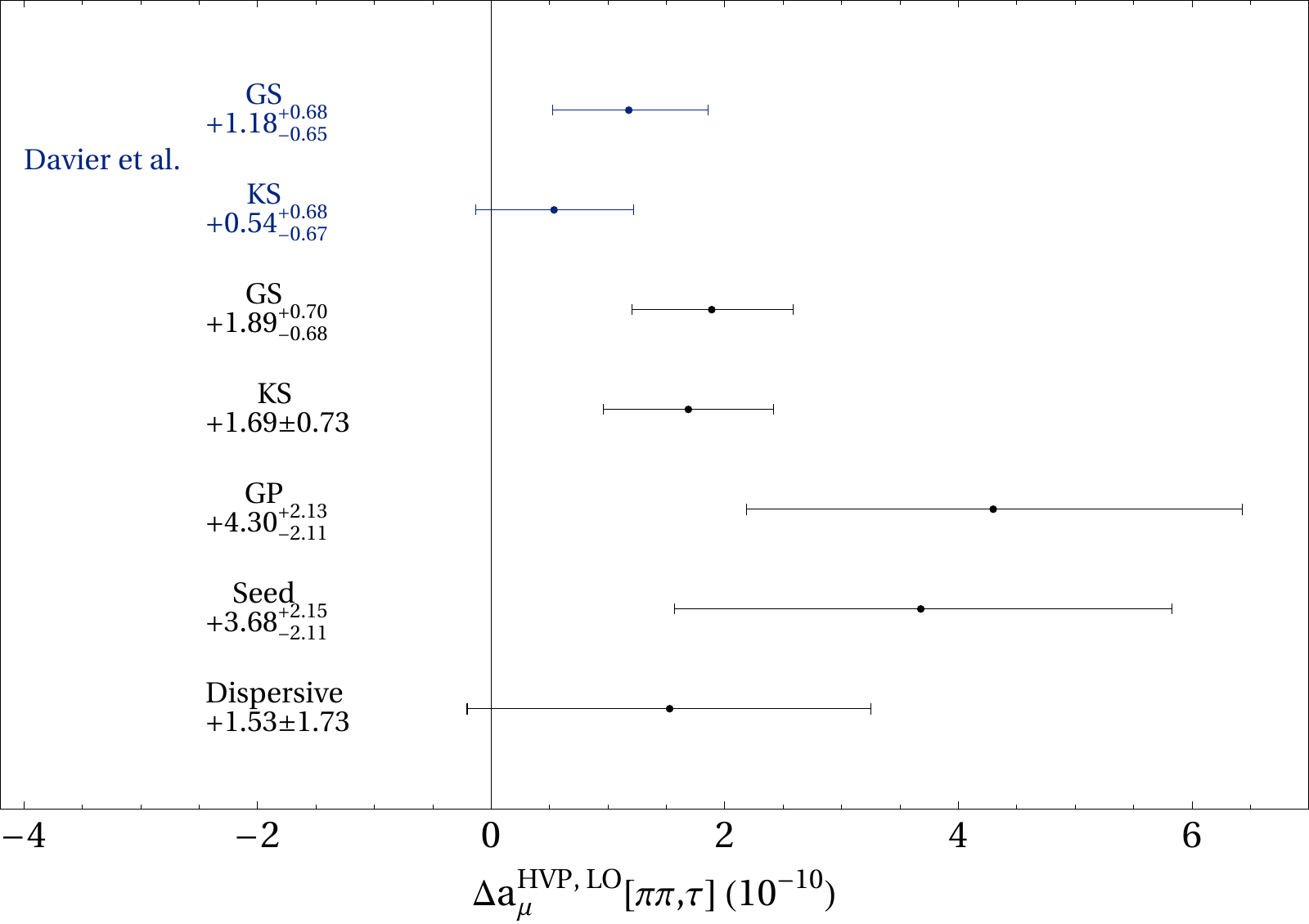}
    \includegraphics[width=0.449\textwidth]{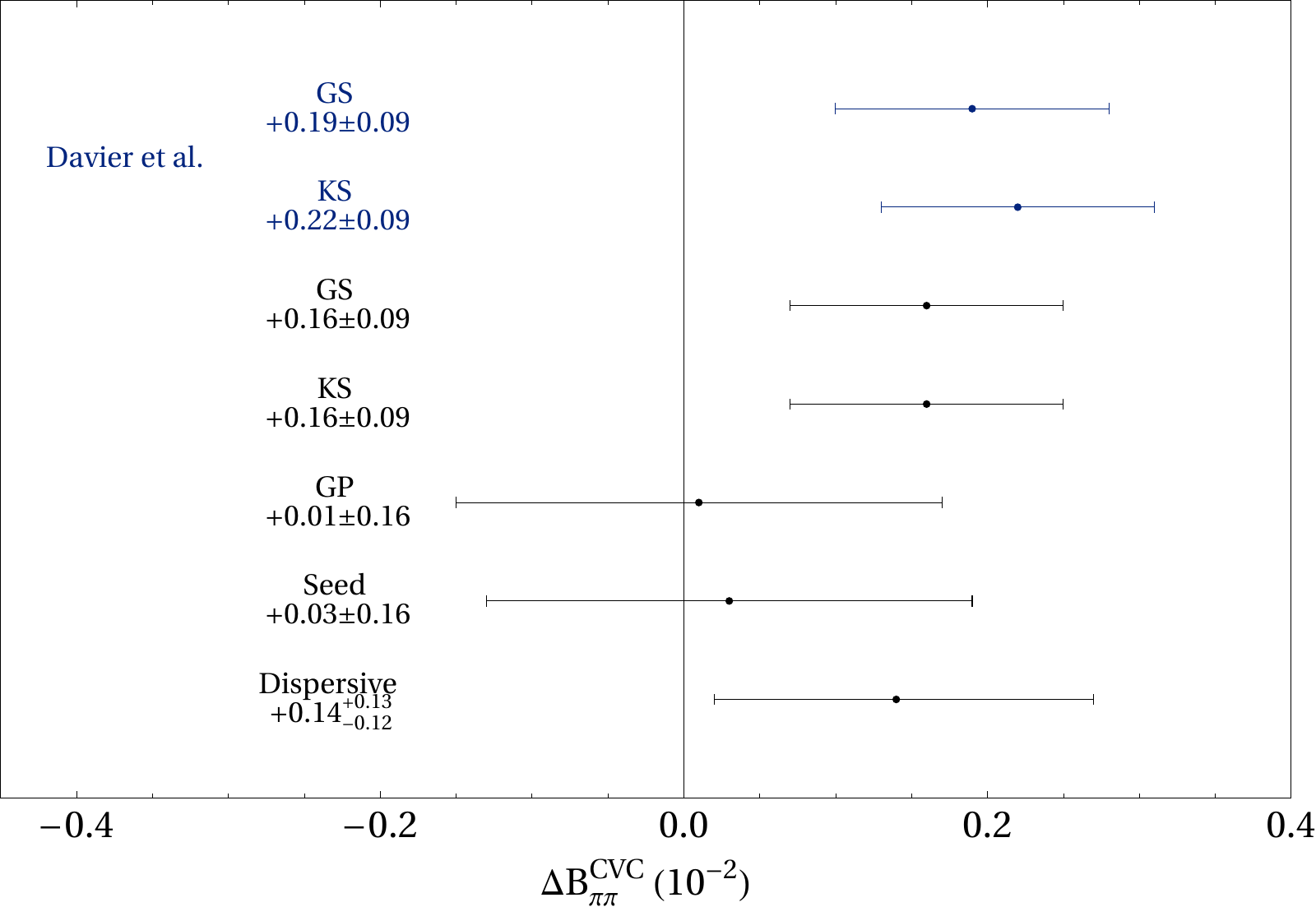}
    \caption{IB corrections in the ratio of the form factors $\vert F_V(s)/f_+(s)\vert$ to $a_\mu^\text{HVP, LO}$ and $\mathcal{B}^\text{CVC}_{\pi\pi}$. {\color{black}The number on the left side of the plots is the sum of the IB corrections in Tables \ref{tab:IB_BR_CMD3} and \ref{tab:IB_amu_CMD3}, from row 5 to 11.} }
    \label{fig:IB_correctionsFF}
\end{figure}

Using the results in the last column of the quoted tables for these models, we update the numbers in Table \ref{tab:IB_BR} for $\Delta\mathcal{B}^\mathrm{CVC}_{\pi\pi}$ and $\Delta a_\mu^{\mathrm{HVP,\,LO}}$. The new outcomes (which also include the results of the dispersive fits in Tables \ref{tab:charge_radius}, \ref{tab:IB_BRFF_disp}, \ref{tab:IB_amuFF_disp}, and Fig.~\ref{fig:Fit2_Dispersivep4v4}, to be discussed below) are summarized in Tables \ref{tab:IB_BR_CMD3} and \ref{tab:IB_amu_CMD3}, and displayed in Fig. \ref{fig:IB_correctionsFF}. 

In an effort to perform a reliable comparison between the diverse models, we use the same input for $\Delta\Gamma_\rho$ and $\Delta M_\rho$, as previously done 
 for the GS and KS models \cite{Davier:2010fmf}~\footnote{Our attempts of fitting also this information from the data yielded only upper bounds for $(\Delta M_{\rho}, \Delta\Gamma_\rho)$, consistent with our inputs.}. The results obtained using the central values for these IB corrections are slightly different, but they agree within uncertainties with Ref.~\cite{Davier:2010fmf}. 
To understand the difference between the GP model with respect to the GS/KS models in Table~\ref{tab:IB_BR}, we split the IB corrections in the ratio of the pion form factors due to $\Delta M_\rho$ accounting for Eq. (\ref{eq:rho_width_GP}). This effect, which is absent in the GS/KS models, is shown in the seventh row of Tables \ref{tab:IB_BR_CMD3} and \ref{tab:IB_amu_CMD3}. A nice agreement is observed among the GS/KS model and the dispersive one concerning the effect of $\Delta M_\rho$ in the eighth row, although better for $\Delta \mathcal{B}_{\pi\pi}^\mathrm{CVC}$ than for $\Delta a_\mu^{\mathrm{HVP,\,LO}}{[\pi\pi,\tau]}$, according to the different relative weights of the low and $\rho$-resonance energy regions. 
{Following Ref. \cite{Davier:2010fmf}, we assign a 10\% uncertainty to the FSR and $\pi\pi\gamma$ IB corrections. In the former case, which is computed using scalar QED, it allows to account for missing structure-dependent corrections. For the latter, the calculated structure-dependent QED corrections within the VMD model turn out to be small; we estimated the associated uncertainty from a comparison of theory to the measured rate for $\rho^0\to \pi^+\pi^-\gamma$. Uncertainties in other IB sources in Tables \ref{tab:IB_BR_CMD3} and \ref{tab:IB_amu_CMD3} are estimated from the errors in input parameters and the experimental di-pion spectrum.}

Additionally, we perform a global fit to all the $e^+e^-$ and $\tau$ data using the dispersive parametrization in Eq.~(\ref{eq:disp02})~\footnote{As an alternative approach, we tried to obtain the IB corrections using a thrice-subtracted dispersion relation, as in Ref.~\cite{Gonzalez-Solis:2019iod}, from a global fit to $e^+e^-$ and $\tau$ data. We included explicitly the IB entering the dispersive form factor in the fit, finding $\Delta \Gamma_\rho=0.99(4)\,\mathrm{MeV}$ and $\Delta M_\rho=0.35(10)\,\mathrm{MeV}$; however, the uncertainties associated with IB effects in the subtraction constants -- which are very correlated with $(\Delta M_\rho, \Delta\Gamma_\rho)$ -- were too large, giving an unrealistic behavior. This is a reason to prefer the dispersive fits including the conformal polynomial.}. These results are partially depicted in Table \ref{tab:charge_radius} with the pion charge radius predicted using Eq. (\ref{eq:radius2}). The results in the second column, $p_0$, correspond to the fit without $G_\mathrm{in}^N$, and the other columns are fits using different approximations for the conformal polynomial. For those where the restriction in Eq. (\ref{eq:in_rest}) was used, we write $p_{N-1}$, where $N$ is the order of $G_\mathrm{in}^N$. In Fig. \ref{fig:Fit2_Dispersivep4v4}, we show the results with the $p_{4-1}$ dispersive model, which yields the smallest reduced $\chi^2$ (and also the best agreement with Ref.~\cite{Colangelo:2018mtw} for the pion charge radius, see Table \ref{tab:charge_radius}), that align well with the previous fits using the GS, KS and Seed models. The IB effects (detailed in Tables \ref{tab:IB_BRFF_disp} and \ref{tab:IB_amuFF_disp}, for $\Delta \mathcal{B}_{\pi\pi}^{\text{CVC}}$ and $\Delta a_\mu^\text{HVP, LO}[\pi\pi,\tau]$, respectively) were included in Fig. \ref{fig:IBCtoFF}, where the $\rho-\omega$ mixing was shown on the left and the other IB corrections on the right. A second peak (bottom-right in Figure \ref{fig:IBCtoFF}), besides the one at the $\rho$ mass, is observed below $0.2\,\mathrm{GeV}^2$. It is related to the IB correction in $\Gamma_\rho$ due to the $\pi$ mass splitting and the electromagnetic decay. The IB corrections from the dispersive model are also included in Tables \ref{tab:IB_BR_CMD3} and \ref{tab:IB_amu_CMD3}. We use as central value the Global fit 2 number in Table \ref{tab:Disp_Fit_v4}  from the $p_{4-1}$ dispersive parametrization. 
The second uncertainty in the total IB effect for the dispersive approach in Tables \ref{tab:IB_BR_CMD3} and \ref{tab:IB_amu_CMD3} takes into account the spread of values obtained using the different approximations for the conformal polynomial (in Tables \ref{tab:IB_BRFF_disp} and \ref{tab:IB_amuFF_disp}). 

\begin{table}[ht]
    \centering
    \resizebox{0.90\textwidth}{!}{\begin{tabular}{|c|cccccc|}
    \hline
         & BABAR12+Belle & KLOE12+Belle & KLOEc+Belle & CMD3+Belle & Global fit 1 & Global fit 2\\[0.3ex]
        data points  & 337 + 62 & 60 + 62 & 85 + 62 & 209 + 62 & 337+60+209+62 & 337+85+209+62 \\[0.3ex]
        $\chi^2_{ee}+\chi^2_\tau$ & $327.4 + 99.5$ & $157.9 + 274.5$ & $138.2 + 377.5$ & $181.6 + 87.3$ & $1686.8 + 143.4$ & $2786.8 + 214.1$ \\[0.3ex]
        $\chi^2$ & $426.9$ & $432.4$ & $515.7$ & $268.9$ & $1830.2$ & $3000.9$ \\[0.3ex]
         \hline
        $\chi^2/\text{d.o.f}$ & $1.1$ & $3.9$ & $3.8$ & $1.0$ & $2.8$ & $4.4$ \\[0.3ex]
        $m_{\rho}$ & $774.0\pm0.1\text{ MeV}$ & $773.4\pm0.2\text{ MeV}$ & $773.5\pm0.1\text{ MeV}$ & $773.4\pm0.1\text{ MeV}$ & $773.5\pm0.1\text{ MeV}$ & $772.9\pm0.1\text{ MeV}$ \\[0.3ex] 
        $\Gamma_{\rho}$ & $148.9\pm0.3\text{ MeV}$ & $144.7\pm0.5\text{ MeV}$ & $147.1\pm0.3\text{ MeV}$ & $147.6\pm0.2\text{ MeV}$ & $146.6\pm0.2\text{ MeV}$ & $146.9\pm0.2\text{ MeV}$\\[0.3ex]
        $\vert\delta_{\rho\omega}\vert$ & $(2.1\pm0.0)\cdot 10^{-3}$ & $(1.9\pm0.1)\cdot 10^{-3}$ & $(1.8\pm0.0)\cdot 10^{-3}$ & $(2.0\pm0.0)\cdot 10^{-3}$ & $(1.9\pm0.0)\cdot 10^{-3}$ & $(2.0\pm0.0)\cdot 10^{-3}$ \\[0.3ex]
        arg[$\delta_{\rho\omega}$] & $(10.8\pm1.1)^\circ$ & $(13.8\pm2.6)^\circ$ & $(18.2\pm2.6)^\circ$ & $(12.0\pm0.5)^\circ$ & $(10.1\pm0.4)^\circ$ & $(5.5\pm0.4)^\circ$ \\[0.3ex]
        $\vert\delta_{\rho\phi}\vert$ & $0^\dagger$ & $0^\dagger$ & $0^\dagger$ & $(2.2\pm0.2)\cdot 10^{-4}$ & $(1.5\pm0.2)\cdot 10^{-4}$ & $(1.4\pm0.2)\cdot 10^{-4}$ \\[0.3ex]
        arg[$\delta_{\rho\phi}$] & $-$ & $-$ & $-$ & $(83.8\pm5.9)^\circ$ & $(68.4\pm7.8)^\circ$ & $(46.6\pm8.6)^\circ$ \\[0.3ex]
        \hline
        $m_{\rho^\prime}$ & $1460.9\pm5.9\text{ MeV}$ & $1412.9\pm7.8\text{ MeV}$ & $1398.9\pm6.7\text{ MeV}$ & $1459.0\pm7.0\text{ MeV}$ & $1433.3\pm4.8\text{ MeV}$ & $1443.7\pm5.2\text{ MeV}$\\[0.3ex] 
        $\Gamma_{\rho^\prime}$ & $444\pm14\text{ MeV}$ & $441\pm20\text{ MeV}$ & $445\pm20\text{ MeV}$ & $450\pm20\text{ MeV}$ & $403\pm12\text{ MeV}$ & $391\pm11\text{ MeV}$\\[0.3ex]
        Re[$c_{\rho^{\prime}}$] & $-0.12\pm0.00$ & $-0.11\pm0.00$ & $-0.11\pm0.01$ & $-0.10\pm0.00$ & $-0.09\pm0.00$ & $-0.11\pm0.00$ \\[0.3ex]
        Im[$c_{\rho^{\prime}}$] & $-0.03\pm0.01$ & $0.03\pm0.01$ & $-0.26\pm0.02$ & $-0.20\pm0.02$ & $-0.17\pm0.01$ & $-0.02\pm0.00$ \\[0.3ex]
        $m_{\rho^{\prime\prime}}$ & $1806.7\pm9.7\text{ MeV}$ & $1730^\dagger\text{ MeV}$ & $1730^\dagger\text{ MeV}$ & $1730^\dagger\text{ MeV}$ & $1784.8\pm10.5\text{ MeV}$ & $1816.8\pm9.2\text{ MeV}$\\[0.3ex] 
        $\Gamma_{\rho^{\prime\prime}}$ & $273\pm16\text{ MeV}$ & $260^\dagger\text{ MeV}$ & $260^\dagger\text{ MeV}$ & $260^\dagger\text{ MeV}$ & $245\pm14\text{ MeV}$ & $245\pm16\text{ MeV}$\\[0.3ex]
        Re[$c_{\rho^{\prime\prime}}$] & $(2.9\pm0.5)\cdot 10^{-2}$ & $(4.9\pm0.6)\cdot 10^{-2}$ & $(-7.7\pm0.7)\cdot 10^{-2}$ & $(-7.5\pm0.6)\cdot 10^{-2}$ & $(-6.6\pm0.6)\cdot 10^{-2}$ & $(2.1\pm0.4)\cdot 10^{-2}$ \\[0.3ex]
        Im[$c_{\rho^{\prime\prime}}$] & $(4.5\pm0.3)\cdot 10^{-2}$ & $(1.7\pm0.3)\cdot 10^{-2}$ & $(-1.7\pm0.6)\cdot 10^{-3}$ & $(5.8\pm6.7)\cdot 10^{-3}$ & $(-2.9\pm0.4)\cdot 10^{-2}$ & $(4.0\pm0.2)\cdot 10^{-2}$ \\[0.3ex]
        \hline
    \end{tabular}}
    \captionsetup{width=0.88\linewidth}
    \caption{Fit results for the GS model using different $e^+e^-$ data sets and Belle $\tau$ decay measurements. Here, and in the following tables, we use `Global fit 1' for the one including KLOE12 data and `Global fit 2' for the one including KLOEc data instead. \textcolor{black}{The only two-sided p-values which exceed $10^{-4}$ are obtained for the BABAR12+Belle and CMD3+Belle fits, which are $0.161$ and $0.525$, respectively.}}
    \label{tab:GS_Fit}
\end{table}

\begin{table}[ht]
    \centering
    \resizebox{0.90\textwidth}{!}{\begin{tabular}{|c|cccccc|}
    \hline
         & BABAR12+Belle & KLOE12+Belle & KLOEc+Belle & CMD3+Belle & Global fit 1 & Global fit 2\\[0.3ex]
        data points  & 337 + 62 & 60 + 62 & 85 + 62 & 209 + 62 & 337+60+209+62 & 337+85+209+62 \\[0.3ex]
        $\chi_{ee}^2+\chi^2_\tau$ & $395.7 + 122.7$ & $181.5 + 314.8$ & $236.1 + 432.2$ & $463.3 + 151.5$ & $2070.6 + 200.0$ & $3100.7 + 318.8$ \\[0.3ex]
        $\chi^2$ & $518.4$ & $496.3$ & $668.3$ & $614.8$ & $2270.6$ & $3419.5$ \\[0.3ex]
         \hline
        $\chi^2/\text{d.o.f}$ & $1.3$ & $4.4$ & $4.9$ & $2.4$ & $3.5$ & $5.0$ \\[0.3ex]
        $m_\rho$ & $772.3\pm0.1\text{ MeV}$ & $771.8\pm0.2\text{ MeV}$ & $771.6\pm0.1\text{ MeV}$ & $771.4\pm0.1\text{ MeV}$ & $771.8\pm0.1\text{ MeV}$ & $771.5\pm0.1\text{ MeV}$ \\[0.3ex] 
        $\Gamma_\rho$ & $147.5\pm0.3\text{ MeV}$ & $142.5\pm0.4\text{ MeV}$ & $145.0\pm0.3\text{ MeV}$ & $146.0\pm0.2\text{ MeV}$ & $145.5\pm0.2\text{ MeV}$ & $145.7\pm0.2\text{ MeV}$\\[0.3ex]
        $\vert\delta_{\rho\omega}\vert$ & $(1.9\pm0.0)\cdot 10^{-3}$ & $(1.7\pm0.1)\cdot 10^{-3}$ & $(1.7\pm0.0)\cdot 10^{-3}$ & $(2.0\pm0.0)\cdot 10^{-3}$ & $(1.9\pm0.0)\cdot 10^{-3}$ & $(1.9\pm0.0)\cdot 10^{-3}$ \\[0.3ex]
        arg[$\delta_{\rho\omega}$] & $(11.7\pm1.1)^\circ$ & $(17.6\pm2.9)^\circ$ & $(17.6\pm2.7)^\circ$ & $(11.3\pm0.5)^\circ$ & $(9.5\pm0.4)^\circ$ & $(6.5\pm0.4)^\circ$ \\[0.3ex]
        $\vert\delta_{\rho\phi}\vert$ & $0^\dagger$ & $0^\dagger$ & $0^\dagger$ & $(2.8\pm0.2)\cdot 10^{-4}$ & $(1.7\pm0.2)\cdot 10^{-4}$ & $(1.4\pm0.2)\cdot 10^{-4}$ \\[0.3ex]
        arg[$\delta_{\rho\phi}$] & $-$ & $-$ & $-$ & $(87.0\pm4.9)^\circ$ & $(78.9\pm7.5)^\circ$ & $(73.2\pm8.1)^\circ$ \\[0.3ex]
        \hline
        $m_{\rho^\prime}$ & $1536.1\pm9.5\text{ MeV}$ & $1425.2\pm8.3\text{ MeV}$ & $1476.7\pm8.0\text{ MeV}$ & $1575.8\pm10.9\text{ MeV}$ & $1564.9\pm9.3\text{ MeV}$ & $1547.5\pm8.1\text{ MeV}$\\[0.3ex] 
        $\Gamma_{\rho^\prime}$ & $538\pm18\text{ MeV}$ & $471\pm18\text{ MeV}$ & $474\pm14\text{ MeV}$ & $717\pm23\text{ MeV}$ & $450\pm17\text{ MeV}$ & $447\pm16\text{ MeV}$\\[0.3ex]
        Re[$c_{\rho^{\prime}}$] & $-0.13\pm0.01$ & $-0.13\pm0.00$ & $-0.13\pm0.00$ & $-0.23\pm0.01$ & $(3.6\pm3.3)\cdot 10^{-2}$ & $(-0.6\pm2.7)\cdot 10^{-2}$ \\[0.3ex]
        Im[$c_{\rho^{\prime}}$] & $-0.28\pm0.02$ & $-0.19\pm0.02$ & $-0.17\pm0.02$ & $-0.23\pm0.01$ & $-0.28\pm0.03$ & $-0.27\pm0.02$ \\[0.3ex]
        $m_{\rho^{\prime\prime}}$ & $1831.4\pm12.6\text{ MeV}$ & $1730^\dagger\text{ MeV}$ & $1730^\dagger\text{ MeV}$ & $1730^\dagger\text{ MeV}$ & $1865.8\pm20.1\text{ MeV}$ & $1868.2\pm16.3\text{ MeV}$\\[0.3ex] 
        $\Gamma_{\rho^{\prime\prime}}$ & $442\pm33\text{ MeV}$ & $260^\dagger\text{ MeV}$ & $260^\dagger\text{ MeV}$ & $260^\dagger\text{ MeV}$ & $721\pm53\text{ MeV}$ & $650\pm50\text{ MeV}$\\[0.3ex]
        Re[$c_{\rho^{\prime\prime}}$] & $-(8.4\pm1.5)\cdot 10^{-2}$ & $(-6.0\pm0.6)\cdot 10^{-2}$ & $(-5.0\pm0.7)\cdot 10^{-2}$ & $(1.9\pm1.0)\cdot 10^{-2}$ & $-0.25\pm0.04$ & $-0.20\pm0.03$ \\[0.3ex]
        Im[$c_{\rho^{\prime\prime}}$] & $0.13\pm0.01$ & $(3.1\pm0.7)\cdot 10^{-2}$ & $(4.6\pm0.7)\cdot 10^{-2}$ & $0.10\pm0.00$ & $0.13\pm0.02$ & $0.12\pm0.01$ \\[0.3ex]
        \hline
    \end{tabular}}
    \captionsetup{width=0.88\linewidth}
    \caption{Fit results for the KS model using the $e^+e^-$ data sets and Belle $\tau$ decay measurements. \textcolor{black}{All two-sided p-values are smaller than $10^{-4}$.}}
    \label{tab:KS_Fit}
\end{table}

\begin{table}[ht]
    \centering
    \resizebox{0.90\textwidth}{!}{\begin{tabular}{|c|cccccc|}
    \hline
         & BABAR12+Belle & KLOE12+Belle & KLOEc+Belle & CMD3+Belle & Global Fit 1 & Global Fit 2 \\[0.3ex]
        data points  & 270 + 19 & 60 + 19 & 85 + 19 & 172 + 19 & 270+60+172+19 & 270+85+172+19 \\[0.3ex]
        $\chi^2_{ee}+\chi^2_\tau$ & $1089.1 + 582.6$ & $53.0 + 519.5$ & $265.1 + 583.8$ & $2607.9 + 671.7$ & $4274.1 + 613.0$ & $4444.5 + 620.9$ \\[0.3ex]
        $\chi^2$ & $1671.7$ & $572.5$ & $848.9$ & $3279.6$ & $4887.1$ & $5065.4$ \\[0.3ex]
         \hline
        $\chi^2/\text{d.o.f}$ & $5.8$ & $7.5$ & $8.4$ & $17.4$ & $9.4$ & $9.3$ \\[0.3ex]
        $m_\rho$ & $776.0\pm0.1\text{ MeV}$ & $777.4\pm0.1\text{ MeV}$ & $776.0\pm0.1\text{ MeV}$ & $774.8\pm0.1\text{ MeV}$ & $775.5\pm0.1\text{ MeV}$ & $775.4\pm0.1\text{ MeV}$\\[0.3ex] 
        $\text{Re}\vert\delta_{\rho\omega}\vert=\text{Re}\left\vert-\frac{\theta_{\rho\omega}}{3m_\rho}\right\vert$ & $(2.5\pm0.0)\cdot 10^{-3}$ & $(2.6\pm0.1)\cdot 10^{-3}$ & $(2.5\pm0.1)\cdot 10^{-3}$ & $(2.5\pm0.0)\cdot 10^{-3}$ & $(2.5\pm0.0)\cdot 10^{-3}$ & $(2.5\pm0.0)\cdot 10^{-3}$ \\[0.3ex]
        $\text{Im}\vert\delta_{\rho\omega}\vert=\text{Im}\left\vert-\frac{\theta_{\rho\omega}}{3m_\rho}\right\vert$ & $(0.3\pm0.0)\cdot 10^{-3}$ & $(0.4\pm0.1)\cdot 10^{-3}$ & $(0.8\pm0.1)\cdot 10^{-3}$ & $(0.1\pm0.0)\cdot 10^{-3}$ & $(0.1\pm0.0)\cdot 10^{-3}$ & $(0.1\pm0.0)\cdot 10^{-3}$ \\[0.3ex]
        \hline
        $\vert\delta_{\rho\omega}\vert$ & $(2.6\pm0.0)\cdot 10^{-3}$ & $(2.6\pm0.1)\cdot 10^{-3}$ & $(2.7\pm0.1)\cdot 10^{-3}$ & $(2.5\pm0.0)\cdot 10^{-3}$ & $(2.5\pm0.0)\cdot 10^{-3}$ & $(2.5\pm0.0)\cdot 10^{-3}$ \\[0.3ex]
        arg[$\delta_{\rho\omega}$] & $(6.2\pm0.9)^\circ$ & $(9.2\pm2.1)^\circ$ & $(18.6\pm1.8)^\circ$ & $(1.4\pm0.4)^\circ$ & $(2.6\pm0.3)^\circ$ & $(2.8\pm0.3)^\circ$ \\[0.3ex]
        \hline
    \end{tabular}}
    \captionsetup{width=0.88\linewidth}
    \caption{ Fit results for the GP model using the $e^+e^-$ data sets and Belle $\tau$ decay measurements below $1.0\text{ GeV}$. \textcolor{black}{All two-sided p-values are smaller than $10^{-4}$.}}
    \label{tab:GP_Fit_All}
\end{table}

\begin{table}[ht]
    \centering
    \resizebox{0.90\textwidth}{!}{\begin{tabular}{|c|cccccc|}
    \hline
         & BABAR12+Belle & KLOE12+Belle & KLOEc+Belle & CMD3+Belle & Global fit 1 & Global fit 2\\[0.3ex]
        data points  & 337 + 62 & 60 + 62 & 85 + 62 & 209 + 62 & 337+60+209+62 & 337+85+209+62 \\[0.3ex]
        $\chi^2_{ee}+\chi^2_\tau$ & $426.7 + 129.7$ & $145.3 + 287.6$ & $134.1 + 405.3$ & $139.0 + 83.5$ & $1455.6 + 207.1$ & $2277.1 + 237.4$ \\[0.3ex]
        $\chi^2$ & $556.4$ & $432.9$ & $539.4$ & $222.5$ & $1662.7$ & $2514.5$ \\[0.3ex]
         \hline
        $\chi^2/\text{d.o.f}$ & $1.4$ & $3.8$ & $3.9$ & $0.9$ & $2.5$ & $3.7$ \\[0.3ex]
        $m_\rho$ & $774.6\pm0.1\text{ MeV}$ & $774.1\pm0.2\text{ MeV}$ & $773.7\pm0.1\text{ MeV}$ & $773.6\pm0.1\text{ MeV}$ & $773.8\pm0.1\text{ MeV}$ & $773.5\pm0.1\text{ MeV}$ \\[0.3ex] 
        $\vert\delta_{\rho\omega}\vert$ & $(2.5\pm0.0)\cdot 10^{-3}$ & $(2.4\pm0.1)\cdot 10^{-3}$ & $(2.3\pm0.1)\cdot 10^{-3}$ & $(2.5\pm0.0)\cdot 10^{-3}$ & $(2.4\pm0.0)\cdot 10^{-3}$ & $(2.4\pm0.0)\cdot 10^{-3}$ \\[0.3ex]
        arg[$\delta_{\rho\omega}$] & $(13.5\pm1.0)^\circ$ & $(10.7\pm2.5)^\circ$ & $(16.8\pm2.3)^\circ$ & $(11.3\pm0.4)^\circ$ & $(9.1\pm0.4)^\circ$ & $(6.5\pm0.4)^\circ$ \\[0.3ex]
        $\vert\delta_{\rho\phi}\vert$ & $0^\dagger$ & $0^\dagger$ & $0^\dagger$ & $(2.8\pm0.2)\cdot 10^{-4}$ & $(2.0\pm0.2)\cdot 10^{-4}$ & $(2.1\pm0.3)\cdot 10^{-4}$ \\[0.3ex]
        arg[$\delta_{\rho\phi}$] & $-$ & $-$ & $-$ & $(63.1\pm5.4)^\circ$ & $(58.2\pm6.3)^\circ$ & $(40.9\pm6.7)^\circ$ \\[0.3ex]
        \hline
        $m_{\rho^\prime}$ & $1397.2\pm8.3\text{ MeV}$ & $1413.8\pm11.7\text{ MeV}$ & $1406.3\pm12.1\text{ MeV}$ & $1449.2\pm11.5\text{ MeV}$ & $1414.5\pm3.3\text{ MeV}$ & $1369.4\pm7.8\text{ MeV}$\\[0.3ex] 
        $\Gamma_{\rho^\prime}$ & $324\pm15\text{ MeV}$ & $386\pm28\text{ MeV}$ & $447\pm33\text{ MeV}$ & $385\pm28\text{ MeV}$ & $231\pm7\text{ MeV}$ & $343\pm16\text{ MeV}$\\[0.3ex]
        Re[$c_{\rho^{\prime}}$] & $(9.2\pm0.5)\cdot10^{-2}$ & $0.11\pm0.01$ & $0.12\pm0.00$ & $0.13\pm0.01$ & $(3.1\pm0.5)\cdot 10^{-2}$ & $(7.9\pm0.5)\cdot 10^{-2}$ \\[0.3ex]
        Im[$c_{\rho^{\prime}}$] & $(-5.1\pm0.6)\cdot10^{-2}$ & $(-8.8\pm0.9)\cdot10^{-2}$ & $-0.12\pm0.01$ & $(-4.7\pm0.8)\cdot10^{-2}$ & $0.19\pm0.01$ & $(-7.2\pm0.5)\cdot10^{-2}$ \\[0.3ex]
        $m_{\rho^{\prime\prime}}$ & $1721.7\pm10.5\text{ MeV}$ & $1730^\dagger\text{ MeV}$ & $1730^\dagger\text{ MeV}$ & $1730^\dagger\text{ MeV}$ & $1793.4\pm6.7\text{ MeV}$ & $1698.2\pm10.6\text{ MeV}$\\[0.3ex] 
        $\Gamma_{\rho^{\prime\prime}}$ & $211\pm12\text{ MeV}$ & $260^\dagger\text{ MeV}$ & $260^\dagger\text{ MeV}$ & $260^\dagger\text{ MeV}$ & $120\pm6\text{ MeV}$ & $203\pm10\text{ MeV}$\\[0.3ex]
        Re[$c_{\rho^{\prime\prime}}$] & $(-8.1\pm0.5)\cdot 10^{-2}$ & $-0.11\pm0.01$ & $-0.12\pm0.00$ & $-0.11\pm0.01$ & $(-2.7\pm0.4)\cdot10^{-2}$ & $(-7.3\pm0.5)\cdot10^{-2}$ \\[0.3ex]
        Im[$c_{\rho^{\prime\prime}}$] & $(2.6\pm0.8)\cdot10^{-2}$ & $(4.1\pm0.8)\cdot 10^{-2}$ & $(9.8\pm0.7)\cdot 10^{-2}$ & $(2.0\pm0.7)\cdot10^{-2}$ & $(4.0\pm0.3)\cdot10^{-2}$ & $(4.5\pm0.8)\cdot10^{-2}$ \\[0.3ex]
        \hline
    \end{tabular}}
    \captionsetup{width=0.88\linewidth}
    \caption{ Fit results with $F_V(s)^{\mathrm{seed}}$  using the $e^+e^-$ data sets and Belle $\tau$ decay measurements. \textcolor{black}{The only two-sided p-value which is larger than $10^{-4}$ is obtained for the CMD-3+Belle fit and is $0.986$.}}
    \label{tab:GPg_Fit}
\end{table}

\begin{table}[ht]
    \centering
    \resizebox{0.90\textwidth}{!}{\begin{tabular}{ccccccccccc}
    \hline
          & Dispersive & Dispersive & Dispersive & Dispersive & Dispersive & Dispersive & Dispersive & Dispersive & Dispersive & Dispersive \\[0.3ex]
          & $p_{0}$ & $p_{1}$ & $p_{2}$ & $p_{2-1}$ & $p_{3}$ & $p_{3-1}$ & $p_{4}$ & $p_{4-1}$ & $p_{5}$ & $p_{5-1}$ \\[0.3ex]
          \hline
        $\chi^2$ & $3616.4$ & $3261.6$ & $3226.5$ & $3307.4$ & $3113.1$ & $3416.7$ & $2695.3$ & $2683.0$ & $2685.7$ & $3173.2$ \\[0.3ex]
        $d.o.f$ & { \color{black}$665$} & { \color{black}$664$} & { \color{black}$663$} & { \color{black}$664$} & { \color{black}$662$} & { \color{black}$663$} & { \color{black}$661$} & { \color{black}$662$} & { \color{black}$660$} & { \color{black}$661$} \\[0.3ex]
        $\chi^2/d.o.f$ & $5.4$ & $4.9$ & $4.9$ & $5.0$ & $4.7$ & $5.2$ & $4.1$ & $4.1$ & $4.1$ & $4.8$ \\[0.3ex]
        \hline
        $\langle r_\pi^2\rangle\,[\text{fm}^2]$ & $0.477$ & $0.399$ & $0.401$ & $0.413$ & $0.441$ & $0.406$ & $0.366$ & $0.428$ & $0.389$ & $0.412$ \\[0.3ex]
        \hline
    \end{tabular}}
    \captionsetup{width=0.88\linewidth}
    \caption{ Fit results (Global Fit 2) using the dispersive framework. 
    {\color{black}In the second row, $d.o.f.=322+85+209+62-x$ corresponds to the sum of data points from Babar, KLOE, CMD-3 and Belle, respectively, and $x$ is the number of fitting parameters.} For comparison, $\langle r_\pi^2\rangle=0.429(4)$ fm$^2$ in Ref.~\cite{Colangelo:2018mtw}. \textcolor{black}{All two-sided p-values are smaller than $10^{-4}$.}}
    \label{tab:charge_radius}
\end{table}

\begin{table}[ht]
    \centering
    \resizebox{0.90\textwidth}{!}{\begin{tabular}{|c|cccccc|}
    \hline
         & BABAR12+Belle & KLOE12+Belle & KLOEc+Belle & CMD3+Belle & Global fit 1 & Global fit 2\\[0.3ex]
        data points  & 322 + 62 & 60 + 62 & 85 + 62 & 209 + 62 & 322+60+209+62 & 322+85+209+62 \\[0.3ex]
        $\chi^2$ & $493.4$ & $356.4$ & $442.9$ & $584.5$ & $2192.2$ & $2683.0$ \\[0.3ex]
         \hline
        $\chi^2/\text{d.o.f}$ & $1.3$ & $2.9$ & $3.0$ & $2.2$ & $3.4$ & $4.0$ \\[0.3ex]
        $m_\rho$ & $774.9\pm0.1\text{ MeV}$ & $774.9\pm0.1\text{ MeV}$ & $774.7\pm0.1\text{ MeV}$ & $774.9\pm0.2\text{ MeV}$ & $774.3\pm0.1\text{ MeV}$ & $774.1\pm0.1\text{ MeV}$ \\[0.3ex] 
        $\vert\delta_{\rho\omega}\vert$ & $(2.0\pm0.0)\cdot 10^{-3}$ & $(2.0\pm0.0)\cdot 10^{-3}$ & $(2.0\pm0.0)\cdot 10^{-3}$ & $(2.1\pm0.0)\cdot 10^{-3}$ & $(2.1\pm0.0)\cdot 10^{-3}$ & $(2.1\pm0.0)\cdot 10^{-3}$ \\[0.3ex]
        arg[$\delta_{\rho\omega}$] & $(17.2\pm0.9)^\circ$ & $(16.1\pm 1.3)^\circ$ & $(25.0\pm2.3)^\circ$ & $(14.3\pm1.2)^\circ$ & $(13.3\pm0.4)^\circ$ & $(10.2\pm0.4)^\circ$ \\[0.3ex]
        $\vert\delta_{\rho\phi}\vert$ & $0^\dagger$ & $0^\dagger$ & $0^\dagger$ & $(2.0\pm0.1)\cdot 10^{-4}$ & $(1.3\pm0.1)\cdot 10^{-4}$ & $(1.7\pm0.2)\cdot 10^{-4}$ \\[0.3ex]
        arg[$\delta_{\rho\phi}$] & $-$ & $-$ & $-$ & $(56.5\pm2.9)^\circ$ & $(22.9\pm5.3)^\circ$ & $(40.4\pm6.3)^\circ$ \\[0.3ex]
        $C_2$ & $-0.63\pm0.00$ & $-0.26\pm0.00 $ & $-0.42\pm0.01$ & $-0.43\pm0.00$ & $(-0.27\pm0.00)\cdot10^{-1}$ & $-0.39\pm0.02$ \\[0.3ex]
        $C_3$ & $0.37\pm0.00$ & $0.17\pm0.00$ & $0.29\pm0.01$ & $0.29\pm0.01$ & $(0.28\pm0.00)\cdot10^{-1}$ & $0.26\pm0.01$ \\[0.3ex]
        $C_4$ & $-0.10\pm0.00$ & $-0.05\pm0.00$ & $-0.09\pm0.00$ & $-0.09\pm0.01$ & $(-0.15\pm0.00)\cdot10^{-1}$ & $(-0.76\pm0.03)\cdot10^{-1}$ \\[0.3ex]
        \hline
        $m_{\rho^\prime}$ & $1603.5\pm3.1\text{ MeV}$ & $1564.8\pm 1.5\text{ MeV}$ & $1636.1\pm7.0\text{ MeV}$ & $1600.5\pm1.6\text{ MeV}$ & $1482.3\pm5.1\text{ MeV}$ & $1552.5\pm6.7 \text{ MeV}$\\[0.3ex] 
        $\Gamma_{\rho^\prime}$ & $426\pm5\text{ MeV}$ & $426\pm7\text{ MeV}$ & $611\pm9\text{ MeV}$ & $405\pm8\text{ MeV}$ & $313\pm9\text{ MeV}$ & $356\pm10\text{ MeV}$\\[0.3ex]
        Re[$c_{\rho^{\prime}}$] & $0.30\pm0.00$ & $0.65\pm0.01$ & $1.85\pm0.02$ & $0.36\pm0.01$ & $0.49\pm0.01$ & $0.26\pm0.01$ \\[0.3ex]
        Im[$c_{\rho^{\prime}}$] & $0.31\pm0.01$ & $0.31\pm0.01$ & $0.25\pm0.01$ & $0.40\pm0.01$ & $0.27\pm0.01$ & $0.34\pm0.02$ \\[0.3ex]
        $m_{\rho^{\prime\prime}}$ & $1899.4\pm6.2\text{ MeV}$ & $1730^\dagger\text{ MeV}$ & $1730^\dagger\text{ MeV}$ & $1730^\dagger\text{ MeV}$ & $1822.2\pm4.0\text{ MeV}$ & $1896.2\pm3.9\text{ MeV}$\\[0.3ex] 
        $\Gamma_{\rho^{\prime\prime}}$ & $406\pm9\text{ MeV}$ & $260^\dagger\text{ MeV}$ & $260^\dagger\text{ MeV}$ & $260^\dagger\text{ MeV}$ & $125\pm10\text{ MeV}$ & $239\pm11\text{ MeV}$\\[0.3ex]
        Re[$c_{\rho^{\prime\prime}}$] & $-0.11\pm0.00$ & $-0.28\pm0.00$ & $-1.00\pm0.04$ & $-0.18\pm0.01$ & $(-0.13\pm0.11)\cdot10^{-1}$ & $(0.78\pm0.14)\cdot10^{-1}$ \\[0.3ex]
        Im[$c_{\rho^{\prime\prime}}$] & $-0.35\pm0.01$ & $-0.32\pm0.01$ & $-0.29\pm0.01$ & $-0.31\pm0.01$ & $-0.25\pm0.01$ & $-0.31\pm0.02$ \\[0.3ex]
        \hline
    \end{tabular}}
    \captionsetup{width=0.88\linewidth}
    \caption{ Fit results with the $p_{4-1}$ dispersive model using the $e^+e^-$ data sets and Belle $\tau$ decay measurements. \textcolor{black}{The largest two-sided p-value is obtained in the BABAR12+Belle fit, which is $10^{-4}$.}}
    \label{tab:Disp_Fit_v4}
\end{table}

\begin{table}[ht]
    \centering
    \resizebox{0.90\textwidth}{!}{\begin{tabular}{ccccccccccc}
    \hline
        Source & \multicolumn{10}{c}{$\Delta \mathcal{B}_{\pi\pi}^{\text{CVC}}\,(10^{-2})$}\\[0.3ex]
          & Dispersive & Dispersive & Dispersive & Dispersive & Dispersive & Dispersive & Dispersive & Dispersive & Dispersive & Dispersive \\[0.3ex]
          & $p_{0}$ & $p_{1}$ & $p_{2}$ & $p_{2-1}$ & $p_{3}$ & $p_{3-1}$ & $p_{4}$ & $p_{4-1}$ & $p_{5}$ & $p_{5-1}$ \\[0.3ex]
        \hline
        $m_{\pi^\pm}-m_{\pi^0}$ effect on $\Gamma_\rho$ & $-0.20$ & $-0.19$ & $-0.19$ & $-0.19$ & $-0.20$ & $-0.20$ & $-0.20$ & $-0.20$ & $-0.20$ & $-0.19$ \\[0.3ex]
        $m_{K^\pm}-m_{K^0}$ effect on $\Gamma_\rho$ & $+0.01$ & $+0.01$ & $+0.01$ & $+0.01$ & $+0.01$ & $+0.01$ & $+0.01$ & $+0.01$ & $+0.01$ & $+0.01$ \\[0.3ex]
        $m_{\rho^\pm}-m_{\rho^0}$ on $\Gamma_\rho$ & $-0.10(9)$ & $-0.10(9)$ & $-0.10(9)$ & $-0.10(9)$ & $-0.09(^{9}_{8})$ & $-0.09(8)$ & $-0.10(9)$ & $-0.10(9)$ & $-0.10(9)$ & $-0.09(^{9}_{8})$ \\[0.3ex]
        $m_{\rho^\pm}-m_{\rho^0}$ & $+0.07(7)$ & $+0.08(7)$ & $+0.08(7)$ & $+0.08(7)$ & $+0.08(7)$ & $+0.08(^{8}_{7})$ & $+0.08(^{8}_{7})$ & $+0.08(^{8}_{7})$ & $+0.08(^{8}_{7})$ & $+0.07(7)$ \\[0.3ex]
        $\rho-\omega$ interference & $+0.00(0)$ & $+0.01(0)$ & $+0.01(0)$ & $+0.00(0)$ & $-0.01(0)$ & $+0.00(0)$ & $-0.01(0)$ & $-0.01(0)$ & $-0.01(0)$ & $+0.00(0)$ \\[0.3ex]
        $\rho-\phi$ interference & $-0.01(0)$ & $-0.01(0)$ & $-0.01(0)$ & $-0.01(0)$ & $-0.01(0)$ & $-0.01(0)$ & $-0.01(0)$ & $-0.01(0)$ & $-0.01(0)$ & $-0.01(0)$ \\[0.3ex]
        $\pi\pi\gamma$, electromagnetic decays & $+0.37(4)$ & $+0.37(4)$ & $+0.37(4)$ & $+0.37(4)$ & $+0.37(4)$ & $+0.37(4)$ & $+0.37(4)$ & $+0.37(4)$ & $+0.37(4)$ & $+0.37(4)$ \\[0.3ex]
        \hline
        TOTAL & $+0.14(12)$ & $+0.17(12)$ & $+0.17(12)$ & $+0.16(12)$ & $+0.15(^{12}_{11})$ & $+0.16(^{12}_{11})$ & $+0.14(^{13}_{12})$ & $+0.14(_{12}^{13})$ & $+0.14(^{13}_{12})$ & $+0.16(^{12}_{11})$ \\[0.3ex]
        \hline
    \end{tabular}}
    \captionsetup{width=0.88\linewidth}
    \caption{IB contributions to $\Delta \mathcal{B}_{\pi\pi}^{\text{CVC}} (10^{-2})$ using a dispersive form factor. }
    \label{tab:IB_BRFF_disp}
\end{table}

\begin{table}[ht]
    \centering
    \resizebox{0.90\textwidth}{!}{\begin{tabular}{ccccccccccc}
    \hline
    Source & \multicolumn{10}{c}{$\Delta a_\mu^\text{HVP, LO}[\pi\pi,\tau]\,(10^{-10})$} \\[0.3ex]
    & Dispersive & Dispersive & Dispersive & Dispersive & Dispersive & Dispersive & Dispersive & Dispersive & Dispersive & Dispersive \\[0.3ex]
      & $p_{0}$ & $p_{1}$ & $p_{2}$ & $p_{2-1}$ & $p_{3}$ & $p_{3-1}$ & $p_{4}$ & $p_{4-1}$ & $p_{5}$ & $p_{5-1}$ \\[0.3ex]
    \hline
    $m_{\pi^\pm}-m_{\pi^0}$ effect on $\Gamma_\rho$ & $+3.59$ & $+3.51$ & $+3.51$ & $+3.53$ & $+3.57$ & $+3.53$ & $+3.51$ & $+3.59$  & $+3.55$ & $+3.51$ \\[0.3ex]
    $m_{K^\pm}-m_{K^0}$ effect on $\Gamma_\rho$ & $-0.09$ & $-0.16$ & $-0.16$ & $-0.16$ & $-0.14$ & $-0.13$ & $-0.21$ & $-0.22$ & $-0.23$ & $-0.16$ \\[0.3ex]
    $m_{\rho^\pm}-m_{\rho^0}$ on $\Gamma_\rho$ & $+1.67^{+1.51}_{-1.50}$ & $+1.66^{+1.50}_{-1.49}$ & $+1.65^{+1.50}_{-1.49}$ & $+1.66^{+1.50}_{-1.49}$ & $+1.65^{+1.49}_{-1.48}$ & $+1.64^{+1.48}_{-1.47}$ & $+1.72^{+1.56}_{-1.55}$ & $+1.72^{+1.56}_{-1.55}$ & $+1.72^{+1.56}_{-1.54}$ & $+1.64^{+1.49}_{-1.48}$ \\[0.3ex]
    $m_{\rho^\pm}-m_{\rho^0}$ & $+0.34^{+0.33}_{-0.31}$ & $+0.29^{+0.29}_{-0.26}$ & $+0.30^{+0.30}_{-0.28}$ & $+0.27^{+0.28}_{-0.25}$ & $+0.26^{+0.26}_{-0.23}$ & $+0.22^{+0.23}_{-0.21}$ & $+0.22^{+0.22}_{-0.20}$ & $+0.23^{+0.23}_{-0.21}$ & $+0.23^{+0.23}_{-0.21}$ & $+0.34^{+0.33}_{-0.30}$ \\[0.3ex]
    $\rho-\omega$ interference & $+2.65(0.04)$ & $+2.57(0.04)$ & $+2.60(0.03)$ & $+2.61(0.04)$ & $+2.79(0.04)$ & $+2.59(0.03)$ & $+2.76(0.04)$ & $+2.75(0.08)$ & $+2.74(0.04)$ & $+2.73(0.04)$ \\[0.3ex]
    $\rho-\phi$ interference & $+0.14(0.02)$ & $+0.13(0.01)$ & $+0.15(0.01)$ & $+0.13(0.01)$ & $+0.12(0.01)$ & $+0.12(0.01)$ & $+0.13(0.01)$ & $+0.12(0.02)$ & $+0.13(0.01)$ & $+0.14(0.02)$ \\[0.3ex]
    $\pi\pi\gamma$ & $-6.63(0.73)$ & $-6.64(0.73)$ & $-6.63(0.73)$ & $-6.65(0.73)$ & $-6.64(0.73)$ & $-6.66(0.73)$ & $-6.65(0.73)$ & $-6.66(0.73)$ & $-6.66(0.73)$ & $-6.61(0.73)$ \\[0.3ex]
    \hline
    TOTAL & $+1.67^{+1.71}_{-1.70}$ & $+1.36^{+1.69}_{-1.68}$ & $+1.42^{+1.70}_{-1.68}$ & $+1.39^{+1.69}_{-1.68}$ & $+1.61^{+1.68}_{-1.67}$ & $+1.31^{+1.67}_{-1.65}$ & $+1.48^{+1.74}_{-1.73}$ & $+1.53^{+1.74}_{-1.73}$ & $+1.48^{+1.74}_{-1.72}$ & $+1.59^{+1.69}_{-1.68}$ \\[0.3ex]
    \hline
    \end{tabular}}
    \captionsetup{width=0.88\linewidth}
    \caption{Contributions to $a_\mu^{\text{HVP, LO}}[\pi\pi,\tau](10^{-10})$ from the isospin-breaking (IB) in the dispersive framework.}
    \label{tab:IB_amuFF_disp}
\end{table}

\begin{table}[ht]
    \centering
    \resizebox{0.80\textwidth}{!}{\begin{tabular}{cccccc}
    \hline
        Source & \multicolumn{5}{c}{$\Delta \mathcal{B}_{\pi\pi}^{\text{CVC}}\,(10^{-2})$}\\[0.3ex]
          & GS & KS & GP & Seed & Dispersive \\[0.3ex]
          &  &  &  &  & $p_{4-1} $\\[0.3ex]
        \hline
        $S_\text{EW}$ & \multicolumn{5}{c}{$+0.57(1)$} \\[0.3ex]
        $G_\text{EM}$ &  \multicolumn{5}{c}{$-0.09(^{3}_{1})$} \\[0.3ex]
        FSR & \multicolumn{5}{c}{$-0.19(2)$} \\[0.3ex]
        $m_{\pi^\pm}-m_{\pi^0}$ effect on $\sigma$ & \multicolumn{5}{c}{$+0.20$} \\[0.3ex]
        $m_{\pi^\pm}-m_{\pi^0}$ effect on $\Gamma_\rho$ & $-0.20$ & $-0.22$ & $-0.22$ & $-0.24$ & $-0.20$ \\[0.3ex]
        $m_{K^\pm}-m_{K^0}$ effect on $\Gamma_\rho$ & $-$ & $-$ & $-0.02$ & $-0.02$ &  $+0.01$\\[0.3ex]
        $m_{\rho^\pm}-m_{\rho^0}$ on $\Gamma_\rho$ & $-$ & $-$ 
        & $-0.12(11)$ & $-0.12(11)$ & $-0.10(9)$ \\[0.3ex]
        $m_{\rho^\pm}-m_{\rho^0}$ & $+0.09(8)$ & $+0.09(8)$ & $+0.12(11)$ & $+0.12(11)$ & $+0.08(^{8}_{7})$ \\[0.3ex]
        $\rho-\omega$ interference & $-0.08(0)$
        & $-0.09(0)$
        & $-0.09(0)$
        & $-0.05(0)$
        & $-0.01(0)$
        \\[0.3ex]
        $\rho-\phi$ interference & $-0.00(0)$
        & $-0.00(0)$
        & $-$ & $-0.01(0)$
        & $-0.01(0)$
        \\[0.3ex]
        $\pi\pi\gamma$, electromagnetic decays & $+0.35(4)$ & $+0.38(4)$ & $+0.34(4)$ & $+0.35(4)$ & $+0.37(4)$ \\[0.3ex]
        \hline
        TOTAL & $+0.65(^{10}_{\,\,9})$ 
        & $+0.65(^{10}_{\,\,9})$ 
        & $+0.50(16)$ 
        & $+0.52(16)$ 
        & $+0.63(^{13}_{12})(^{3}_{0})$ 
        \\[0.3ex]
        \hline
    \end{tabular}}
    \captionsetup{width=0.88\linewidth}
    \caption{IB contributions to BR($\tau^-\to\pi^-\pi^0\nu_\tau$) according to the different form factor inputs. 
    The uncertainties are mostly systematic. 
    The errors in the last row were obtained adding quadratically uncertainties from each IB contribution. 
    In the last entry, we take as an additional uncertainty (last shown) the difference between $p_{4-1}$ and the other dispersive results in Table~\ref{tab:IB_BRFF_disp}.}
    \label{tab:IB_BR_CMD3}
\end{table}



\begin{table}[ht]
    \centering
    \resizebox{0.80\textwidth}{!}{\begin{tabular}{cccccc}
    \hline
    Source & \multicolumn{5}{c}{$\Delta a_\mu^\text{HVP, LO}[\pi\pi,\tau]\,(10^{-10})$} \\[0.3ex]
     & GS & KS & GP & Seed & Dispersive\\[0.3ex]
      &  &  &  &  & $p_{4-1}$ \\[0.3ex]
    \hline
    $S_\text{EW}$ & \multicolumn{5}{c}{$-12.16(0.15)$} \\[0.3ex]
    $G_\text{EM}$ & \multicolumn{5}{c}{$-1.67(^{0.60}_{1.39})$} \\[0.3ex]
    FSR & \multicolumn{5}{c}{$+4.62(0.46)$} \\[0.3ex]
    $m_{\pi^\pm}-m_{\pi^0}$ effect on $\sigma$ & \multicolumn{5}{c}{$-7.52$} \\[0.3ex]
    $m_{\pi^\pm}-m_{\pi^0}$ effect on $\Gamma$ & $+3.75$ 
    & $+4.13$ 
    & $+4.11$ & $+4.15$ & $+3.59$ \\[0.3ex]
    $m_{K^\pm}-m_{K^0}$ effect on $\Gamma$ & $-$ & $-$ & $+0.37$ & $+0.37$ & $-0.22$ \\[0.3ex]
    $m_{\rho^\pm}-m_{\rho^0}$ on $\Gamma_\rho$ & $-$ 
    & $-$ 
    & $+2.18(^{1.97}_{1.96})$ & $+2.18(^{1.98}_{1.96})$ & $+1.72(1.54)$ \\[0.3ex]
    $m_{\rho^\pm}-m_{\rho^0}$ & $+0.14(^{0.22}_{0.13})$ 
    & $-0.01(^{0.08}_{0.00})$ 
    & $-0.49(^{0.44}_{0.35})$ & $-0.50(^{0.45}_{0.36})$ & $+0.23(^{0.23}_{0.20})$ \\[0.3ex]
    $\rho-\omega$ interference & $+3.87(0.08)$
    & $+4.03(0.08)$
    & $+4.35(0.07)$
    & $+3.60(0.07)$
    & $+2.75(0.08)$
    \\[0.3ex]
    $\rho-\phi$ interference & $+0.09(0.03)$
    & $+0.03(0.03)$
    & $-$ & $+0.13(0.03)$
    & $+0.12(0.02)$
    \\[0.3ex]
    $\pi\pi\gamma$ & $-5.96(0.66)$ 
    & $-6.49(0.72)$ 
    & $-6.22(0.69)$ & $-6.25(0.69)$ & $-6.66(0.73)$  \\[0.3ex]
    \hline
    TOTAL & $-14.84(^{1.04}_{1.62})$ 
    & $-15.04(_{1.64}^{1.06})$ 
    & $-12.43(^{2.27}_{2.57})$ 
    & $-13.05(^{2.28}_{2.57})$ 
    & $-15.20(^{1.89}_{2.26})(^{0.14}_{0.22})$ 
    \\[0.3ex]
    \hline
    \end{tabular}}
    \captionsetup{width=\linewidth}
    \caption{Contributions to $a_\mu^{\text{HVP, LO}}[\pi\pi,\tau](10^{-10})$ from the isospin-breaking (IB) corrections according to the different form factor inputs. The uncertainties presented here align closely with those in Table~\ref{tab:IB_BR_CMD3}. In the last entry, we take as an additional uncertainty (last shown) the difference between $p_{4-1}$ and the other dispersive results in Table~\ref{tab:IB_amuFF_disp}.   } 
    \label{tab:IB_amu_CMD3}
\end{table}

\begin{figure}[ht]
    \centering 
    \includegraphics[width=0.88\textwidth]{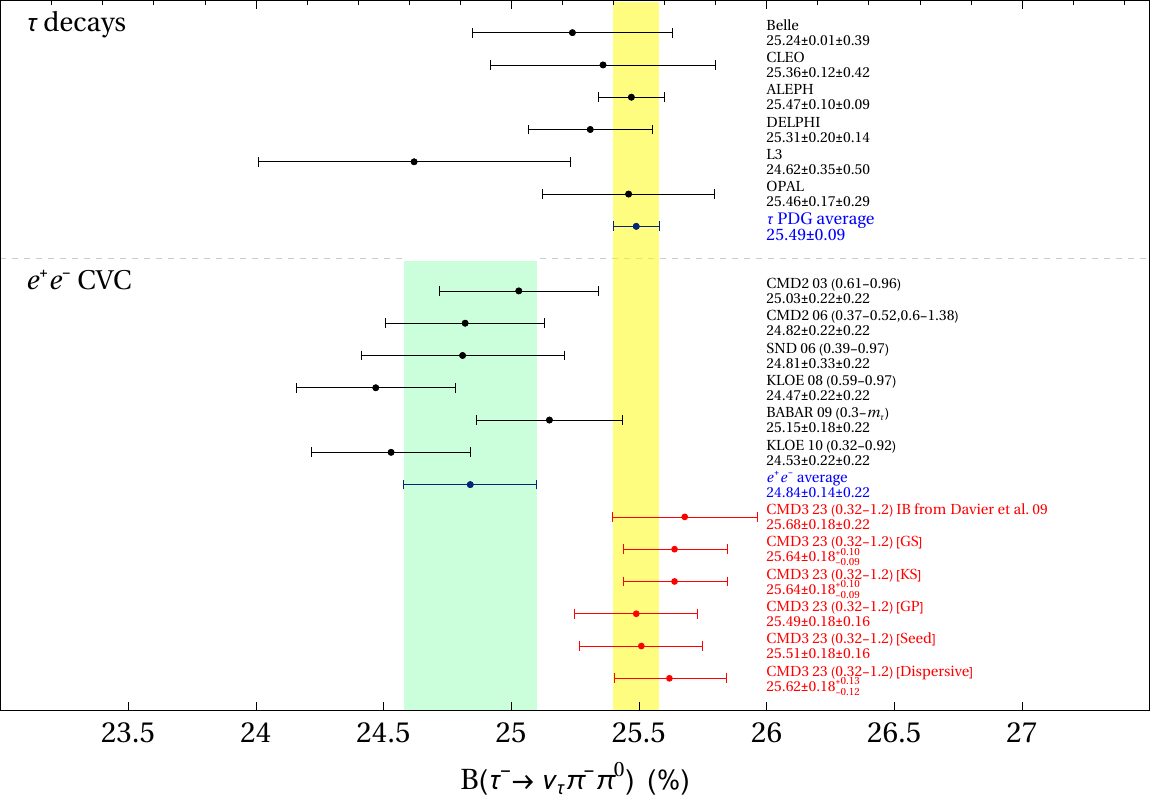}
    \captionsetup{width=0.88\linewidth}
    \caption{ Comparison between the measured branching fractions for $\tau^-\to\pi^-\pi^0\nu_\tau$  (their weighted average is the yellow band) and the predictions from the $e^+e^-\to\pi^+\pi^-$ spectral functions, applying the isospin-breaking corrections given in Table \ref{tab:IB_BR_CMD3} (their weighted average --before CMD-3-- is indicated by the green band). As discussed in the main text, our reference result comes from the dispersive evaluation, and we add linearly to the final theory uncertainty (second error shown, preceded by the statistical one) its difference with the GS value, that also complies well with analyticity.    }
    \label{fig:BR_CVC}
\end{figure}

Finally, the results obtained for the branching fractions of two-pion tau lepton decays using the CMD-3 data and  including IB in different models for the form factors, are shown in Fig.~\ref{fig:BR_CVC}. The uncertainties in the numbers in red are an estimation based on our results. We highlight the compatibility of our results obtained using different form factor models and CMD-3 data. This prediction based on CVC  agrees with the average of $\tau$ branching fractions~\cite{ParticleDataGroup:2024cfk}, although exceeds slightly this experimental value. Our results are also consistent with those obtained using BaBar data~\cite{BaBar:2012bdw}.

The analyticity test on the different form factors shown in the Appendix \ref{app_A} clearly favors the GS model over the KS, GP and Seed models, so that we take GS along with the dispersive one as our reference results. We use the dispersive result as the central value and take the difference with respect to GS as an additional uncertainty, which is added linearly. 
We thereby obtain 
\begin{equation}\label{eq_FinalResults}
\Delta\mathcal{B}_{\pi\pi}^\mathrm{CVC}=(+0.63_{-0.14}^{+0.15})\times10^{-2}\,,\quad \Delta a_\mu^{\mathrm{HVP,\,LO}}[\pi\pi,\tau]=(-15.20^{+2.26}_{-2.63})\times10^{-10}\,,
\end{equation}
as our final result for the IB corrections studied in this work. For comparison, Ref.~\cite{Davier:2010fmf} obtained $\Delta\mathcal{B}_{\pi\pi}^\mathrm{CVC}=(+0.69\pm0.22)\times10^{-2}$ and $\Delta a_\mu^{\mathrm{HVP,\,LO}}{[\pi\pi,\tau]}=(-16.07\pm1.85)\times10^{-10}$ using the GS model, and the most recent evaluation by the Orsay group \cite{Davier:2023fpl} yielded $\Delta a_\mu^{\mathrm{HVP,\,LO}}{[\pi\pi,\tau]}=(-14.9\pm1.9)\times10^{-10}$
. We agree nicely with these results and we observe that, 
while we have reduced the uncertainty on $\Delta\mathcal{B}_{\pi\pi}^\mathrm{CVC}$, we have a larger error on $\Delta a_\mu^{\mathrm{HVP,\,LO}}{[\pi\pi,\tau]}$. According to our preceding discussion, we have managed to reduce the uncertainty around the rho resonance, but not in the low di-pion invariant mass region, where our uncertainty on the $G_\mathrm{EM}(s)$ correction (see Ref.~\cite{Miranda:2020wdg}) penalizes us.

Our $a_\mu$ IB corrections imply that 
$a_\mu^{\mathrm{HVP,\,LO}}{[\pi\pi,\tau]}=(517.0\pm2.8[\mathrm{exp}]^{+2.3}_{-2.6}[\mathrm{IB}])\times10^{-10}$
, which in turn corresponds to \cite{Miranda:2020wdg}
\begin{equation}
 a_\mu^{\mathrm{HVP,\,LO}}{[\tau\,\mathrm{data}]}=\left(703.1\pm2.8[\mathrm{exp}]^{+2.3}_{-2.6}[\mathrm{IB}]\pm2.0[e^+e^-]\pm0.1[\mathrm{narrow\,res}]\pm0.7[\mathrm{QCD}]\right)\times10^{-10}\,,
 \end{equation}
 adding (in quadrature) to 
 $a_\mu^{\mathrm{HVP,\,LO}}{[\tau\,\mathrm{data}]}=\left(703.1^{+4.2}_{-4.4}\right)\times10^{-10}$
 . When accounting for the remaining SM contributions, this corresponds to 
 \begin{equation}
 \Delta a_\mu=a_\mu^{\mathrm{exp}}-a_\mu^{\mathrm{SM}}=\left(14.9^{+5.0}_{-5.2}\right)\times10^{-10},
 \end{equation}
 a $2.9\,\sigma$ deviation. The difference between the SM prediction and the $a_\mu$ world average is slightly reduced by combining compatible results, as illustrated in Ref.~\cite{Davier:2023fpl} with the BaBar, CMD-3 and tau data-driven determinations. In any case, the best agreement is found with the BMW evaluation \cite{Borsanyi:2020mff}, whose improvement with long-distance data-driven results \cite{Boccaletti:2024guq} accords with the $a_\mu$ measurement at $0.9\,\sigma$.

\section{Conclusions}\label{sec_Concl}
In this paper we have revisited the IB corrections relating the $e^+e^-$ cross-section and the tau decays spectral function, focusing on the $\pi\pi$ channel. For the IB corrections relating the weak and electromagnetic pion form factors we have considered different parametrizations: Gounaris-Sakurai (GS), K\"uhn-Santamar\'ia (KS), Guerrero-Pich (GP) and a dispersive (Disp) approach (whose seed form factor was also taken into account to extend the applicability of GP to larger energies). 

Since the GS and the Disp form factors are those satisfying better analyticity constraints, we consider them most reliable and use these in our reference results. Through fits including different $e^+e^-$ data sets (in addition to the Belle $\tau$ data), we have verified (particularly for GS and Disp) that the exclusion of the KLOE measurements improves notably the description of the measurements (a similar observation was made in Ref.~\cite{Davier:2023fpl}). This procedure leads to our final results for the IB corrections studied in this work, Eq.~(\ref{eq_FinalResults}), 
$\Delta\mathcal{B}_{\pi\pi}^\mathrm{CVC}=(+0.63_{-0.14}^{+0.15})\times10^{-2}\,,\; \Delta a_\mu^{\mathrm{HVP,\,LO}}{[\pi\pi,\tau]}=(-15.20^{+2.26}_{-2.63})\times10^{-10}$
, corresponding to 
$\Delta a_\mu=a_\mu^{\mathrm{exp}}-a_\mu^{\mathrm{SM}}=\left(14.9^{+5.0}_{-5.2}\right)\times10^{-10}$
, a $2.9\,\sigma$ difference. Our results agree nicely with previous determinations, with similar precision. Then, as our main conclusion, we have assessed and confirmed the soundness of the IB corrections that relate $e^+e^-$ and tau decay data into two pions,  owing to the reduced model-dependence of these effects. Consequently -while the incompatibility between KLOE and CMD-3 data is not properly understood-, the nice agreement among tau measurements (and particularly between the most precise ones, by ALEPH and Belle), emphasizes the \textcolor{black}{usefulness} of taking them into account in the updated SM prediction of the muon $g-2$ that will be published in the second White Paper by the Muon $g-2$ Theory Initiative.

\section{Acknowledgements}
We thank very useful discussions on this topic over the years with Michel Davier, Zhiqing Zhang, Bogdan Malaescu, Vincenzo Cirigliano, Toni Pich, Rafel Escribano and Pere Masjuan. We are particularly grateful to the Orsay group for insightful comments on our results, which helped us to improve our work. G.~L.~C. acknowledges Conahcyt support, from project CBF2023-2024-3226. 
A.~M. is supported by the European Union’s Horizon 2020 Research and Innovation Programme under grant 824093 (H2020-INFRAIA-2018-1), the Ministerio de Ciencia e Innovación under grant PID2020-112965GB-I00, by the
Secretaria d’Universitats i Recerca del Departament d’Empresa i Coneixement de la Generalitat
de Catalunya under grant 2021 SGR 00649, by MICINN with funding from European
Union NextGenerationEU (PRTR-C17.I1) and by Generalitat de Catalunya. IFAE is partially funded by the CERCA program of
the Generalitat de Catalunya.  
P.~R. is partly funded by Conahcyt (México) by project CBF2023-2024-3226, and by MCIN/AEI/10.13039/501100011033 (Spain), grants PID2020-114473GB-I00 and PID2023-146220NB-I00, and by Generalitat Valenciana (Spain), grant PROMETEO/2021/071.
\appendix
\section{Analyticity tests} \label{app_A}
Analyticity allows us to relate the real part of the form factor to its discontinuity or imaginary part. The unsubtracted dispersion relation is written as~\cite{Gan:2020aco}

\begin{equation}
    F(s)=\frac{1}{\pi}\int_{s_\text{thr}}^\infty ds^\prime \frac{\text{Im}\,F(s^\prime)}{s^\prime-s-i\epsilon}.
\end{equation}

As was already mentioned above, this relation is very powerful since it implies that the form factor $F(s)$ can be rebuilt anywhere in the complex plane once we know its imaginary part along the branch cut, which is in turn given by unitarity.

\begin{figure}[ht]
    \centering
    \includegraphics[width=0.44\textwidth]{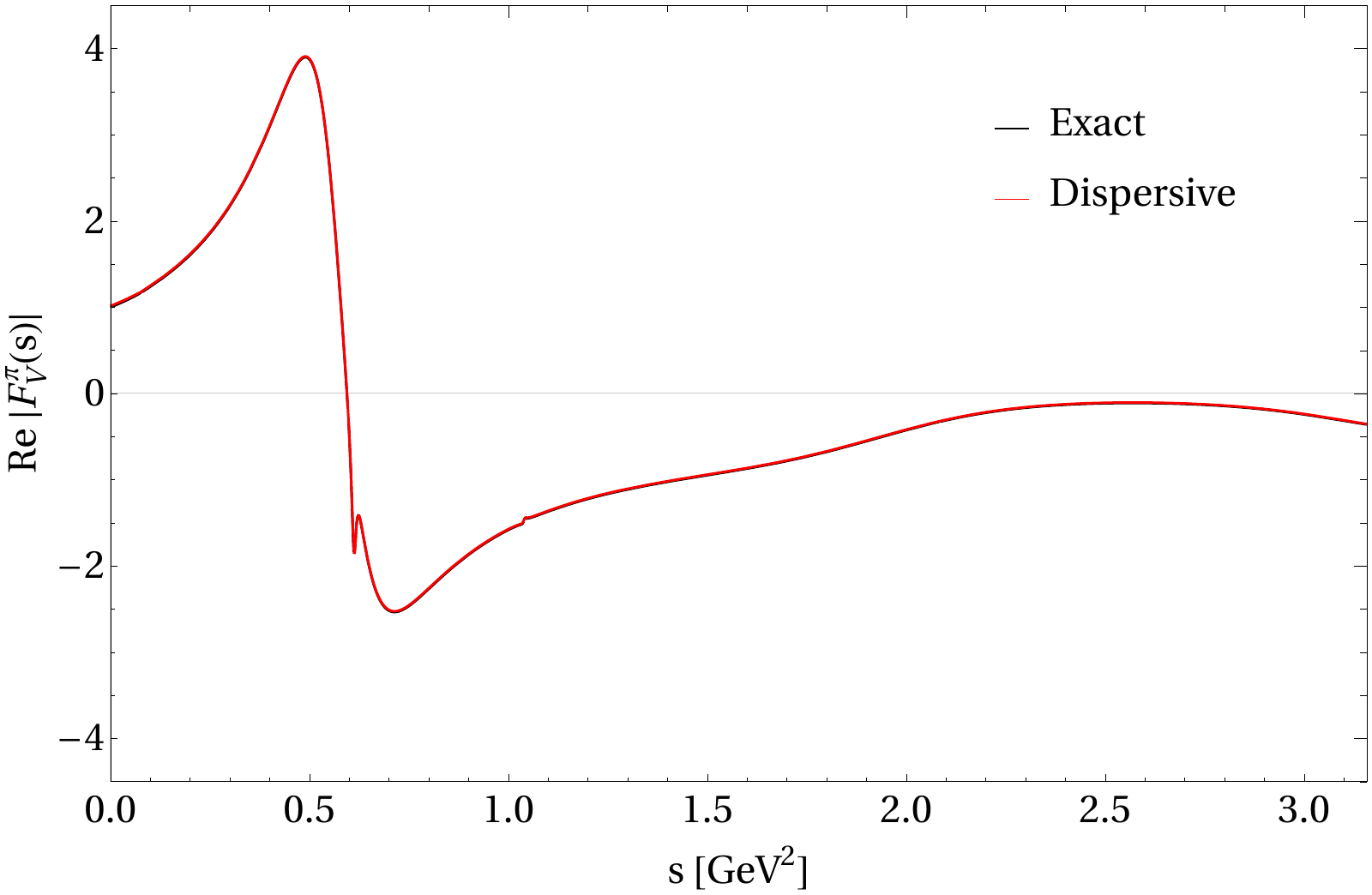}
    \includegraphics[width=0.44\textwidth]{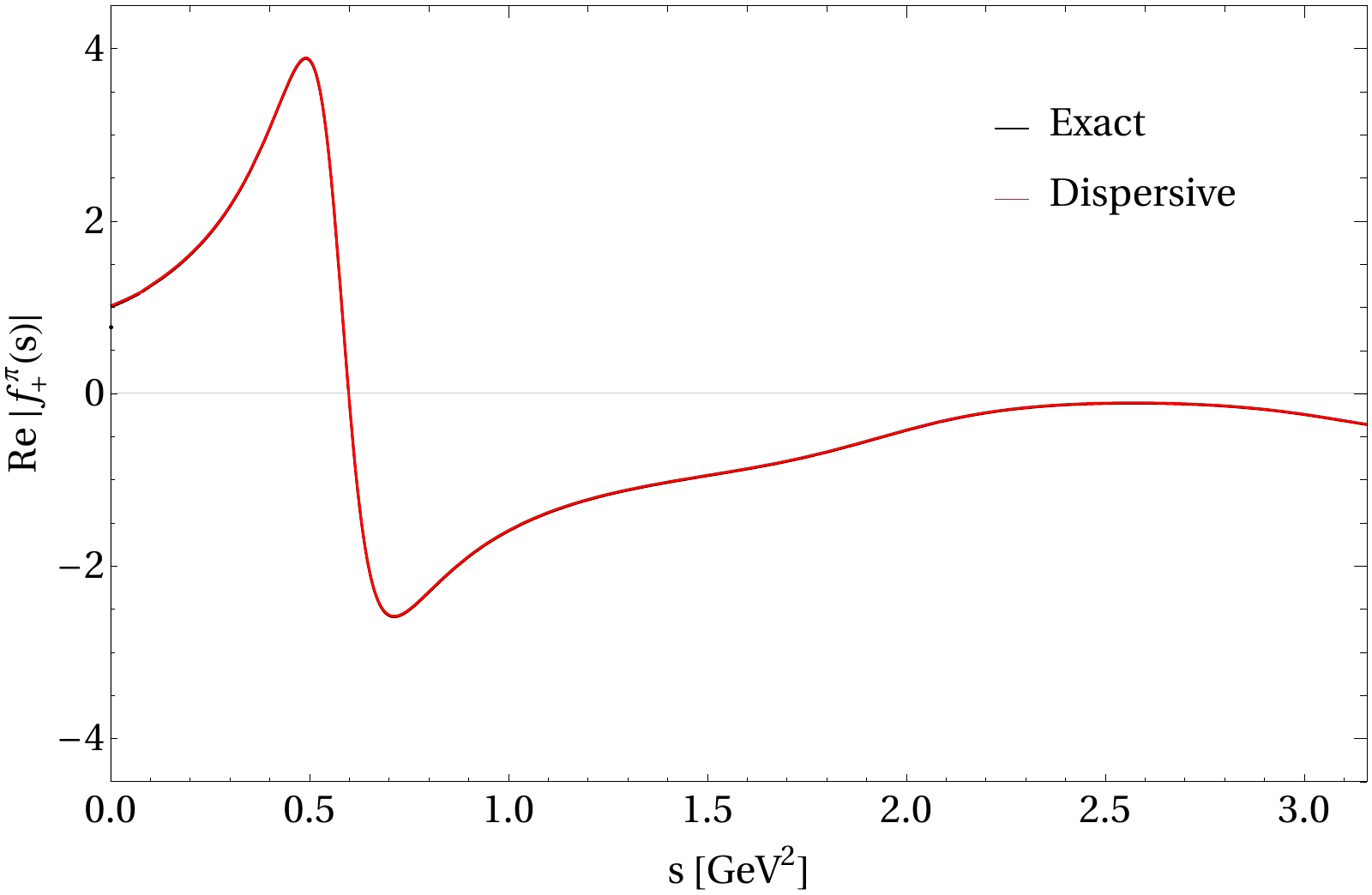}
    \includegraphics[width=0.44\textwidth]{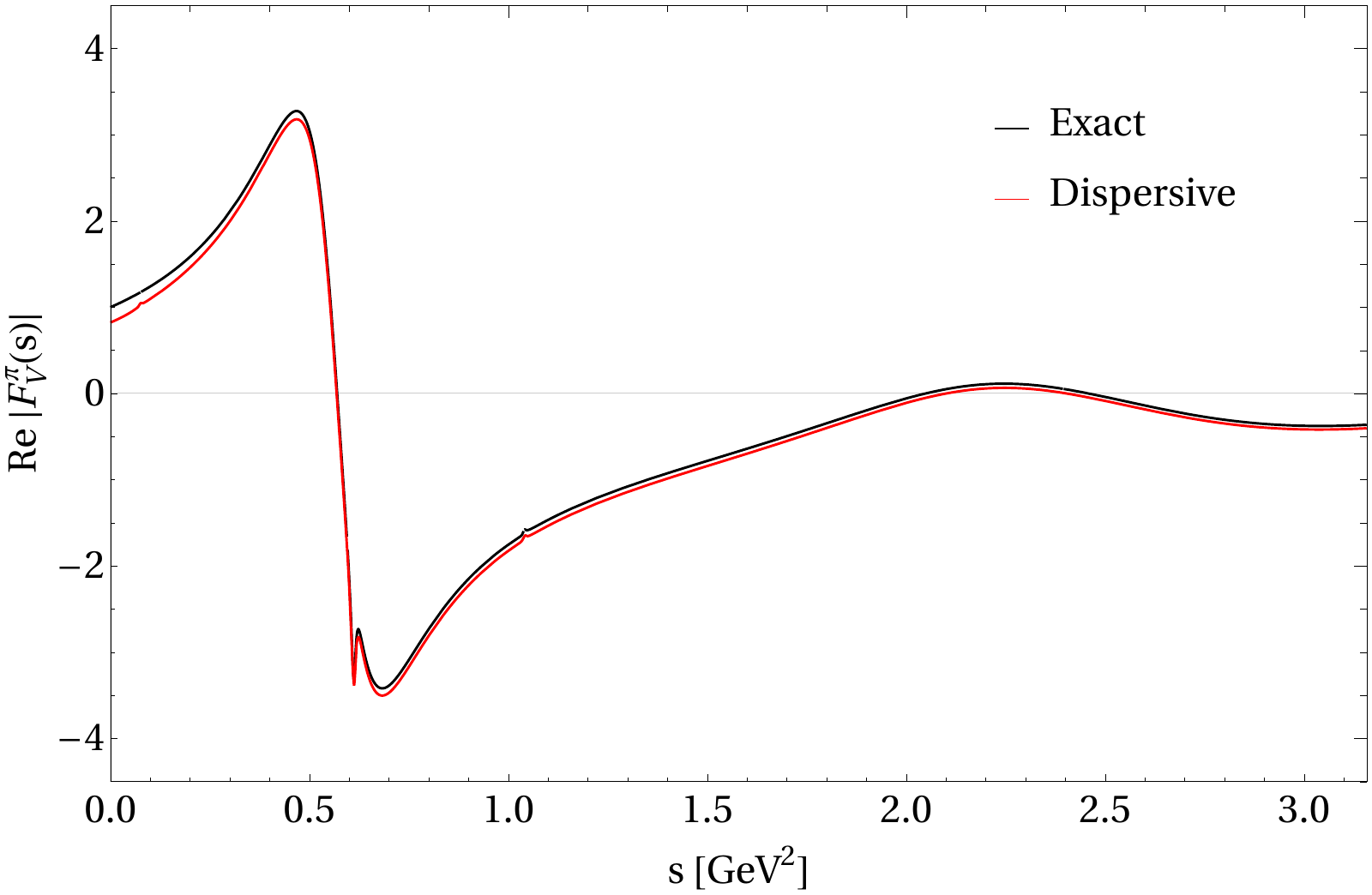}
    \includegraphics[width=0.44\textwidth]{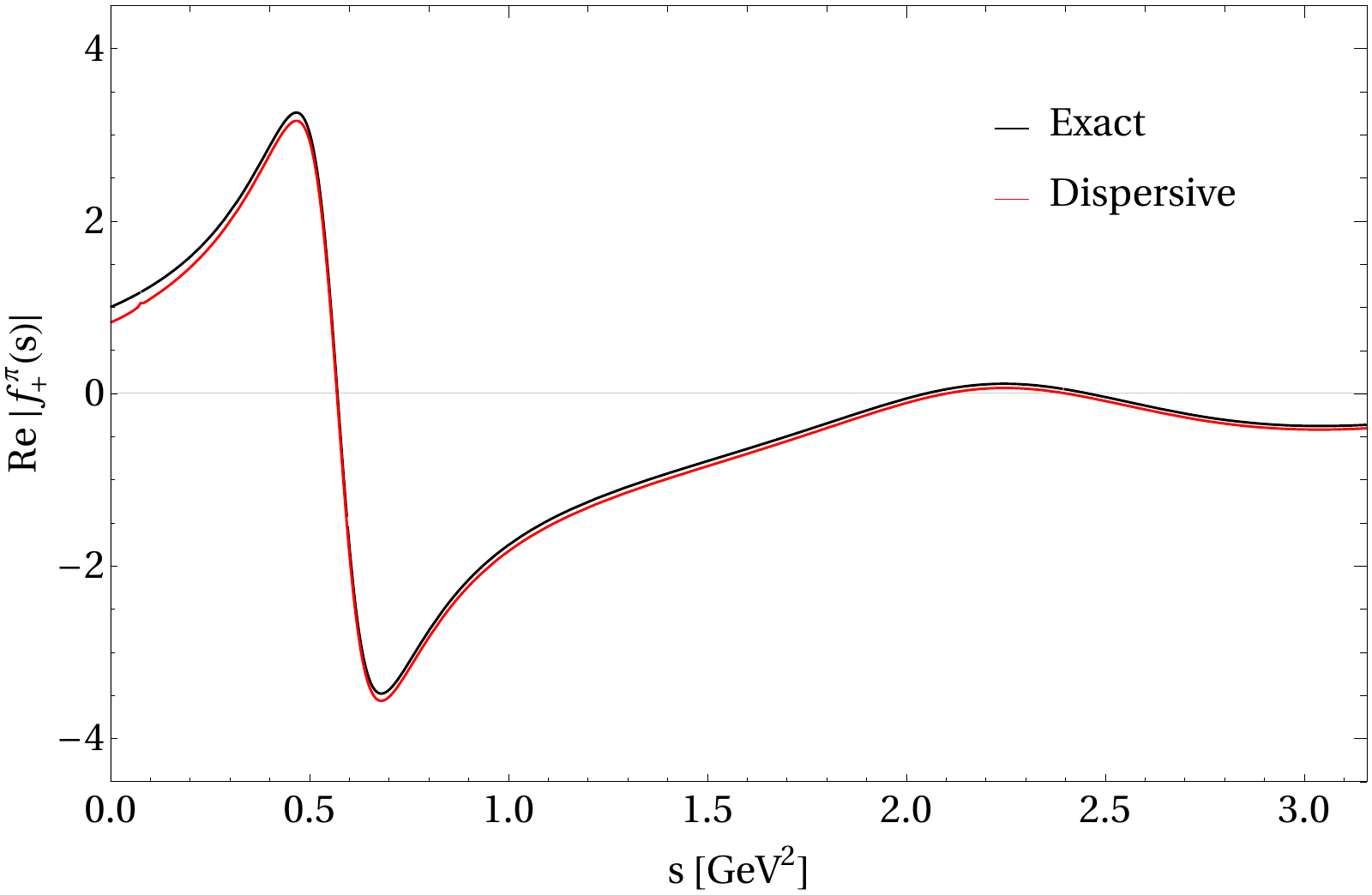}
    \includegraphics[width=0.44\textwidth]{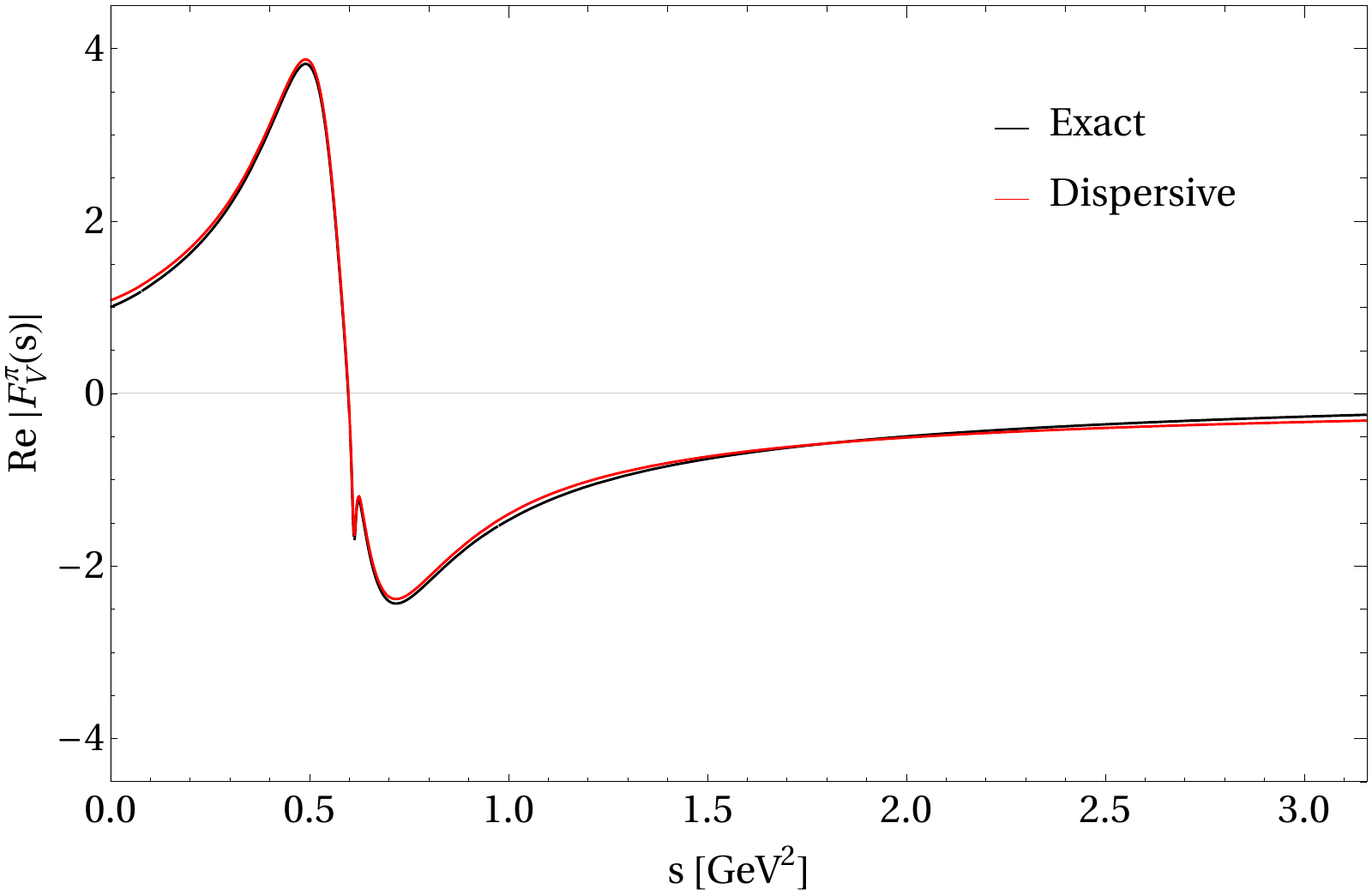}
    \includegraphics[width=0.44\textwidth]{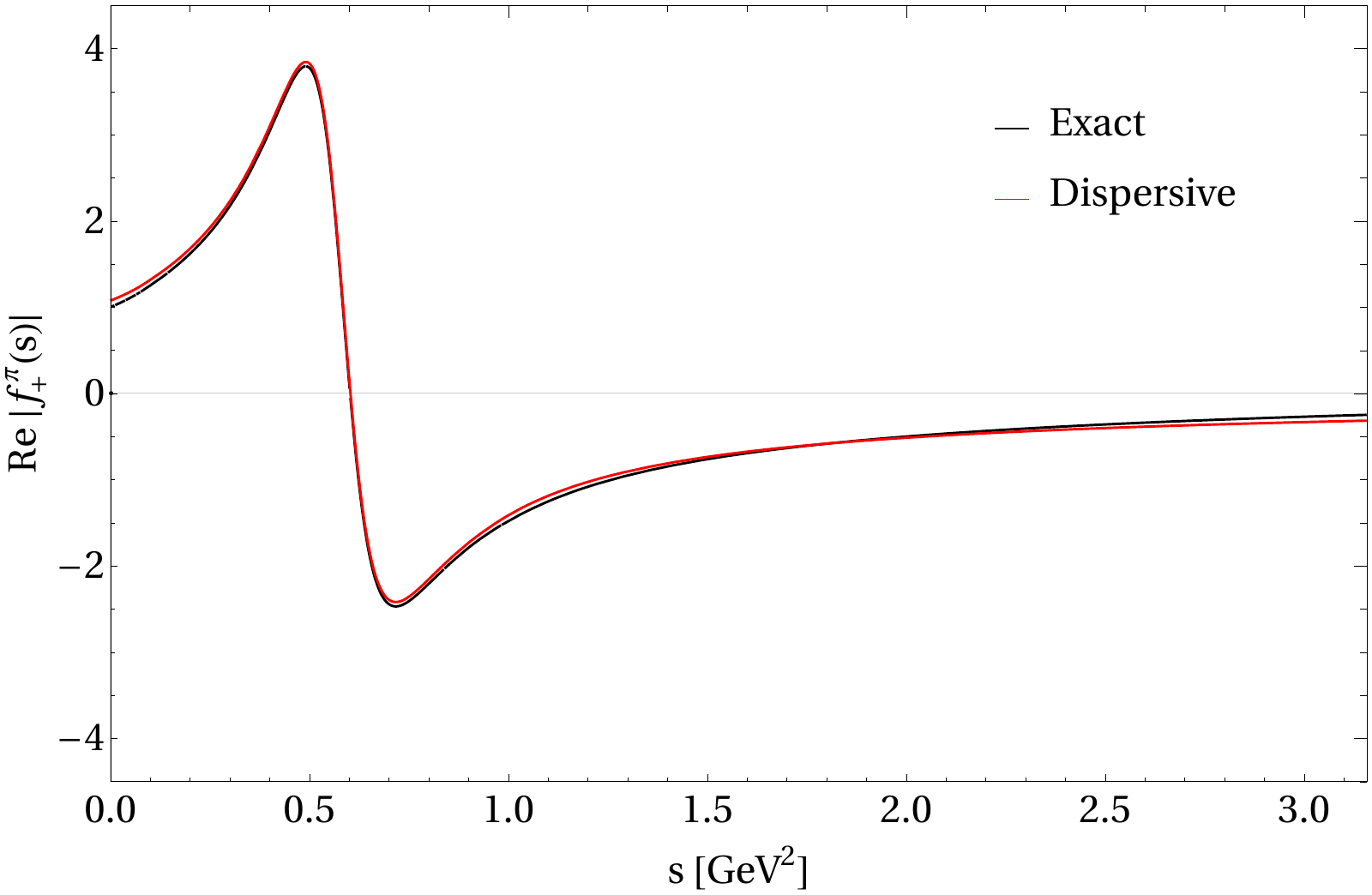}
    \includegraphics[width=0.44\textwidth]{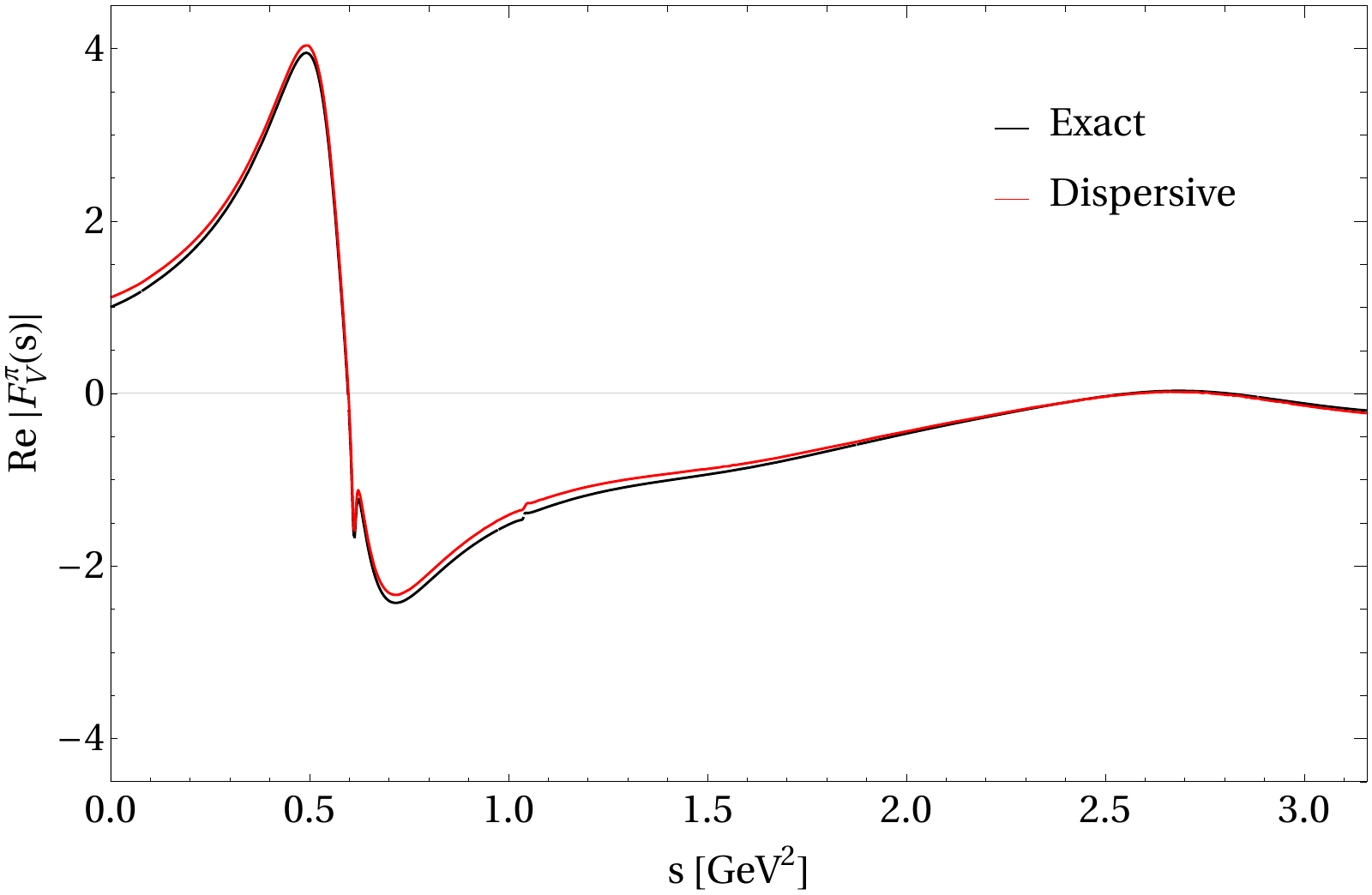}
    \includegraphics[width=0.44\textwidth]{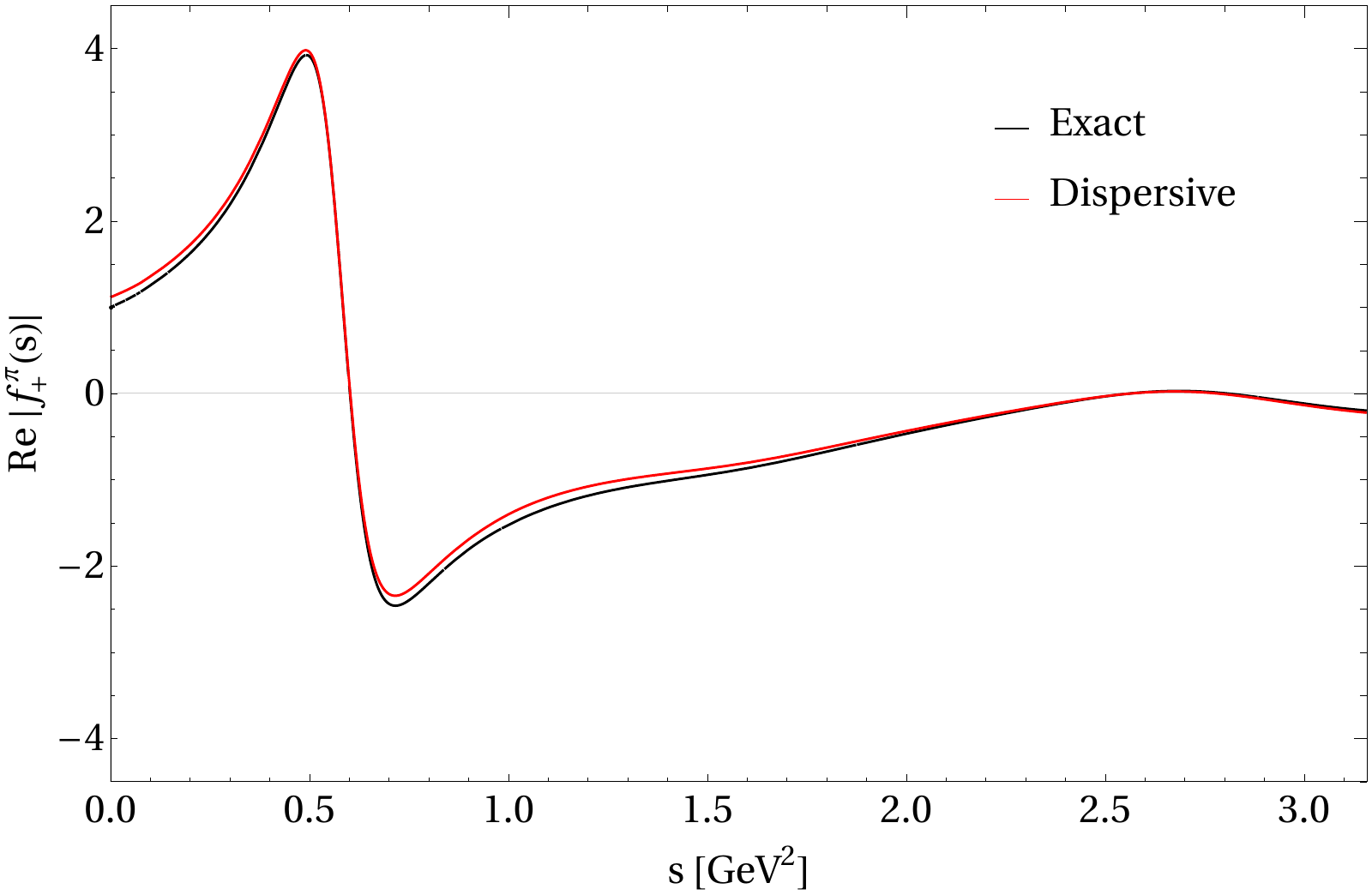}
    \captionsetup{width=0.88\linewidth}
    \caption{ From top to bottom: Comparison between the exact expression for the $\text{Re}\vert {(F/f)}^\pi_{V/+}(s)\vert$ and its estimation from Eq. (\ref{eq:DR01}) for the GS, KS, GP and Seed models. 
    }
    \label{fig:test_analyticity}
\end{figure}

Thus, the form factors in Eqs. (\ref{eq:KSGS}), (\ref{eq:FFGP}), (\ref{eq:FFGP2}) and (\ref{eq:FVseed}) must satisfy 

\begin{equation}\label{eq:DR01}
    \text{Re}\, F(s)=\frac{1}{\pi}\,\mathcal{P}\int_{s_\text{thr}}^\infty ds^\prime\,\frac{\text{Im}F(s^\prime)}{s^\prime-s},
\end{equation}

which can be used to evaluate the analyticity of these form factors. Our findings are depicted in Fig. \ref{fig:test_analyticity} for the GS, KS, GP and Seed parametrizations of the FFs. 

\begin{figure}[ht]
    \centering
    \includegraphics[width=0.44\textwidth]{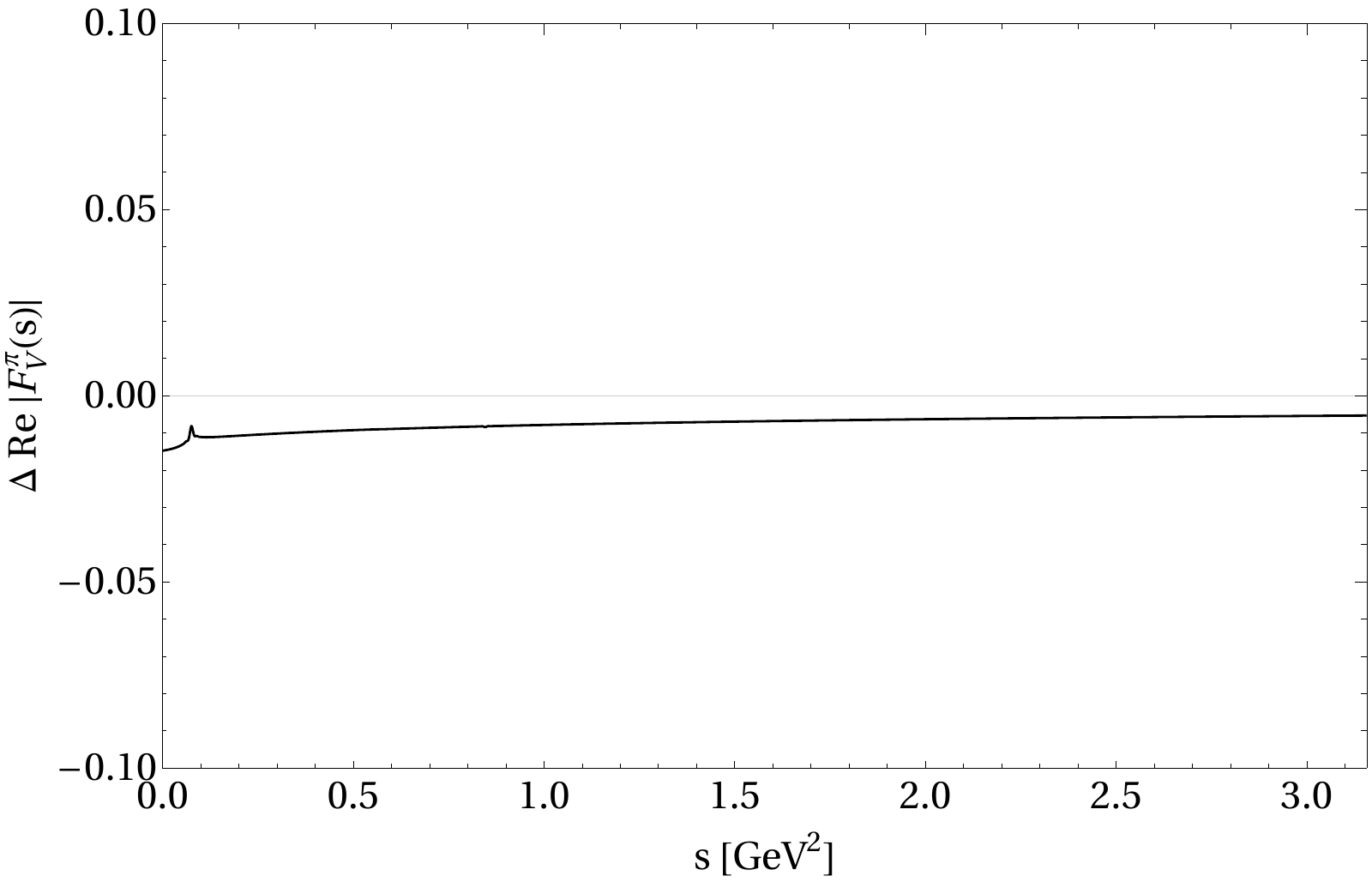}
    \includegraphics[width=0.44\textwidth]{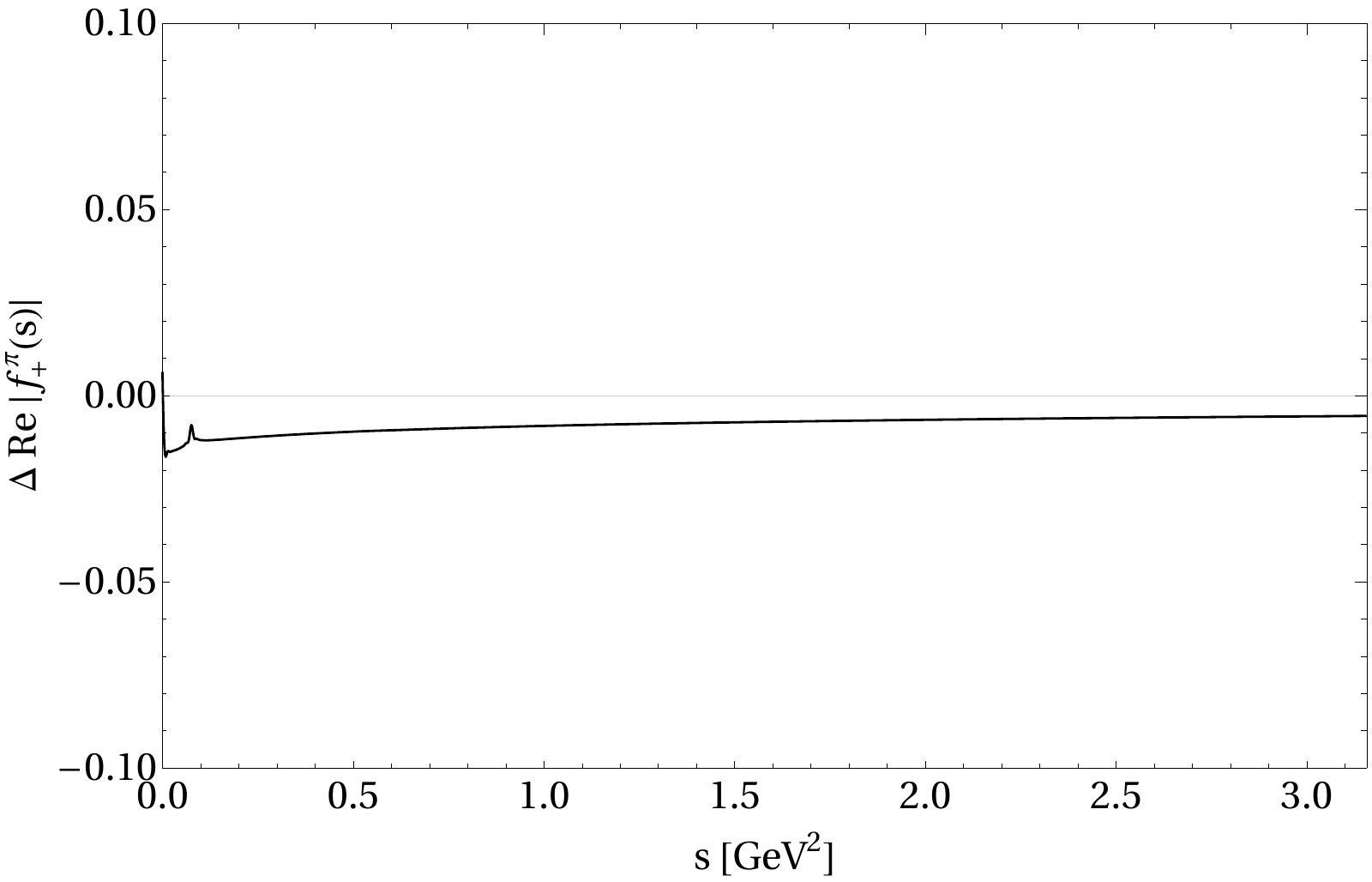}
    \includegraphics[width=0.44\textwidth]{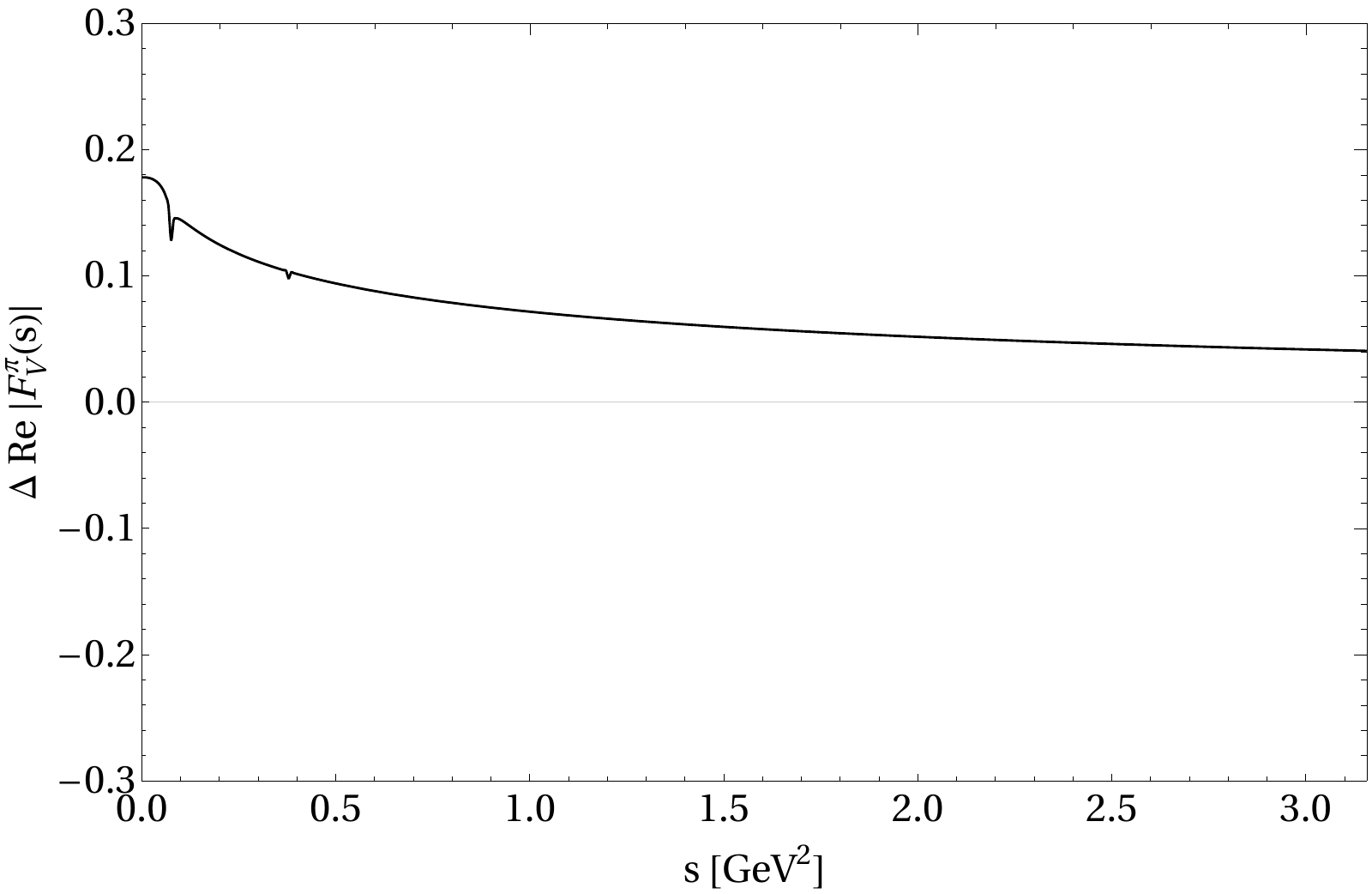}
    \includegraphics[width=0.44\textwidth]{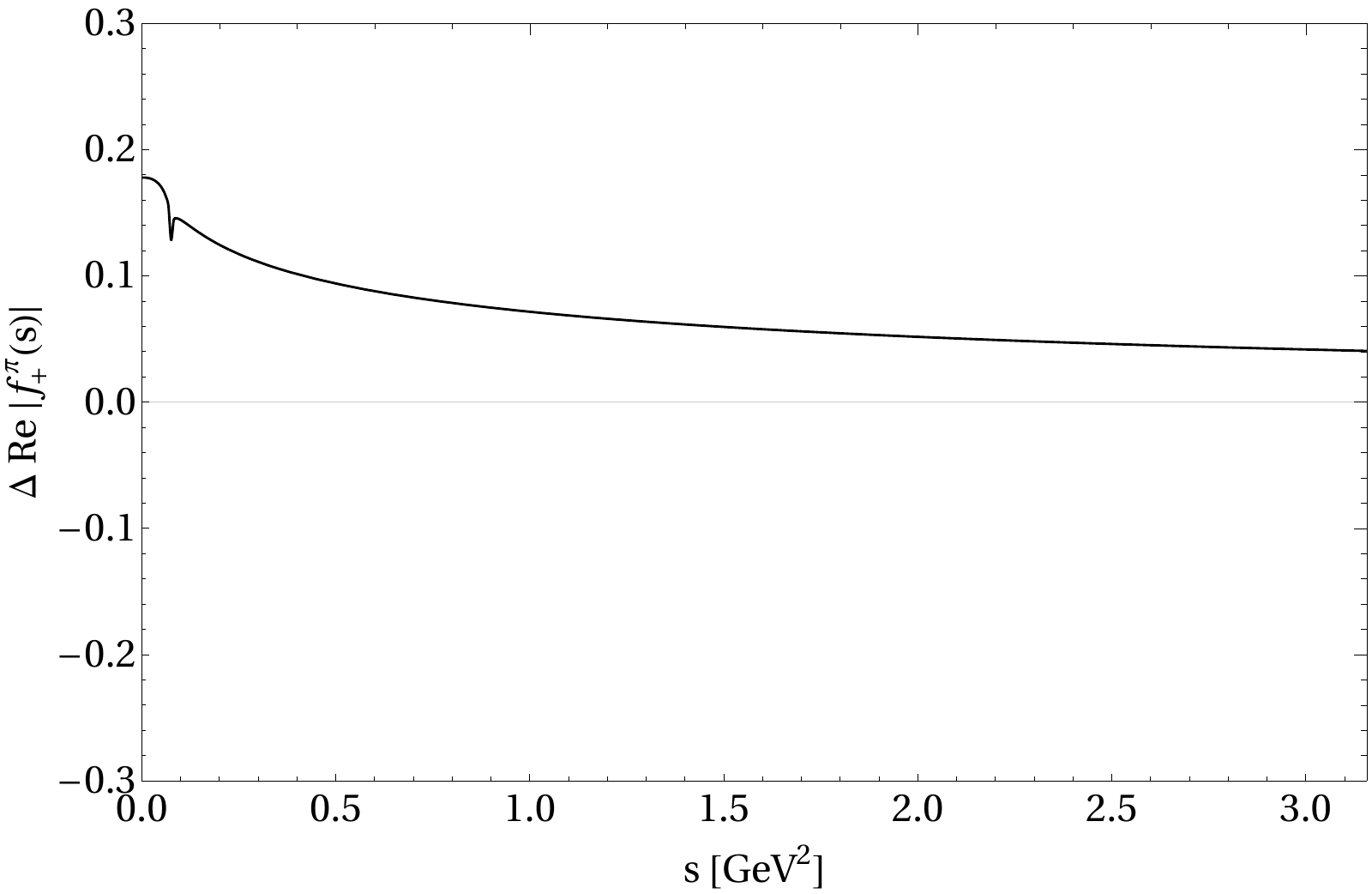}
    \includegraphics[width=0.44\textwidth]{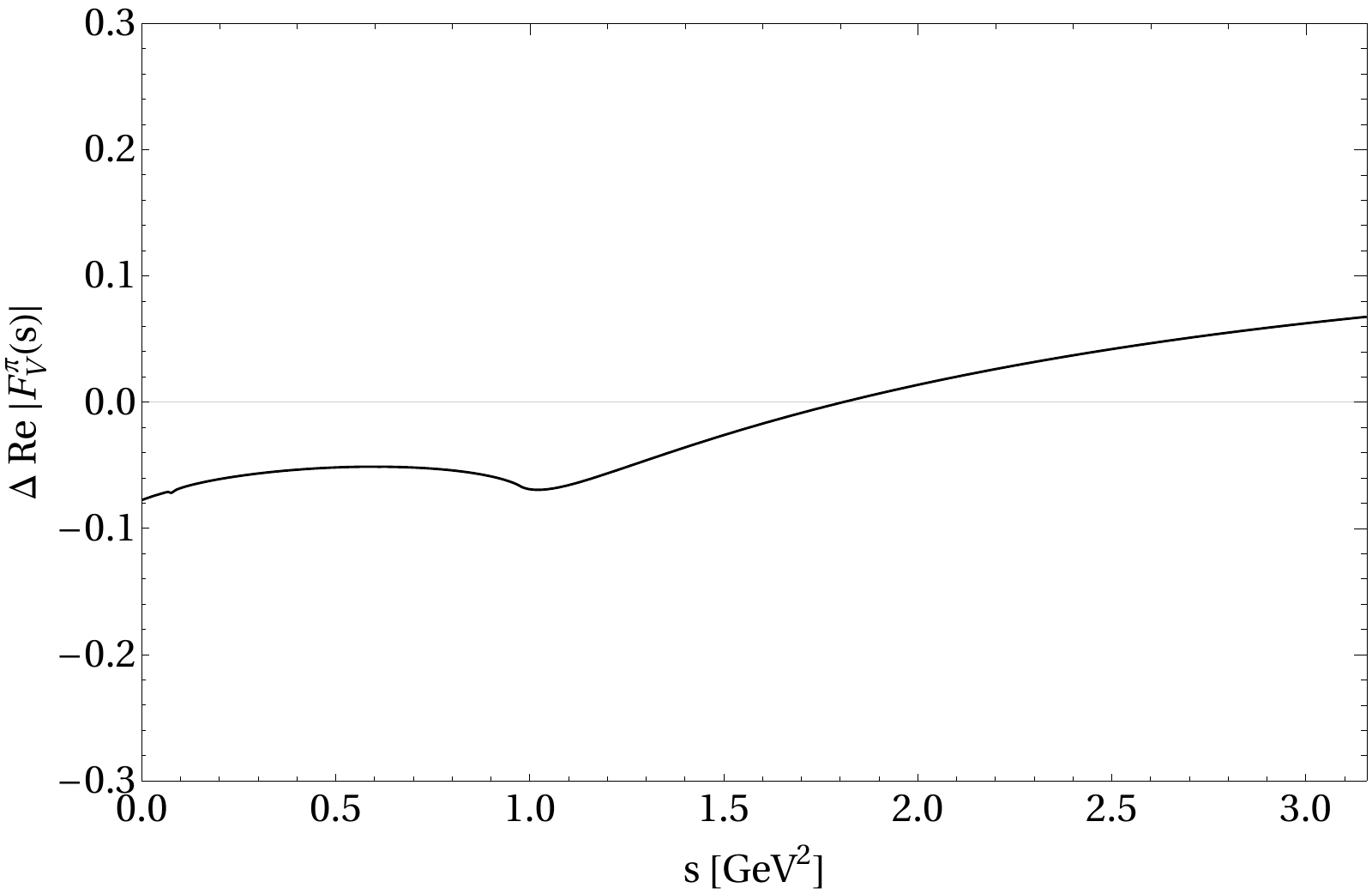}
    \includegraphics[width=0.44\textwidth]{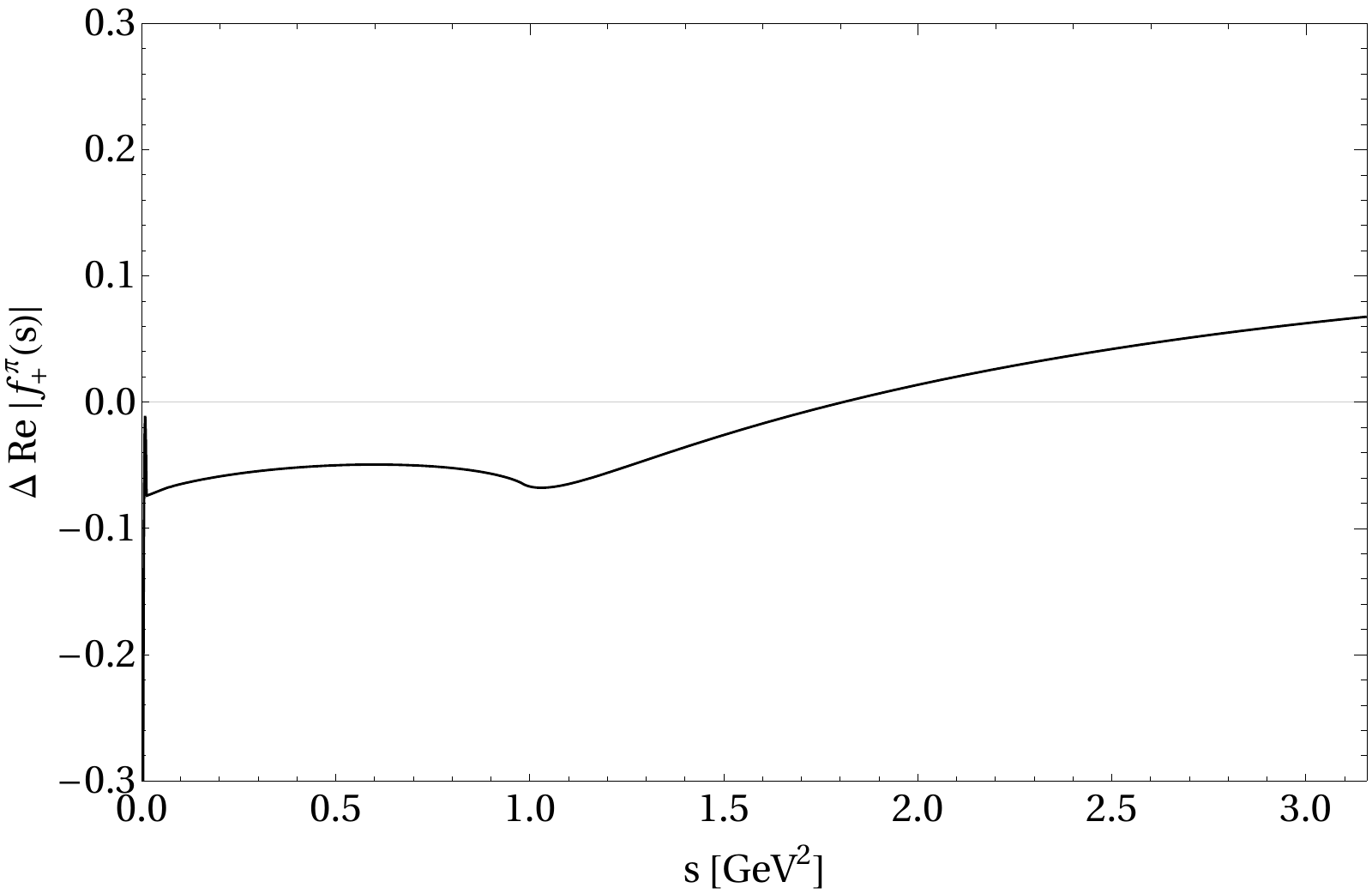}
    \includegraphics[width=0.44\textwidth]{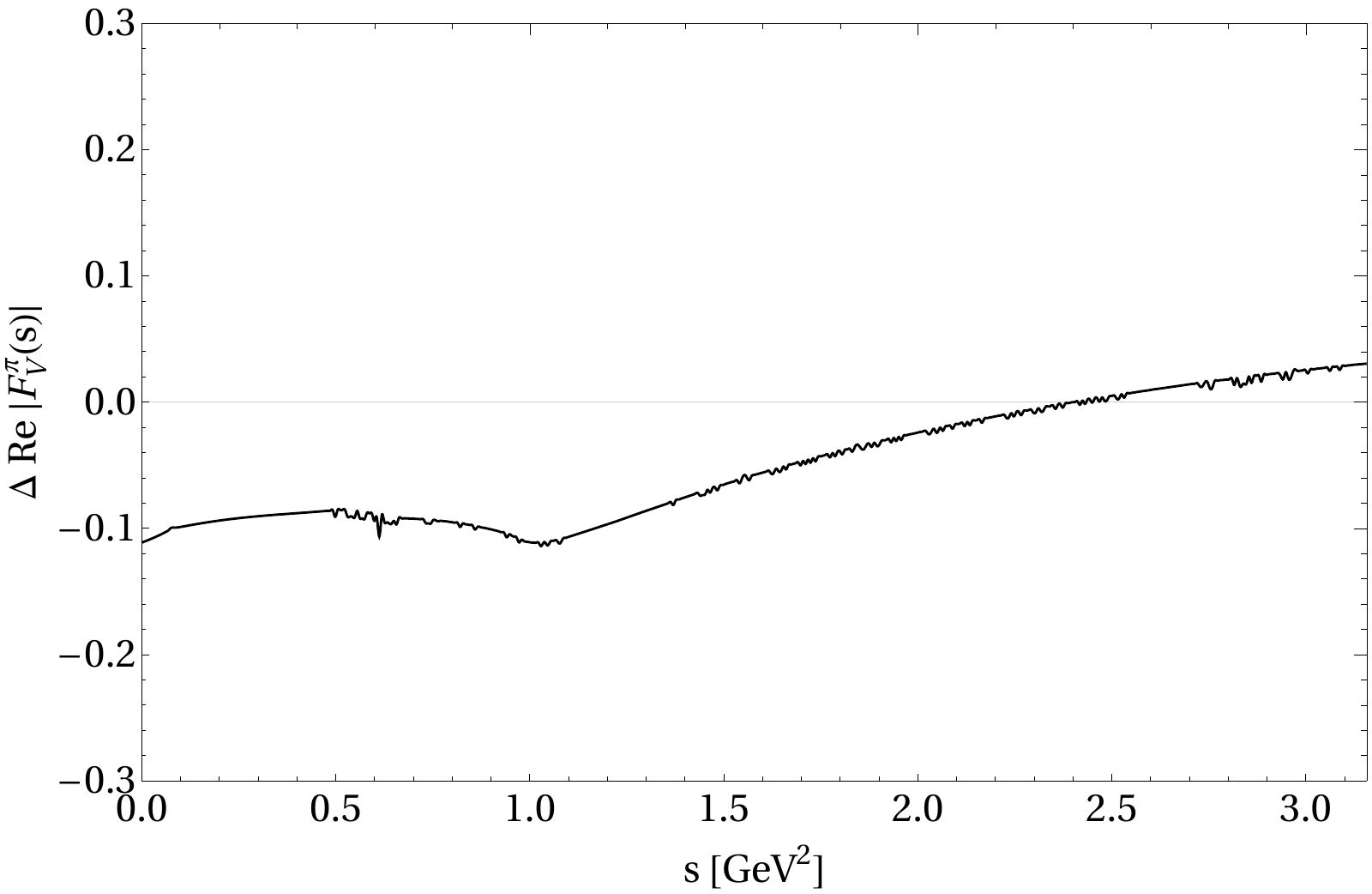}
    \includegraphics[width=0.44\textwidth]{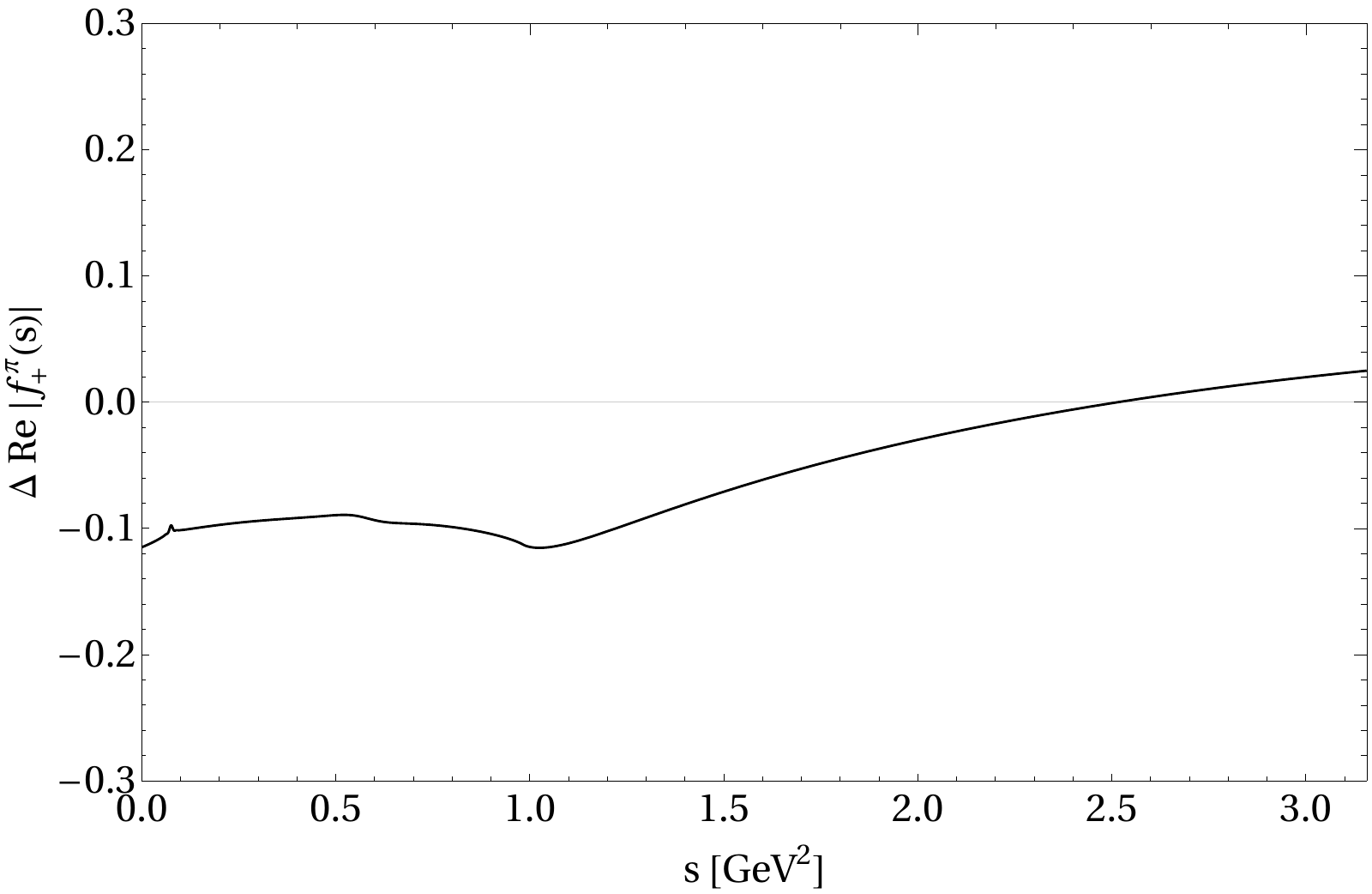}
    \captionsetup{width=0.88\linewidth}
    \caption{ From top to bottom: $\Delta \text{Re}[{(F/f)}_{V,+}(s)]$ in Eq. (\ref{eq:DR02}) for the GS, KS, GP and Seed models. 
    }
    \label{fig:test_analyticity2}
\end{figure}

In Fig. \ref{fig:test_analyticity2}, we plot the difference between the form factor and its prediction using Eq. (\ref{eq:DR01}), which reads

\begin{equation}\label{eq:DR02}
    \Delta \text{Re} [F(s)]\equiv\text{Re} F(s)-\frac{1}{\pi}\,\mathcal{P}\int_{s_\text{thr}}^\infty ds^\prime\,\frac{\text{Im}F(s^\prime)}{s^\prime-s}.
\end{equation}
 
These results show clearly that GS complies with analyticity constraints much better than the other FFs (KS, GP and Seed). The fulfillment of these restrictions is even better using the dispersive form factor, by construction. This justifies why we take GS and Disp as our reference results.

\bibliographystyle{unsrt}
\bibliography{bib}

\begin{thebibliography}{100}

\bibitem{ParticleDataGroup:2024cfk}
S.~Navas et~al.
\newblock {Review of particle physics}.
\newblock {\em Phys. Rev. D}, 110(3):030001, 2024.

\bibitem{Alemany:1997tn}
Ricard Alemany, Michel Davier, and Andreas Höcker.
\newblock {Improved determination of the hadronic contribution to the muon $(g-2)$ and to $\alpha(M_Z^2)$ using new data from hadronic $\tau$ decays}.
\newblock {\em Eur. Phys. J. C}, 2:123--135, 1998.

\bibitem{Cirigliano:2001er}
V.~Cirigliano, G.~Ecker, and H.~Neufeld.
\newblock {Isospin violation and the magnetic moment of the muon}.
\newblock {\em Phys. Lett. B}, 513:361--370, 2001.

\bibitem{Cirigliano:2002pv}
V.~Cirigliano, G.~Ecker, and H.~Neufeld.
\newblock {Radiative tau decay and the magnetic moment of the muon}.
\newblock {\em JHEP}, 08:002, 2002.

\bibitem{Davier:2002dy}
M.~Davier, S.~Eidelman, Andreas Höcker, and Z.~Zhang.
\newblock {Confronting spectral functions from $e^+ e^-$ annihilation and $\tau$ decays: Consequences for the muon magnetic moment}.
\newblock {\em Eur. Phys. J. C}, 27:497--521, 2003.

\bibitem{Davier:2003pw}
M.~Davier, S.~Eidelman, Andreas Höcker, and Z.~Zhang.
\newblock {Updated estimate of the muon magnetic moment using revised results from $e^+ e^-$ annihilation}.
\newblock {\em Eur. Phys. J. C}, 31:503--510, 2003.

\bibitem{Davier:2010fmf}
M.~Davier, A.~Hoecker, G.~López~Castro, B.~Malaescu, X.~H. Mo, G.~Toledo~Sánchez, P.~Wang, C.~Z. Yuan, and Z.~Zhang.
\newblock {The Discrepancy Between $\tau$ and $e^+e^-$ Spectral Functions Revisited and the Consequences for the Muon Magnetic Anomaly}.
\newblock {\em Eur. Phys. J. C}, 66:127--136, 2010.

\bibitem{Jegerlehner:2011ti}
Fred Jegerlehner and Robert Szafron.
\newblock {$\rho^0 - \gamma$ mixing in the neutral channel pion form factor $F_{\pi}^{(e)}$ and its role in comparing $e^+ e^-$ with $\tau$ spectral functions}.
\newblock {\em Eur. Phys. J. C}, 71:1632, 2011.

\bibitem{Muong-2:2006rrc}
G.~W. Bennett et~al.
\newblock {Final Report of the Muon E821 Anomalous Magnetic Moment Measurement at BNL}.
\newblock {\em Phys. Rev. D}, 73:072003, 2006.

\bibitem{Muong-2:2021ojo}
B.~Abi et~al.
\newblock {Measurement of the Positive Muon Anomalous Magnetic Moment to 0.46 ppm}.
\newblock {\em Phys. Rev. Lett.}, 126(14):141801, 2021.

\bibitem{Muong-2:2023cdq}
D.~P. Aguillard et~al.
\newblock {Measurement of the Positive Muon Anomalous Magnetic Moment to 0.20~ppm}.
\newblock {\em Phys. Rev. Lett.}, 131(16):161802, 2023.

\bibitem{Aoyama:2020ynm}
T.~Aoyama et~al.
\newblock {The anomalous magnetic moment of the muon in the Standard Model}.
\newblock {\em Phys. Rept.}, 887:1--166, 2020.

\bibitem{Colangelo:2022jxc}
G.~Colangelo et~al.
\newblock {Prospects for precise predictions of $a_\mu$ in the Standard Model}.
\newblock 3 2022.

\bibitem{Davier:2017zfy}
Michel Davier, Andreas Hoecker, Bogdan Malaescu, and Zhiqing Zhang.
\newblock {Reevaluation of the hadronic vacuum polarisation contributions to the Standard Model predictions of the muon $g-2$ and ${\alpha (m_Z^2)}$ using newest hadronic cross-section data}.
\newblock {\em Eur. Phys. J. C}, 77(12):827, 2017.

\bibitem{Keshavarzi:2018mgv}
Alexander Keshavarzi, Daisuke Nomura, and Thomas Teubner.
\newblock {Muon $g-2$ and $\alpha(M_Z^2)$: a new data-based analysis}.
\newblock {\em Phys. Rev. D}, 97(11):114025, 2018.

\bibitem{Colangelo:2018mtw}
Gilberto Colangelo, Martin Hoferichter, and Peter Stoffer.
\newblock {Two-pion contribution to hadronic vacuum polarization}.
\newblock {\em JHEP}, 02:006, 2019.

\bibitem{Hoferichter:2019mqg}
Martin Hoferichter, Bai-Long Hoid, and Bastian Kubis.
\newblock {Three-pion contribution to hadronic vacuum polarization}.
\newblock {\em JHEP}, 08:137, 2019.

\bibitem{Davier:2019can}
M.~Davier, A.~Hoecker, B.~Malaescu, and Z.~Zhang.
\newblock {A new evaluation of the hadronic vacuum polarisation contributions to the muon anomalous magnetic moment and to $\alpha(m_Z^2)$}.
\newblock {\em Eur. Phys. J. C}, 80(3):241, 2020.
\newblock [Erratum: Eur.Phys.J.C 80, 410 (2020)].

\bibitem{Keshavarzi:2019abf}
Alexander Keshavarzi, Daisuke Nomura, and Thomas Teubner.
\newblock {$g-2$ of charged leptons, $\alpha (M^2_Z)$ , and the hyperfine splitting of muonium}.
\newblock {\em Phys. Rev. D}, 101(1):014029, 2020.

\bibitem{Kurz:2014wya}
Alexander Kurz, Tao Liu, Peter Marquard, and Matthias Steinhauser.
\newblock {Hadronic contribution to the muon anomalous magnetic moment to next-to-next-to-leading order}.
\newblock {\em Phys. Lett. B}, 734:144--147, 2014.

\bibitem{FermilabLattice:2017wgj}
B.~Chakraborty et~al.
\newblock {Strong-Isospin-Breaking Correction to the Muon Anomalous Magnetic Moment from Lattice QCD at the Physical Point}.
\newblock {\em Phys. Rev. Lett.}, 120(15):152001, 2018.

\bibitem{Budapest-Marseille-Wuppertal:2017okr}
Sz. Borsanyi et~al.
\newblock {Hadronic vacuum polarization contribution to the anomalous magnetic moments of leptons from first principles}.
\newblock {\em Phys. Rev. Lett.}, 121(2):022002, 2018.

\bibitem{RBC:2018dos}
T.~Blum, P.~A. Boyle, V.~G\"ulpers, T.~Izubuchi, L.~Jin, C.~Jung, A.~J\"uttner, C.~Lehner, A.~Portelli, and J.~T. Tsang.
\newblock {Calculation of the hadronic vacuum polarization contribution to the muon anomalous magnetic moment}.
\newblock {\em Phys. Rev. Lett.}, 121(2):022003, 2018.

\bibitem{Giusti:2019xct}
D.~Giusti, V.~Lubicz, G.~Martinelli, F.~Sanfilippo, and S.~Simula.
\newblock {Electromagnetic and strong isospin-breaking corrections to the muon $g - 2$ from Lattice QCD+QED}.
\newblock {\em Phys. Rev. D}, 99(11):114502, 2019.

\bibitem{Shintani:2019wai}
Eigo Shintani and Yoshinobu Kuramashi.
\newblock {Hadronic vacuum polarization contribution to the muon $g-2$ with 2+1 flavor lattice QCD on a larger than (10 fm$)^4$ lattice at the physical point}.
\newblock {\em Phys. Rev. D}, 100(3):034517, 2019.

\bibitem{FermilabLattice:2019ugu}
C.~T.~H. Davies et~al.
\newblock {Hadronic-vacuum-polarization contribution to the muon\textquoteright{}s anomalous magnetic moment from four-flavor lattice QCD}.
\newblock {\em Phys. Rev. D}, 101(3):034512, 2020.

\bibitem{Gerardin:2019rua}
Antoine G\'erardin, Marco C\`e, Georg von Hippel, Ben H\"orz, Harvey~B. Meyer, Daniel Mohler, Konstantin Ottnad, Jonas Wilhelm, and Hartmut Wittig.
\newblock {The leading hadronic contribution to $(g-2)_\mu$ from lattice QCD with $N_{\rm f}=2+1$ flavours of O($a$) improved Wilson quarks}.
\newblock {\em Phys. Rev. D}, 100(1):014510, 2019.

\bibitem{Aubin:2019usy}
Christopher Aubin, Thomas Blum, Cheng Tu, Maarten Golterman, Chulwoo Jung, and Santiago Peris.
\newblock {Light quark vacuum polarization at the physical point and contribution to the muon $g-2$}.
\newblock {\em Phys. Rev. D}, 101(1):014503, 2020.

\bibitem{Giusti:2019hkz}
D.~Giusti and S.~Simula.
\newblock {Lepton anomalous magnetic moments in Lattice QCD+QED}.
\newblock {\em PoS}, LATTICE2019:104, 2019.

\bibitem{Melnikov:2003xd}
Kirill Melnikov and Arkady Vainshtein.
\newblock {Hadronic light-by-light scattering contribution to the muon anomalous magnetic moment revisited}.
\newblock {\em Phys. Rev. D}, 70:113006, 2004.

\bibitem{Masjuan:2017tvw}
Pere Masjuan and Pablo Sánchez-Puertas.
\newblock {Pseudoscalar-pole contribution to the $(g_{\mu}-2)$: a rational approach}.
\newblock {\em Phys. Rev. D}, 95(5):054026, 2017.

\bibitem{Colangelo:2017fiz}
Gilberto Colangelo, Martin Hoferichter, Massimiliano Procura, and Peter Stoffer.
\newblock {Dispersion relation for hadronic light-by-light scattering: two-pion contributions}.
\newblock {\em JHEP}, 04:161, 2017.

\bibitem{Hoferichter:2018kwz}
Martin Hoferichter, Bai-Long Hoid, Bastian Kubis, Stefan Leupold, and Sebastian~P. Schneider.
\newblock {Dispersion relation for hadronic light-by-light scattering: pion pole}.
\newblock {\em JHEP}, 10:141, 2018.

\bibitem{Gerardin:2019vio}
Antoine G\'erardin, Harvey~B. Meyer, and Andreas Nyffeler.
\newblock {Lattice calculation of the pion transition form factor with $N_f=2+1$ Wilson quarks}.
\newblock {\em Phys. Rev. D}, 100(3):034520, 2019.

\bibitem{Bijnens:2019ghy}
Johan Bijnens, Nils Hermansson-Truedsson, and Antonio Rodr\'\i{}guez-S\'anchez.
\newblock {Short-distance constraints for the HLbL contribution to the muon anomalous magnetic moment}.
\newblock {\em Phys. Lett. B}, 798:134994, 2019.

\bibitem{Colangelo:2019uex}
Gilberto Colangelo, Franziska Hagelstein, Martin Hoferichter, Laetitia Laub, and Peter Stoffer.
\newblock {Longitudinal short-distance constraints for the hadronic light-by-light contribution to $(g-2)_\mu$ with large-$N_c$ Regge models}.
\newblock {\em JHEP}, 03:101, 2020.

\bibitem{Pauk:2014rta}
Vladyslav Pauk and Marc Vanderhaeghen.
\newblock {Single meson contributions to the muon`s anomalous magnetic moment}.
\newblock {\em Eur. Phys. J. C}, 74(8):3008, 2014.

\bibitem{Danilkin:2016hnh}
Igor Danilkin and Marc Vanderhaeghen.
\newblock {Light-by-light scattering sum rules in light of new data}.
\newblock {\em Phys. Rev. D}, 95(1):014019, 2017.

\bibitem{Jegerlehner:2017gek}
Friedrich Jegerlehner.
\newblock {\em {The Anomalous Magnetic Moment of the Muon}}, volume 274.
\newblock Springer, Cham, 2017.

\bibitem{Knecht:2018sci}
M.~Knecht, S.~Narison, A.~Rabemananjara, and D.~Rabetiarivony.
\newblock {Scalar meson contributions to $a_\mu$ from hadronic light-by-light scattering}.
\newblock {\em Phys. Lett. B}, 787:111--123, 2018.

\bibitem{Eichmann:2019bqf}
Gernot Eichmann, Christian~S. Fischer, and Richard Williams.
\newblock {Kaon-box contribution to the anomalous magnetic moment of the muon}.
\newblock {\em Phys. Rev. D}, 101(5):054015, 2020.

\bibitem{Roig:2019reh}
Pablo Roig and Pablo Sánchez-Puertas.
\newblock {Axial-vector exchange contribution to the hadronic light-by-light piece of the muon anomalous magnetic moment}.
\newblock {\em Phys. Rev. D}, 101(7):074019, 2020.

\bibitem{Colangelo:2014qya}
Gilberto Colangelo, Martin Hoferichter, Andreas Nyffeler, Massimo Passera, and Peter Stoffer.
\newblock {Remarks on higher-order hadronic corrections to the muon g\ensuremath{-}2}.
\newblock {\em Phys. Lett. B}, 735:90--91, 2014.

\bibitem{Blum:2019ugy}
Thomas Blum, Norman Christ, Masashi Hayakawa, Taku Izubuchi, Luchang Jin, Chulwoo Jung, and Christoph Lehner.
\newblock {Hadronic Light-by-Light Scattering Contribution to the Muon Anomalous Magnetic Moment from Lattice QCD}.
\newblock {\em Phys. Rev. Lett.}, 124(13):132002, 2020.

\bibitem{Aoyama:2012wk}
Tatsumi Aoyama, Masashi Hayakawa, Toichiro Kinoshita, and Makiko Nio.
\newblock {Complete Tenth-Order QED Contribution to the Muon $g-2$}.
\newblock {\em Phys. Rev. Lett.}, 109:111808, 2012.

\bibitem{Aoyama:2019ryr}
Tatsumi Aoyama, Toichiro Kinoshita, and Makiko Nio.
\newblock {Theory of the Anomalous Magnetic Moment of the Electron}.
\newblock {\em Atoms}, 7(1):28, 2019.

\bibitem{Czarnecki:2002nt}
Andrzej Czarnecki, William~J. Marciano, and Arkady Vainshtein.
\newblock {Refinements in electroweak contributions to the muon anomalous magnetic moment}.
\newblock {\em Phys. Rev. D}, 67:073006, 2003.
\newblock [Erratum: Phys.Rev.D 73, 119901 (2006)].

\bibitem{Gnendiger:2013pva}
C.~Gnendiger, D.~St\"ockinger, and H.~St\"ockinger-Kim.
\newblock {The electroweak contributions to $(g-2)_\mu$ after the Higgs boson mass measurement}.
\newblock {\em Phys. Rev. D}, 88:053005, 2013.

\bibitem{KLOE:2012anl}
D.~Babusci et~al.
\newblock {Precision measurement of $\sigma(e^+e^-\rightarrow \pi^+\pi^-\gamma)/ \sigma(e^+e^-\rightarrow \mu^+\mu^-\gamma)$ and determination of the $\pi^+\pi^-$ contribution to the muon anomaly with the KLOE detector}.
\newblock {\em Phys. Lett. B}, 720:336--343, 2013.

\bibitem{BaBar:2012bdw}
J.~P. Lees et~al.
\newblock {Precise Measurement of the $e^+ e^- \to \pi^+\pi^- (\gamma)$ Cross Section with the Initial-State Radiation Method at BABAR}.
\newblock {\em Phys. Rev. D}, 86:032013, 2012.

\bibitem{KLOE-2:2017fda}
A.~Anastasi et~al.
\newblock {Combination of KLOE $\sigma\big(e^+e^-\rightarrow\pi^+\pi^-\gamma(\gamma)\big)$ measurements and determination of $a_{\mu}^{\pi^+\pi^-}$ in the energy range $0.10 < s < 0.95$ GeV$^2$}.
\newblock {\em JHEP}, 03:173, 2018.

\bibitem{CMD-2:2003gqi}
R.~R. Akhmetshin et~al.
\newblock {Reanalysis of hadronic cross-section measurements at CMD-2}.
\newblock {\em Phys. Lett. B}, 578:285--289, 2004.

\bibitem{CMD-2:2005mvb}
V.~M. Aul'chenko et~al.
\newblock {Measurement of the pion form-factor in the range $1.04-1.38$ GeV with the CMD-2 detector}.
\newblock {\em JETP Lett.}, 82:743--747, 2005.

\bibitem{Aulchenko:2006dxz}
V.~M. Aul'chenko et~al.
\newblock {Measurement of the $e^+ e^- \to \pi^+ \pi^-$ cross section with the CMD-2 detector in the 370 - 520-MeV c.m. energy range}.
\newblock {\em JETP Lett.}, 84:413--417, 2006.

\bibitem{CMD-2:2006gxt}
R.~R. Akhmetshin et~al.
\newblock {High-statistics measurement of the pion form factor in the rho-meson energy range with the CMD-2 detector}.
\newblock {\em Phys. Lett. B}, 648:28--38, 2007.

\bibitem{Achasov:2006vp}
M.~N. Achasov et~al.
\newblock {Update of the $e^+ e^- \to \pi^+ \pi^-$ cross-section measured by SND detector in the energy region 400 \ensuremath{<} $\sqrt{s}$ \ensuremath{<} 1000 MeV}.
\newblock {\em J. Exp. Theor. Phys.}, 103:380--384, 2006.

\bibitem{SND:2020nwa}
M.~N. Achasov et~al.
\newblock {Measurement of the $e^+e^- \to\pi^+\pi^- $ process cross section with the SND detector at the VEPP-2000 collider in the energy region $0.525<\sqrt{s}<0.883$ GeV}.
\newblock {\em JHEP}, 01:113, 2021.

\bibitem{KLOE:2008fmq}
F.~Ambrosino et~al.
\newblock {Measurement of $\sigma(e^+ e^- \to \pi^+ \pi^- \gamma(\gamma)$ and the dipion contribution to the muon anomaly with the KLOE detector}.
\newblock {\em Phys. Lett. B}, 670:285--291, 2009.

\bibitem{KLOE:2010qei}
F.~Ambrosino et~al.
\newblock {Measurement of $\sigma(e^+ e^- \to \pi^+ \pi^-)$ from threshold to 0.85 GeV$^2$ using Initial State Radiation with the KLOE detector}.
\newblock {\em Phys. Lett. B}, 700:102--110, 2011.

\bibitem{BESIII:2015equ}
M.~Ablikim et~al.
\newblock {Measurement of the $e^+ e^- \to \pi^+ \pi^-$ cross section between 600 and 900 MeV using initial state radiation}.
\newblock {\em Phys. Lett. B}, 753:629--638, 2016.
\newblock [Erratum: Phys.Lett.B 812, 135982 (2021)].

\bibitem{CMD-3:2023alj}
F.~V. Ignatov et~al.
\newblock {Measurement of the $e^+e^-$\textrightarrow{}\ensuremath{\pi^+}\ensuremath{\pi^-} cross section from threshold to 1.2~GeV with the CMD-3 detector}.
\newblock {\em Phys. Rev. D}, 109(11):112002, 2024.

\bibitem{CMD-3:2023rfe}
F.~V. Ignatov et~al.
\newblock {Measurement of the Pion Form Factor with CMD-3 Detector and its Implication to the Hadronic Contribution to Muon $(g-2)$}.
\newblock {\em Phys. Rev. Lett.}, 132(23):231903, 2024.

\bibitem{ALEPH:2005qgp}
S.~Schael et~al.
\newblock {Branching ratios and spectral functions of tau decays: Final ALEPH measurements and physics implications}.
\newblock {\em Phys. Rept.}, 421:191--284, 2005.

\bibitem{Belle:2008xpe}
M.~Fujikawa et~al.
\newblock {High-Statistics Study of the $\tau^- \to \pi^- \pi^0 \nu_\tau$ Decay}.
\newblock {\em Phys. Rev. D}, 78:072006, 2008.

\bibitem{CLEO:1999dln}
S.~Anderson et~al.
\newblock {Hadronic structure in the decay $\tau^- \to \pi^- \pi^0 \nu_\tau$}.
\newblock {\em Phys. Rev. D}, 61:112002, 2000.

\bibitem{OPAL:1998rrm}
K.~Ackerstaff et~al.
\newblock {Measurement of the strong coupling constant $\alpha_S$ and the vector and axial-vector spectral functions in hadronic tau decays}.
\newblock {\em Eur. Phys. J. C}, 7:571--593, 1999.

\bibitem{Flores-Baez:2006yiq}
F.~Flores-Báez, A.~Flores-Tlalpa, G.~López~Castro, and G.~Toledo~Sánchez.
\newblock {Long-distance radiative corrections to the di-pion tau lepton decay}.
\newblock {\em Phys. Rev. D}, 74:071301, 2006.

\bibitem{Flores-Baez:2007vnd}
F.~V. Flores-Báez, G.~López Castro, and G.~Toledo~Sánchez.
\newblock {The Width difference of $\rho$ vector mesons}.
\newblock {\em Phys. Rev. D}, 76:096010, 2007.

\bibitem{Miranda:2020wdg}
J.~A. Miranda and P.~Roig.
\newblock {New $\tau$-based evaluation of the hadronic contribution to the vacuum polarization piece of the muon anomalous magnetic moment}.
\newblock {\em Phys. Rev. D}, 102:114017, 2020.

\bibitem{Masjuan:2023qsp}
Pere Masjuan, Alejandro Miranda, and Pablo Roig.
\newblock {\ensuremath{\tau} data-driven evaluation of Euclidean windows for the hadronic vacuum polarization}.
\newblock {\em Phys. Lett. B}, 850:138492, 2024.

\bibitem{Davier:2023fpl}
Michel Davier, Andreas Hoecker, Anne-Marie Lutz, Bogdan Malaescu, and Zhiqing Zhang.
\newblock {Tensions in $e^+e^-\rightarrow \pi ^+\pi ^-(\gamma )$ measurements: the new landscape of data-driven hadronic vacuum polarization predictions for the muon $g - 2$}.
\newblock {\em Eur. Phys. J. C}, 84(7):721, 2024.

\bibitem{Maltman:2005yk}
Kim Maltman.
\newblock {Constraints on the $I=1$ hadronic $\tau$ decay and $e^+e^- \to$ hadrons data sets and implications for $(g-2)_\mu$}.
\newblock {\em Phys. Lett. B}, 633:512--518, 2006.

\bibitem{Maltman:2005qq}
Kim Maltman and Carl~E. Wolfe.
\newblock {Isospin breaking in the relation between the $\tau^- \to \nu_\tau \pi^- \pi^0$ and $e^+ e^- \to \pi^+ \pi^-$ versions of $|F_\pi(s)|^2$ and implications for $(g-2)_\mu$}.
\newblock {\em Phys. Rev. D}, 73:013004, 2006.

\bibitem{Davier:2010nc}
Michel Davier, Andreas Hoecker, Bogdan Malaescu, and Zhiqing Zhang.
\newblock {Reevaluation of the Hadronic Contributions to the Muon $g-2$ and to $\alpha(M_Z^2)$}.
\newblock {\em Eur. Phys. J. C}, 71:1515, 2011.
\newblock [Erratum: Eur.Phys.J.C 72, 1874 (2012)].

\bibitem{Davier:2013sfa}
Michel Davier, Andreas H\"ocker, Bogdan Malaescu, Chang-Zheng Yuan, and Zhiqing Zhang.
\newblock {Update of the ALEPH non-strange spectral functions from hadronic $\tau$ decays}.
\newblock {\em Eur. Phys. J. C}, 74(3):2803, 2014.

\bibitem{Bruno:2018ono}
Mattia Bruno, Taku Izubuchi, Christoph Lehner, and Aaron Meyer.
\newblock {On isospin breaking in $\tau$ decays for $(g-2)_\mu$ from Lattice QCD}.
\newblock {\em PoS}, LATTICE2018:135, 2018.

\bibitem{Narison:2023srj}
Stephan Narison.
\newblock {QCD parameters and SM-high precisions from e$^+$e$^-$ $\to$ Hadrons}.
\newblock {\em Nucl. Phys. A}, 1039:122744, 2023.

\bibitem{Esparza-Arellano:2023dps}
Leonardo Esparza-Arellano, Antonio Rojas, and Genaro Toledo.
\newblock {Sizing the double pole resonant enhancement in $e^+e^-\to\ensuremath{\pi}^0\ensuremath{\pi}^0\ensuremath{\gamma}$ cross section and $\ensuremath{\tau}^-\to\ensuremath{\pi}^-\ensuremath{\pi}^0\ensuremath{\nu}_\ensuremath{\tau}\ensuremath{\gamma}$ decay}.
\newblock {\em Phys. Rev. D}, 108(11):116020, 2023.

\bibitem{Boccaletti:2024guq}
A.~Boccaletti et~al.
\newblock {High precision calculation of the hadronic vacuum polarisation contribution to the muon anomaly}.
\newblock 7 2024.

\bibitem{Borsanyi:2020mff}
Sz. Borsanyi et~al.
\newblock {Leading hadronic contribution to the muon magnetic moment from lattice QCD}.
\newblock {\em Nature}, 593(7857):51--55, 2021.

\bibitem{Djukanovic:2024cmq}
Dalibor Djukanovic, Georg von Hippel, Simon Kuberski, Harvey~B. Meyer, Nolan Miller, Konstantin Ottnad, Julian Parrino, Andreas Risch, and Hartmut Wittig.
\newblock {The hadronic vacuum polarization contribution to the muon $g-2$ at long distances}.
\newblock 11 2024.

\bibitem{Bazavov:2024eou}
Alexei Bazavov et~al.
\newblock {Hadronic vacuum polarization for the muon $g-2$ from lattice QCD: Long-distance and full light-quark connected contribution}.
\newblock 12 2024.

\bibitem{Miranda:2018cpf}
J.~A. Miranda and P.~Roig.
\newblock {Effective-field theory analysis of the $\tau^-\to \pi^-\pi^0\nu_\tau$ decays}.
\newblock {\em JHEP}, 11:038, 2018.

\bibitem{Cirigliano:2018dyk}
Vincenzo Cirigliano, Adam Falkowski, Mart\'\i{}n Gonz\'alez-Alonso, and Antonio Rodr\'\i{}guez-S\'anchez.
\newblock {Hadronic \ensuremath{\tau} Decays as New Physics Probes in the LHC Era}.
\newblock {\em Phys. Rev. Lett.}, 122(22):221801, 2019.

\bibitem{Gonzalez-Solis:2020jlh}
Sergi Gonz\`alez-Sol\'\i{}s, Alejandro Miranda, Javier Rend\'on, and Pablo Roig.
\newblock {Exclusive hadronic tau decays as probes of non-SM interactions}.
\newblock {\em Phys. Lett. B}, 804:135371, 2020.

\bibitem{Cirigliano:2021yto}
Vincenzo Cirigliano, David D\'\i{}az-Calder\'on, Adam Falkowski, Mart\'\i{}n Gonz\'alez-Alonso, and Antonio Rodr\'\i{}guez-S\'anchez.
\newblock {Semileptonic tau decays beyond the Standard Model}.
\newblock {\em JHEP}, 04:152, 2022.

\bibitem{Buchmuller:1985jz}
W.~Buchmuller and D.~Wyler.
\newblock {Effective Lagrangian Analysis of New Interactions and Flavor Conservation}.
\newblock {\em Nucl. Phys. B}, 268:621--653, 1986.

\bibitem{Grzadkowski:2010es}
B.~Grzadkowski, M.~Iskrzynski, M.~Misiak, and J.~Rosiek.
\newblock {Dimension-Six Terms in the Standard Model Lagrangian}.
\newblock {\em JHEP}, 10:085, 2010.

\bibitem{Passera:2008jk}
M.~Passera, W.~J. Marciano, and A.~Sirlin.
\newblock {The Muon $g-2$ and the bounds on the Higgs boson mass}.
\newblock {\em Phys. Rev. D}, 78:013009, 2008.

\bibitem{Gounaris:1968mw}
G.~J. Gounaris and J.~J. Sakurai.
\newblock {Finite width corrections to the vector meson dominance prediction for $\rho \to e^+ e^-$}.
\newblock {\em Phys. Rev. Lett.}, 21:244--247, 1968.

\bibitem{Bouchiat:1961lbg}
Claude Bouchiat and Louis Michel.
\newblock {La r\'esonance dans la diffusion m\'eson \ensuremath{\pi}\textemdash{} m\'eson \ensuremath{\pi} et le moment magn\'etique anormal du m\'eson \ensuremath{\mu}}.
\newblock {\em J. Phys. Radium}, 22(2):121--121, 1961.

\bibitem{Durand:1962zzb}
Loyal Durand.
\newblock {Pionic Contributions to the Magnetic Moment of the Muon}.
\newblock {\em Phys. Rev.}, 128:441--448, 1962.

\bibitem{Brodsky:1967sr}
Stanley~J. Brodsky and Eduardo De~Rafael.
\newblock {SUGGESTED BOSON - LEPTON PAIR COUPLINGS AND THE ANOMALOUS MAGNETIC MOMENT OF THE MUON}.
\newblock {\em Phys. Rev.}, 168:1620--1622, 1968.

\bibitem{Gourdin:1969dm}
M.~Gourdin and E.~De~Rafael.
\newblock {Hadronic contributions to the muon g-factor}.
\newblock {\em Nucl. Phys. B}, 10:667--674, 1969.

\bibitem{Eidelman:1995ny}
S.~Eidelman and F.~Jegerlehner.
\newblock {Hadronic contributions to $g-2$ of the leptons and to the effective fine structure constant $\alpha (M_Z^2)$}.
\newblock {\em Z. Phys. C}, 67:585--602, 1995.

\bibitem{Lautrup:1968tdb}
B.~E. Lautrup and E.~De~Rafael.
\newblock {Calculation of the sixth-order contribution from the fourth-order vacuum polarization to the difference of the anomalous magnetic moments of muon and electron}.
\newblock {\em Phys. Rev.}, 174:1835--1842, 1968.

\bibitem{Sirlin:1974ni}
A.~Sirlin.
\newblock {Radiative corrections to $G_V/G_\mu$ in simple extensions of the $SU(2) \times U(1)$ gauge model}.
\newblock {\em Nucl. Phys. B}, 71:29--51, 1974.

\bibitem{Sirlin:1977sv}
A.~Sirlin.
\newblock {Current Algebra Formulation of Radiative Corrections in Gauge Theories and the Universality of the Weak Interactions}.
\newblock {\em Rev. Mod. Phys.}, 50:573, 1978.
\newblock [Erratum: Rev.Mod.Phys. 50, 905 (1978)].

\bibitem{Sirlin:1981ie}
A.~Sirlin.
\newblock {Large $m_W$, $m_Z$ Behavior of the $O(\alpha)$ Corrections to Semileptonic Processes Mediated by $W$}.
\newblock {\em Nucl. Phys. B}, 196:83--92, 1982.

\bibitem{Marciano:1985pd}
W.~J. Marciano and A.~Sirlin.
\newblock {Radiative Corrections to beta Decay and the Possibility of a Fourth Generation}.
\newblock {\em Phys. Rev. Lett.}, 56:22, 1986.

\bibitem{Marciano:1988vm}
W.~J. Marciano and A.~Sirlin.
\newblock {Electroweak Radiative Corrections to $\tau$ decay}.
\newblock {\em Phys. Rev. Lett.}, 61:1815--1818, 1988.

\bibitem{Marciano:1993sh}
William~J. Marciano and A.~Sirlin.
\newblock {Radiative corrections to $\pi_{\ell 2}$ decays}.
\newblock {\em Phys. Rev. Lett.}, 71:3629--3632, 1993.

\bibitem{Braaten:1990ef}
Eric Braaten and Chong-Sheng Li.
\newblock {Electroweak radiative corrections to the semihadronic decay rate of the tau lepton}.
\newblock {\em Phys. Rev. D}, 42:3888--3891, 1990.

\bibitem{Erler:2002mv}
Jens Erler.
\newblock {Electroweak radiative corrections to semileptonic tau decays}.
\newblock {\em Rev. Mex. Fis.}, 50:200--202, 2004.

\bibitem{Cirigliano:2023fnz}
Vincenzo Cirigliano, Wouter Dekens, Emanuele Mereghetti, and Oleksandr Tomalak.
\newblock {Effective field theory for radiative corrections to charged-current processes: Vector coupling}.
\newblock {\em Phys. Rev. D}, 108(5):053003, 2023.

\bibitem{Schwinger:1989ix}
Julian~S. Schwinger.
\newblock {\em {PARTICLES, SOURCES, AND FIELDS. VOL. 3}}.
\newblock 1989.

\bibitem{Drees:1990te}
Manuel Drees and Ken-ichi Hikasa.
\newblock {Scalar top production in $e^+e^-$ annihilation}.
\newblock {\em Phys. Lett. B}, 252:127--134, 1990.

\bibitem{Flores-Tlalpa:2006snz}
A.~Flores-Tlalpa, F.~Flores-Báez, G.~López~Castro, and G.~Toledo~Sánchez.
\newblock {Model-dependent radiative corrections to $\tau^- \to \pi^- \pi^0 \nu$ revisited}.
\newblock {\em Nucl. Phys. B Proc. Suppl.}, 169:250--254, 2007.

\bibitem{Ecker:1988te}
G.~Ecker, J.~Gasser, A.~Pich, and E.~de~Rafael.
\newblock {The Role of Resonances in Chiral Perturbation Theory}.
\newblock {\em Nucl. Phys. B}, 321:311--342, 1989.

\bibitem{Ecker:1989yg}
G.~Ecker, J.~Gasser, H.~Leutwyler, A.~Pich, and E.~de~Rafael.
\newblock {Chiral Lagrangians for Massive Spin 1 Fields}.
\newblock {\em Phys. Lett. B}, 223:425--432, 1989.

\bibitem{Cirigliano:2006hb}
V.~Cirigliano, G.~Ecker, M.~Eidemüller, Roland Kaiser, A.~Pich, and J.~Portolés.
\newblock {Towards a consistent estimate of the chiral low-energy constants}.
\newblock {\em Nucl. Phys. B}, 753:139--177, 2006.

\bibitem{Kampf:2011ty}
Karol Kampf and Jiri Novotny.
\newblock {Resonance saturation in the odd-intrinsic parity sector of low-energy QCD}.
\newblock {\em Phys. Rev. D}, 84:014036, 2011.

\bibitem{Roig:2013baa}
Pablo Roig and Juan~Jos\'e Sanz~Cillero.
\newblock {Consistent high-energy constraints in the anomalous QCD sector}.
\newblock {\em Phys. Lett. B}, 733:158--163, 2014.

\bibitem{GomezDumm:2013sib}
D.~G\'omez~Dumm and P.~Roig.
\newblock {Dispersive representation of the pion vector form factor in $\tau\to\pi\pi\nu_\tau$ decays}.
\newblock {\em Eur. Phys. J. C}, 73(8):2528, 2013.

\bibitem{Gonzalez-Solis:2019iod}
Sergi Gonz\`alez-Sol\'\i{}s and Pablo Roig.
\newblock {A dispersive analysis of the pion vector form factor and $\tau ^{-}\rightarrow K^{-}K_{S}\nu _{\tau }$ decay}.
\newblock {\em Eur. Phys. J. C}, 79(5):436, 2019.

\bibitem{Ademollo:1964sr}
M.~Ademollo and Raoul Gatto.
\newblock {Nonrenormalization Theorem for the Strangeness Violating Vector Currents}.
\newblock {\em Phys. Rev. Lett.}, 13:264--265, 1964.

\bibitem{Kuhn:1990ad}
Johann~H. Kühn and A.~Santamaría.
\newblock {$\tau$ Decays to pions}.
\newblock {\em Z. Phys. C}, 48:445--452, 1990.

\bibitem{Guerrero:1997ku}
Francisco Guerrero and Antonio Pich.
\newblock {Effective field theory description of the pion form-factor}.
\newblock {\em Phys. Lett. B}, 412:382--388, 1997.

\bibitem{RuizArriola:2024gwb}
Enrique Ruiz~Arriola and Pablo Sánchez-Puertas.
\newblock {Phase of the electromagnetic form factor of the pion}.
\newblock {\em Phys. Rev. D}, 110(5):054003, 2024.

\bibitem{Kirk:2024oyl}
Matthew Kirk, Bastian Kubis, M\'eril Reboud, and Danny van Dyk.
\newblock {A Simple Parametrisation of the Pion Form Factor}.
\newblock 10 2024.

\bibitem{Omnes:1958hv}
R.~Omnès.
\newblock {On the Solution of certain singular integral equations of quantum field theory}.
\newblock {\em Nuovo Cim.}, 8:316--326, 1958.

\bibitem{Leutwyler:2002hm}
H.~Leutwyler.
\newblock {Electromagnetic form-factor of the pion}.
\newblock In {\em {Continuous Advances in QCD 2002 / ARKADYFEST (honoring the 60th birthday of Prof. Arkady Vainshtein)}}, pages 23--40, 12 2002.

\bibitem{Colangelo:2022prz}
Gilberto Colangelo, Martin Hoferichter, Bastian Kubis, and Peter Stoffer.
\newblock {Isospin-breaking effects in the two-pion contribution to hadronic vacuum polarization}.
\newblock {\em JHEP}, 10:032, 2022.

\bibitem{Hoferichter:2023sli}
Martin Hoferichter, Gilberto Colangelo, Bai-Long Hoid, Bastian Kubis, Jacobo~Ruiz de~Elvira, Dominic Schuh, Dominik Stamen, and Peter Stoffer.
\newblock {Phenomenological Estimate of Isospin Breaking in Hadronic Vacuum Polarization}.
\newblock {\em Phys. Rev. Lett.}, 131(16):161905, 2023.

\bibitem{Hoferichter:2023bjm}
Martin Hoferichter, Bai-Long Hoid, Bastian Kubis, and Dominic Schuh.
\newblock {Isospin-breaking effects in the three-pion contribution to hadronic vacuum polarization}.
\newblock {\em JHEP}, 08:208, 2023.

\bibitem{Stoffer:2023gba}
Peter Stoffer, Gilberto Colangelo, and Martin Hoferichter.
\newblock {Puzzles in the hadronic contributions to the muon anomalous magnetic moment}.
\newblock {\em JINST}, 18(10):C10021, 2023.

\bibitem{Gilman:1984ry}
Frederick~J. Gilman and Sun~Hong Rhie.
\newblock {Calculation of Exclusive Decay Modes of the tau}.
\newblock {\em Phys. Rev. D}, 31:1066, 1985.

\bibitem{Sobie:1995kp}
R.~J. Sobie.
\newblock {The CVC prediction for the $\tau^- \to h^- \pi^0\nu_\tau$ branching ratio}.
\newblock {\em Z. Phys. C}, 65:79--86, 1995.

\bibitem{Bernicha:1995rh}
A.~Bernicha, G.~López~Castro, and J.~Pestieau.
\newblock {$S$-matrix approach to two pion production in $e^+ e^-$ annihilation and $\tau$ decay}.
\newblock {\em Phys. Rev. D}, 53:4089--4091, 1996.

\bibitem{Wang:2023njt}
Shi-Jia Wang, Zhen Fang, and Ling-Yun Dai.
\newblock {Two body final states production in electron-positron annihilation and their contributions to (g \ensuremath{-} 2)$_{\mu}$}.
\newblock {\em JHEP}, 07:037, 2023.

\bibitem{Qin:2024ulb}
Bing-Hai Qin, Wen Qin, and Ling-Yun Dai.
\newblock {Study of electron-positron annihilation into $K\bar{K}\pi$ within resonance chiral theory}.
\newblock 3 2024.

\bibitem{Roig:2014uja}
P.~Roig, A.~Guevara, and G.~L\'opez~Castro.
\newblock {$VV^\prime P$ form factors in resonance chiral theory and the $\pi-\eta-\eta^\prime$ light-by-light contribution to the muon $g-2$}.
\newblock {\em Phys. Rev. D}, 89(7):073016, 2014.

\bibitem{Guevara:2018rhj}
A.~Guevara, P.~Roig, and J.~J. Sanz-Cillero.
\newblock {Pseudoscalar pole light-by-light contributions to the muon $(g-2)$ in Resonance Chiral Theory}.
\newblock {\em JHEP}, 06:160, 2018.

\bibitem{Estrada:2024cfy}
Emilio~J. Estrada, Sergi Gonz\`alez-Sol\'\i{}s, Adolfo Guevara, and Pablo Roig.
\newblock {Improved $\pi^0,\eta,\eta^{\prime}$ transition form factors in resonance chiral theory and their $a_\mu^{\rm{HLbL}}$ contribution}.
\newblock 9 2024.

\bibitem{Arbuzov:1997pj}
A.~B. Arbuzov, G.~V. Fedotovich, E.~A. Kuraev, N.~P. Merenkov, V.~D. Rushai, and L.~Trentadue.
\newblock {Large angle QED processes at $e^+ e^-$ colliders at energies below 3 GeV}.
\newblock {\em JHEP}, 10:001, 1997.

\bibitem{web:Fedor}
F.~Ignatov.
\newblock https://cmd.inp.nsk.su/$\sim$ignatov/vpl/.

\bibitem{Gan:2020aco}
Liping Gan, Bastian Kubis, Emilie Passemar, and Sean Tulin.
\newblock {Precision tests of fundamental physics with \ensuremath{\eta} and \ensuremath{\eta}' mesons}.
\newblock {\em Phys. Rept.}, 945:1--105, 2022.

\end{thebibliography}

\end{document}